\documentclass[seceq]{ptptex}

\usepackage{amsmath}
\usepackage{amsfonts}
\usepackage{latexsym}
\usepackage{amssymb} 
\usepackage{verbatim}
\usepackage{amsthm}
\usepackage{graphicx}
\usepackage{wrapft}

\newcommand{\RF}{{{\mathbb R}}}

\newtheorem{conjecture}{Conjecture}[section]

\title{
  Decomposition of linear metric perturbations\\
  on generic background spacetime
}

\subtitle{
  Toward higher-order general-relativistic \\
  gauge-invariant perturbation theory
}

\author{
  Kouji \textsc{Nakamura}%
}

\inst{
  Optical and Infrared Astronomy Division,\\
  National Astronomical Observatory of Japan,\\
  Osawa 2-21-1, Mitaka 181-8588, Japan
}

\recdate{\today}

\abst{
  The decomposition of the linear-order metric perturbation is 
  discussed in the context of the higher-order gauge-invariant
  perturbation theory.
  We show that the linear order metric perturbation is
  decomposed into gauge-invariant and gauge-variant parts on the
  general background spacetime which admits ADM decomposition.
  This decomposition was an important premise of the general
  framework of the higher order gauge-invariant perturbation
  theory proposed in the papers [K.~Nakamura,
  Prog.~Theor.~Phys.\ {\bf 110} (2003), 723; {\it ibid}.
  {\bf 113} (2005), 481].
  This implies that we can develop the higher-order
  gauge-invariant perturbation theory on generic background
  spacetime.
  Remaining issues to complete the general-framework of the
  higher-order gauge-invariant perturbation theories are also
  discussed.
}

\begin{document}

\maketitle

\section{Introduction}
\label{sec:intro}


Perturbation theories are powerful techniques in many area of
physics and the developments of perturbation theories lead
physically fruitful results and interpretations of natural
phenomena.


In physics, physicists want to describe realistic situations in
a compact manner.
Exact solutions in a theory for physical situations are
candidates which can describe realistic natural phenomena. 
However, in many theories of physics, realistic situations are
too complicated and often difficult to describe by an exact
solution of a theory.
This difficulty may be due to the fact that exact solutions only
describe special cases even if the theory is appropriate to
describe the natural phenomena, or may be due to the lack of the
applicability of the theory itself.
Even in the case where an exact solution of a theory well
describes a physical situation, the properties of the physical
system will not be completely described only through the exact
solution. 
In natural phenomena, there always exist ``fluctuations''.
In this case, perturbative treatments of the theory is a
powerful tools and physicists investigate perturbative approach
within the theory to clarify the properties of fluctuations.


General relativity is a theory in which the construction of
exact solutions is not so easy.
Although there are many exact solutions to the Einstein
equation\cite{H.Stephani-D.Kramer-M.A.N.MacCallum-C.Hoenselaers-E.Herlt-2003}
these are often too idealized.
Of course, there are some exact solutions to the Einstein
equation which well-describe our universe, or gravitational
field of stars and black holes.
These exact solutions by itself do not describe the fluctuations
around these exact solutions.
To describe these fluctuations, we have to consider the
perturbations around these exact solutions. 
Therefore, general relativistic linear perturbation theory is a 
useful technique to investigate the properties of fluctuations
around exact solutions\cite{Bardeen-1980}.


On the other hand, higher-order general-relativistic
perturbations also have very wide applications.
In these applications, second-order cosmological perturbations 
are topical
subject\cite{Tomita-1967-Non-Gaussianity,M.Bruni-S.Soonego-CQG1997,S.Sonego-M.Bruni-CMP1998,Nakamura:2010yg,kouchan-cosmo-second}
due to the precise measurements in recent
cosmology\cite{Non-Gaussianity-observation}. 
Higher-order black hole perturbations are also discussed in some
literature\cite{Gleiser-Nicasio}.
Moreover, as a special example of higher-order perturbation
theory, there are researches on perturbations of a spherical
star\cite{Kojima} motivated by the researches on the oscillatory
behaviors of a rotating neutron star.
Thus, there are many physical situations to which general
relativistic higher-order perturbation theory should be
applied.


As well-known, general relativity is based on the concept of
general covariance.
Intuitively speaking, the principle of general covariance states
that there is no preferred coordinate system in nature, though
the notion of general covariance is mathematically included in
the definition of a spacetime manifold in a trivial way.
This is based on the philosophy that coordinate systems are
originally chosen by us, and that natural phenomena have nothing
to do with our coordinate system.
Due to this general covariance, the ``gauge degree of freedom'', 
which is an unphysical degree of freedom of perturbations,
arises in general-relativistic perturbations.
To obtain physically results, we have to fix this gauge degrees
of freedom or to extract some invariant quantities of
perturbations.
This situation becomes more complicated in higher-order
perturbation theory.
In some linear perturbation theories on some background
spacetimes, there are so-called {\it gauge-invariant}
perturbation theories.
In these theories, one may treat only variables which are
independent of gauge degree of freedom without any gauge
fixing. 
Therefore, it is worthwhile to investigate higher-order
gauge-invariant perturbation theory from a general point of
view to avoid gauge issues.


According to these motivation, the general framework of
higher-order general-relativistic gauge-invariant perturbation
theory has been discussed in some 
papers\cite{kouchan-gauge-inv,kouchan-second} by the present
author.
We refer these works as KN2003\cite{kouchan-gauge-inv} and
KN2005\cite{kouchan-second}. 
Although these development of higher-order perturbation theory
was originally motivated by the research on the oscillatory
behavior of a self-gravitating Nambu-Goto
membrane\cite{kouchan-papers}, these works are applicable to
cosmological perturbations and we clarified the gauge-invariance
of the second-order perturbations of the Einstein
equations\cite{kouchan-cosmo-second,kouchan-second-cosmo-matter,kouchan-second-cosmo-consistency}. 
In this paper, we refer these works as
KN2007\cite{kouchan-cosmo-second} and
KN2009\cite{kouchan-second-cosmo-matter}.


In KN2003\cite{kouchan-gauge-inv}, we proposed the procedure to
find gauge-invariant variables for higher-order perturbations on
a generic background spacetime.
This proposal is based on the single assumption that {\it we
  have already known the procedure to find gauge-invariant
  variables for the linear-order metric perturbation.} 
Under the same assumption, we summarize some formulae for the
second-order perturbations of the curvatures and energy-momentum
tensor for the matter fields in KN2005\cite{kouchan-second} and 
KN2009\cite{kouchan-second-cosmo-matter}.
In KN2007\cite{kouchan-cosmo-second}, we develop the
second-order gauge-invariant cosmological perturbation theory
after confirming that the above assumption is correct in the
case of cosmological perturbations.
Through these works, we find that our general framework of
higher-order gauge-invariant perturbation theory is well-defined  
except for the above assumption for linear-order metric
perturbations.
Therefore, we proposed the above assumption as a conjecture in
KN2009\cite{kouchan-second-cosmo-matter}.
If this conjecture is true, higher-order general-relativistic
gauge-invariant perturbation theory is completely formulated on
generic background spacetime and has very wide applications.


The main purpose of this paper is to give a proof of this
conjecture using the premise that the background spacetime
admits ADM decomposition.
Although some special modes are excluded in the proof in this
paper, we may say that the above conjecture is almost correct
for linear-order perturbations on generic background
spacetimes.
This paper is the complete version of our previous short
letter\cite{kouchan-decomp-letter-version}.


The organization of this paper is as follows.
In
\S\ref{sec:General-framework-of-the-gauge-invariant-perturbation-theory},
we review the general framework of the second-order
gauge-invariant perturbation theory developed in  
KN2003\cite{kouchan-gauge-inv} and KN2005\cite{kouchan-second}
with some additional explanations.
In the context of this general framework, the above conjecture
is also declared as Conjecture
\ref{conjecture:decomposition-conjecture} in this section.
In
\S\ref{sec:K.Nakamura-2010-3}, we give a proof of Conjecture
\ref{conjecture:decomposition-conjecture}.
From pedagogical point of view, we consider three different 
situations of the geometry of the background spacetime in terms 
of ADM decomposition.
The first situation is trivial
(\S\ref{sec:trivial-simple-case}), in which we may choose the
unit lapse function $\alpha=1$, the vanishing shift vector
$\beta^{i}=0$, and the vanishing extrinsic curvature $K_{ij}=0$.
Through this trivial case, we give the essential outline of the
proof in more generic situations.
The second situation is the case where $\alpha=1$,
$\beta^{i}=0$, but $K_{ij}\neq 0$
(\S\ref{sec:synchorous-comoving-BG-case}).
Through this second case, we show a technical issue to prove
the conjecture \ref{conjecture:decomposition-conjecture} in
terms of ADM decomposition.
The final situations is most generic case, in which 
$\alpha\neq 1$, $\beta^{i}\neq 0$, $K_{ij}\neq 0$
(\S\ref{sec:most-generic-case}). 
The calculations in this case is complicated.
However the essential outline of the proof is same as in the
first trivial case, and the essential technique using in the
proof is already given in the second case.
We also note that we assume that the existence of Green functions 
for two elliptic differential operators in these proofs.
The comparison with the proof in the case of cosmological
perturbations shown in KN2007\cite{kouchan-cosmo-second} is
discussed in \S\ref{sec:K.Nakamura-2010-4}.
The final section is devoted to summary and discussion.


We employ the notation of KN2003 and KN2005 and use abstract
index notation\cite{Wald-book}. 
We also employ natural units in which Newton's gravitational 
constant is denoted by $G$ and the velocity of light satisfies
$c=1$.


\section{General framework of the higher-order gauge-invariant perturbation theory}
\label{sec:General-framework-of-the-gauge-invariant-perturbation-theory}


In this section, we review the general framework of the
gauge-invariant perturbation theory developed in
KN2003\cite{kouchan-gauge-inv} and KN2005\cite{kouchan-second}
to emphasize that Conjecture
\ref{conjecture:decomposition-conjecture} is the important
premise of our general framework.
In \S\ref{sec:Gauge-degree-of-freedom-in-perturbation-theory},
we review the basic understanding of the {\it gauge degree of freedom}
in general relativistic perturbation theory based on the work of
Stewart et al.\cite{J.M.Stewart-M.Walker11974} and Bruni et
al.\cite{M.Bruni-S.Soonego-CQG1997}.
When we consider perturbations in any theory with general
covariance, we have to exclude these gauge degrees of freedom in
the perturbations.
To accomplish this, {\it gauge-invariant variables} of
perturbations are useful, and these are regarded as physically
meaningful quantities.
In \S\ref{sec:gauge-invariant-variables}, we review the
procedure for finding gauge-invariant variables of
perturbations, which was developed in
KN2003\cite{kouchan-gauge-inv}.
After the introduction of gauge-invariant variables, in
\S\ref{sec:gauge-invariant-perturbation-theory},
we review the general issue of the gauge-invariant formulation
for the second-order perturbation of the Einstein equation
developed in KN2005\cite{kouchan-second}.
We emphasize that the ingredients of this section do not
depend on the details of the background spacetime, if the
decomposition conjecture
\ref{conjecture:decomposition-conjecture} for the linear-order
metric perturbation is correct.


\subsection{Gauge degree of freedom in perturbation theory}
\label{sec:Gauge-degree-of-freedom-in-perturbation-theory}


\subsubsection{Basic idea}
\label{sec:Gauge-degree-of-freedom-in-perturbation-theory-basic}


Here, we explain the concept of gauge in general relativistic 
perturbation theory.
To explain this, we first point out that, in any perturbation
theory, we always treat two spacetime manifolds. 
One is the physical spacetime ${\cal M}$, which we attempt to
describe in terms of perturbations, and the other is the
background spacetime ${\cal M}_{0}$, which is a fictitious
manifold prepared for perturbative analyses by hand.
We emphasize that these two spacetime manifolds
${\cal M}$ and ${\cal M}_{0}$ are distinct.
Let us denote the physical spacetime by $({\cal M},\bar{g}_{ab})$
and the background spacetime by $({\cal M}_{0},g_{ab})$, where 
$\bar{g}_{ab}$ is the metric on ${\cal M}$, and $g_{ab}$ is the
metric on ${\cal M}_{0}$.
Further, we formally denote the spacetime metric and the other
physical tensor fields on the physical spacetime by $Q$ and its
background value on the background spacetime by $Q_{0}$.


Second, in any perturbation theories, we always write equations
for the perturbation of the physical variable $Q$ in the form
\begin{equation}
  \label{eq:variable-symbolic-perturbation}
  Q(``p\mbox{''}) = Q_{0}(p) + \delta Q(p).
\end{equation}
Usually, this equation is simply regarded as a relation
between the physical variable $Q$ and its background value
$Q_{0}$, or as the definition of the deviation $\delta Q$ of
the physical variable $Q$ from its background value $Q_{0}$.
However, Eq.~(\ref{eq:variable-symbolic-perturbation}) has
deeper implications.
Keeping in our mind the above fact that we always treat two
different spacetimes, $({\cal M},\bar{g}_{ab})$ and 
$({\cal M}_{0},g_{ab})$, in perturbation theory,
Eq.~(\ref{eq:variable-symbolic-perturbation}) is a rather 
curious equation in the following sense: 
The variable on the left-hand side of
Eq.~(\ref{eq:variable-symbolic-perturbation}) is a variable on
the physical spacetime $({\cal M},\bar{g}_{ab})$, while the
variables on the right-hand 
side of Eq.~(\ref{eq:variable-symbolic-perturbation}) are
variables on the background spacetime, $({\cal M}_{0},g_{ab})$.
Hence, Eq.~(\ref{eq:variable-symbolic-perturbation}) gives a
relation between variables on two different manifolds.


Further, we point out the fact that, through
Eq.~(\ref{eq:variable-symbolic-perturbation}), we have
implicitly identified points in two different manifolds 
$({\cal M},\bar{g}_{ab})$ and $({\cal M}_{0},g_{ab})$.
More specifically, $Q(``p\mbox{''})$ on the left-hand side of
Eq.~(\ref{eq:variable-symbolic-perturbation}) is a field on 
${\cal M}$, and $``p\mbox{''}\in{\cal M}$.
Similarly, we should regard the background value
$Q_{0}(p)$ of $Q(``p\mbox{''})$ and its deviation $\delta Q(p)$
of $Q(``p\mbox{''})$ from $Q_{0}(p)$, which are on the
right-hand side of
Eq.~(\ref{eq:variable-symbolic-perturbation}), as fields on
${\cal M}_{0}$, and $p\in{\cal M}_{0}$.
Because Eq.~(\ref{eq:variable-symbolic-perturbation}) is
regarded as an equation for field variables, it implicitly
states that the points $``p\mbox{''}\in{\cal M}$ and 
$p\in{\cal M}_{0}$ are same.
Therefore, through
Eq.~(\ref{eq:variable-symbolic-perturbation}), we implicitly
assume the existence of a map 
${\cal M}_{0}\rightarrow{\cal M}$ $:$ $p\in{\cal M}_{0}\mapsto
``p\mbox{''}\in{\cal M}$, which is called a {\it gauge choice}
in perturbation theory\cite{J.M.Stewart-M.Walker11974}.


Further, we have to note that the correspondence between points
on ${\cal M}_{0}$ and ${\cal M}$, which is established by such a
relation as Eq.~(\ref{eq:variable-symbolic-perturbation}), is
not unique to the perturbation theory with general covariance.
Rather, Eq.~(\ref{eq:variable-symbolic-perturbation}) involves
the degree of freedom corresponding to the choice of the map
${\cal X}$ $:$ ${\cal M}_{0}\mapsto{\cal M}$.
This is called the {\it gauge degree of freedom} in general
relativistic perturbation theory.
Such a degree of freedom always exists in perturbations of a
theory with general covariance.
``General covariance'' intuitively means that there is no
preferred coordinate system in the theory.
If general covariance is not imposed on the theory, there is a
preferred coordinate system in our nature, and we naturally
introduce this coordinate system onto both ${\cal M}_{0}$ and
${\cal M}$.
Then, through this preferred coordinate system, we can choose
the identification map ${\cal X}$.
However, due to general covariance, there is no such coordinate
system in general relativity, and we have no guiding principle
to choose the identification map ${\cal X}$.
Actually, we may identify $``p\mbox{''}\in{\cal M}$ with 
$q\in{\cal M}_{0}$ ($q\neq p$) instead of $p\in{\cal M}_{0}$.
In the above understanding of the concept of ``gauge'' in
general-relativistic perturbation theory, a gauge transformation
is simply a change of the identification map ${\cal X}$.


These are the basic ideas necessary to understand 
{\it gauge degree of freedom} in the general relativistic
perturbation theory proposed by Stewart and
Walker\cite{J.M.Stewart-M.Walker11974}.
This understanding has been developed by Bruni et
al.\cite{M.Bruni-S.Soonego-CQG1997}, and by the present
author\cite{kouchan-gauge-inv,kouchan-second}.


\subsubsection{Formulation of perturbations}
\label{sec:Gauge-degree-of-freedom-in-perturbation-theory-formulation}


To formulate the above understanding in more detail, we
introduce an infinitesimal parameter $\lambda$ for the
perturbation.
Further, we consider the $(n+1)+1$-dimensional manifold 
${\cal N}={\cal M}\times\RF$, where $n+1=\dim{\cal M}$ and
$\lambda\in\RF$.
The background spacetime 
${\cal M}_{0}=\left.{\cal N}\right|_{\lambda=0}$ and the
physical spacetime
${\cal M}={\cal M}_{\lambda}=\left.{\cal N}\right|_{\RF=\lambda}$ are
also submanifolds embedded in the extended manifold ${\cal N}$.
Each point on ${\cal N}$ is identified by a pair, $(p,\lambda)$,
where $p\in{\cal M}_{\lambda}$, and each point in the background
spacetime ${\cal M}_{0}$ in ${\cal N}$ is identified by
$\lambda=0$.


Through this construction, the manifold ${\cal N}$ is foliated by
$(n+1)$-dimensional submanifolds ${\cal M}_{\lambda}$ of each
$\lambda$, and these are diffeomorphic to the physical
spacetime ${\cal M}$ and the background spacetime 
${\cal M}_{0}$. 
The manifold ${\cal N}$ has a natural differentiable structure
consisting of the direct product of ${\cal M}$ and $\RF$.
Further, the perturbed spacetimes ${\cal M}_{\lambda}$ for each
$\lambda$ must have the same differential structure with this
construction.
In other words, we require that perturbations be continuous in
the sense that $({\cal M},\bar{g}_{ab})$ and 
$({\cal M}_{0},g_{ab})$ are connected by a continuous curve
within the extended manifold ${\cal N}$.
Hence, the changes of the differential structure resulting from
the perturbation, for example the formation of singularities,
are excluded from our consideration.


Let us consider the set of field equations 
\begin{equation}
  \label{eq:field-eq-for-Q}
  {\cal E}[Q_{\lambda}] = 0
\end{equation}
on the physical spacetime ${\cal M}_{\lambda}$ for the physical
variables $Q_{\lambda}$ on ${\cal M}_{\lambda}$.
The field equation (\ref{eq:field-eq-for-Q}) formally represents
the Einstein equation for the metric on ${\cal M}_{\lambda}$ and
the equations for matter fields on ${\cal M}_{\lambda}$.
If a tensor field $Q_{\lambda}$ is given on each 
${\cal M}_{\lambda}$, $Q_{\lambda}$ is automatically extended to a
tensor field on ${\cal N}$ by $Q(p,\lambda):=Q_{\lambda}(p)$,
where $p\in{\cal M}_{\lambda}$.
In this extension, the field equation (\ref{eq:field-eq-for-Q})
is regarded as an equation on the extended manifold ${\cal N}$.
Thus, we have extended an arbitrary tensor field and the field
equations (\ref{eq:field-eq-for-Q}) on each ${\cal M}_{\lambda}$
to those on the extended manifold ${\cal N}$.


Tensor fields on ${\cal N}$ obtained through the above construction
are necessarily ``tangent'' to each ${\cal M}_{\lambda}$, i.e.,
their normal component to each ${\cal M}_{\lambda}$ identically
vanishes.
To consider the basis of the tangent space of ${\cal N}$, we
introduce the normal form and its dual, which are normal to each
${\cal M}_{\lambda}$ in ${\cal N}$.
These are denoted by $(d\lambda)_{a}$ and
$(\partial/\partial\lambda)^{a}$, respectively, and they satisfy
$(d\lambda)_{a} \left(\partial/\partial\lambda\right)^{a} = 1$.
The form $(d\lambda)_{a}$ and its dual,
$(\partial/\partial\lambda)^{a}$, are normal to any tensor field
extended from the tangent space on each ${\cal M}_{\lambda}$
through the above construction. 
The set consisting of $(d\lambda)_{a}$,
$(\partial/\partial\lambda)^{a}$, and the basis of the tangent 
space on each ${\cal M}_{\lambda}$ is regarded as the basis of
the tangent space of ${\cal N}$.


To define the perturbation of an arbitrary tensor field $Q$, we
compare $Q$ on the physical spacetime ${\cal M}_{\lambda}$ with
$Q_{0}$ on the background spacetime, and it is necessary to
identify the points of ${\cal M}_{\lambda}$ with those of 
${\cal M}_{0}$.
This point identification map is the so-called 
{\it gauge choice} in the context of perturbation theories,
as mentioned above.
The gauge choice is made by assigning a diffeomorphism
${\cal X}_{\lambda}$ $:$ ${\cal N}$ $\rightarrow$ ${\cal N}$
such that ${\cal X}_{\lambda}$ $:$ ${\cal M}_{0}$ $\rightarrow$
${\cal M}_{\lambda}$.
Following the paper of Bruni et
al.\cite{M.Bruni-S.Soonego-CQG1997}, we introduce a gauge choice
${\cal X}_{\lambda}$ as an exponential map on ${\cal N}$, for
simplicity. 
We denote the generator of this exponential map by 
${}^{{\cal X}}\!\eta^{a}$.
This generator ${}^{{\cal X}}\!\eta^{a}$ is decomposed by the 
basis on the tangent space of ${\cal N}$ which are constructed
above.
The arbitrariness of the gauge choice ${\cal X}_{\lambda}$ is
represented by the tangential component of 
${}^{{\cal X}}\!\eta^{a}$ to ${\cal M}_{\lambda}$.


The pull-back ${\cal X}_{\lambda}^{*}Q$, which is induced by the
exponential map ${\cal X}_{\lambda}$, maps a tensor field $Q$
on ${\cal M}_{\lambda}$ to a tensor field 
${\cal X}_{\lambda}^{*}Q$ on ${\cal M}_{0}$.
In terms of this generator ${}^{{\cal X}}\!\eta^{a}$, the
pull-back ${\cal X}_{\lambda}^{*}Q$ is represented by the Taylor
expansion
\begin{eqnarray}
  Q(r)
  =
  Q({\cal X}_{\lambda}(p))
  =
  {\cal X}_{\lambda}^{*}Q(p)
  =
  Q(p)
  + \lambda \left.{\pounds}_{{}^{{\cal X}}\!\eta}Q \right|_{p}
  + \frac{1}{2} \lambda^{2} 
  \left.{\pounds}_{{}^{{\cal X}}\!\eta}^{2}Q\right|_{p}
  + O(\lambda^{3}),
  \label{eq:Taylor-expansion-of-calX-org}
\end{eqnarray}
where $r={\cal X}_{\lambda}(p)\in{\cal M}_{\lambda}$.
Because $p\in{\cal M}_{0}$, we may regard the equation
\begin{eqnarray}
  {\cal X}_{\lambda}^{*}Q(p)
  =
  Q_{0}(p)
  + \lambda \left.{\pounds}_{{}^{{\cal X}}\!\eta}Q\right|_{{\cal M}_{0}}(p)
  + \frac{1}{2} \lambda^{2} 
  \left.{\pounds}_{{}^{{\cal X}}\!\eta}^{2}Q\right|_{{\cal M}_{0}}(p)
  + O(\lambda^{3})
  \label{eq:Taylor-expansion-of-calX}
\end{eqnarray}
as an equation on the background spacetime ${\cal M}_{0}$,
where $Q_{0}=\left.Q\right|_{{\cal M}_{0}}$ is the background
value of the physical variable of $Q$.
Once the definition of the pull-back of the gauge choice 
${\cal X}_{\lambda}$ is given, the perturbations of a tensor
field $Q$ under the gauge choice ${\cal X}_{\lambda}$ are simply
defined by the evaluation of the expansion
(\ref{eq:Taylor-expansion-of-calX}) on ${\cal M}_{0}$
\begin{equation}
  \label{eq:Bruni-35}
  \left.{\cal X}^{*}_{\lambda}Q_{\lambda}\right|_{{\cal M}_{0}}
  =
  Q_{0}
  + \lambda {}^{(1)}_{\;\cal X}\!Q
  + \frac{1}{2} \lambda^{2} {}^{(2)}_{\;\cal X}\!Q
  + O(\lambda^{3}),
\end{equation}
i.e.,
\begin{eqnarray}
  {}^{(1)}_{\;\cal X}\!Q := 
  \left.{\pounds}_{{}^{\cal X}\!\eta} Q\right|_{{\cal M}_{0}},
  \quad
  {}^{(2)}_{\;\cal X}\!Q := 
  \left.{\pounds}_{{}^{\cal X}\!\eta}^{2} Q\right|_{{\cal M}_{0}}.
  \label{eq:representation-of-each-order-perturbation}
\end{eqnarray}
We note that all variables in this definition are defined on
${\cal M}_{0}$.


\subsubsection{Gauge transformation}
\label{sec:Gauge-degree-of-freedom-in-perturbation-theory-gauge-trans}


Here, we consider two {\it different gauge choices}.
Suppose that ${\cal X}_{\lambda}$ and ${\cal Y}_{\lambda}$ are
two exponential maps with the generators ${}^{\cal X}\eta^{a}$
and ${}^{\cal Y}\eta^{a}$ on ${\cal N}$, respectively.
The integral curves of each ${}^{\cal X}\!\eta^{a}$ and
${}^{\cal Y}\!\eta^{a}$ in ${\cal N}$ are the orbits of the
actions of the gauge choices ${\cal X}_{\lambda}$ and 
${\cal Y}_{\lambda}$, respectively.
Since we choose ${}^{\cal X}\!\eta^{a}$ and 
${}^{\cal Y}\!\eta^{a}$ so that these are transverse to each 
${\cal M}_{\lambda}$ everywhere on ${\cal N}$, the integral
curves of these vector fields intersect with each 
${\cal M}_{\lambda}$.
Therefore, points lying on the same integral curve of either of
the two are to be regarded as {\it the same point} within the 
respective gauges.
When these curves are not identical, i.e., the tangential
components of ${}^{\cal X}\!\eta^{a}$ and 
${}^{\cal Y}\!\eta^{a}$ to each ${\cal M}_{\lambda}$ are
different, these point identification maps ${\cal X}_{\lambda}$
and ${\cal Y}_{\lambda}$ are regarded as
{\it two different gauge choices}.
When we have two different gauge choice ${\cal X}_{\lambda}$ and
${\cal Y}_{\lambda}$, we have two different representations of
the perturbative expansion of the pulled-backed variables
$\left.{\cal X}^{*}_{\lambda}Q_{\lambda}\right|_{{\cal M}_{0}}$
and 
$\left.{\cal Y}^{*}_{\lambda}Q_{\lambda}\right|_{{\cal M}_{0}}$:
\begin{eqnarray}
  \label{eq:Xgaueg-perturbation}
  \left.{\cal X}^{*}_{\lambda}Q_{\lambda}\right|_{{\cal M}_{0}}
  &=&
  Q_{0}
  + \lambda {}^{(1)}_{\;\cal X}\!Q
  + \frac{1}{2} \lambda^{2} {}^{(2)}_{\;\cal X}\!Q
  + O(\lambda^{3})
  , \\
  \label{eq:Ygaueg-perturbation}
  \left.{\cal Y}^{*}_{\lambda}Q_{\lambda}\right|_{{\cal M}_{0}}
  &=&
  Q_{0}
  + \lambda {}^{(1)}_{\;\cal Y}\!Q
  + \frac{1}{2} \lambda^{2} {}^{(2)}_{\;\cal Y}\!Q
  + O(\lambda^{3}),
\end{eqnarray}
Although these two representations of the perturbations are
different from each other, these should be equivalent because of
general covariance.


Now, we consider the {\it gauge-transformation rules}
between two different gauge choices.
In general, the representation ${}^{\cal X}Q_{\lambda}$ on
${\cal M}_{0}$ of the perturbed variable $Q$ on 
${\cal M}_{\lambda}$ depends on the gauge choice 
${\cal X}_{\lambda}$.
If we employ a different gauge choice, the representation of
$Q_{\lambda}$ on ${\cal M}_{0}$ may change.
Suppose that ${\cal X}_{\lambda}$ and ${\cal Y}_{\lambda}$ are
two different gauge choices and the generators of these gauge
choices are given by ${}^{\cal X}\!\eta^{a}$ and 
${}^{\cal Y}\!\eta^{a}$, respectively.
In this situation, the change of the gauge choice from 
${\cal X}_{\lambda}$ to ${\cal Y}_{\lambda}$ is represented by
the diffeomorphism 
\begin{equation}
  \label{eq:diffeo-def-from-Xinv-Y}
  \Phi_{\lambda} :=
  ({\cal X}_{\lambda})^{-1}\circ{\cal Y}_{\lambda}.
\end{equation}
This diffeomorphism $\Phi_{\lambda}$ is the map $\Phi_{\lambda}$
$:$ ${\cal M}_{0}$ $\rightarrow$ ${\cal M}_{0}$ for each value
of $\lambda\in\RF$.
The diffeomorphism $\Phi_{\lambda}$ does change the point
identification, as expected from the understanding of the gauge
choice discussed above.
Therefore, the diffeomorphism $\Phi_{\lambda}$ is regarded as
the gauge transformation $\Phi_{\lambda}$ $:$
${\cal X}_{\lambda}$ $\rightarrow$ ${\cal Y}_{\lambda}$.


The gauge transformation $\Phi_{\lambda}$ induces a pull-back
from the representation ${}^{\cal X}Q_{\lambda}$ of the
perturbed tensor field $Q$ in the gauge choice 
${\cal X}_{\lambda}$ to the representation 
${}^{\cal Y}Q_{\lambda}$ in the gauge choice 
${\cal Y}_{\lambda}$.
Actually, the tensor fields ${}^{\cal X}Q_{\lambda}$ and
${}^{\cal Y}Q_{\lambda}$, which are defined on ${\cal M}_{0}$,
are connected by the linear map $\Phi^{*}_{\lambda}$ as
\begin{eqnarray}
  {}^{\cal Y}Q_{\lambda}
  &=&
  \left.{\cal Y}^{*}_{\lambda}Q\right|_{{\cal M}_{0}}
  =
  \left.\left(
      {\cal Y}^{*}_{\lambda}
      \left({\cal X}_{\lambda}
      {\cal X}_{\lambda}^{-1}\right)^{*}Q\right)
  \right|_{{\cal M}_{0}}
  \nonumber\\
  &=&
  \left.
    \left(
      {\cal X}^{-1}_{\lambda}
      {\cal Y}_{\lambda}
    \right)^{*}
    \left(
      {\cal X}^{*}_{\lambda}Q
    \right)
  \right|_{{\cal M}_{0}}
  =  \Phi^{*}_{\lambda} {}^{\cal X}Q_{\lambda}.
  \label{eq:Bruni-45-one}
\end{eqnarray}
According to generic arguments concerning the Taylor expansion
of the pull-back of a tensor field on the same
manifold\cite{S.Sonego-M.Bruni-CMP1998,Nakamura:2010yg}, it
should be expressed the gauge transformation 
$\Phi^{*}_{\lambda} {}^{\cal X}Q_{\lambda}$ in the form
\begin{eqnarray}
  \Phi^{*}_{\lambda}{}^{\cal X}\!Q = {}^{\cal X}\!Q
  + \lambda {\pounds}_{\xi_{1}} {}^{\cal X}\!Q
  + \frac{\lambda^{2}}{2} \left\{
    {\pounds}_{\xi_{2}} + {\pounds}_{\xi_{1}}^{2}
  \right\} {}^{\cal X}\!Q
  + O(\lambda^{3}),
  \label{eq:Bruni-46-one} 
\end{eqnarray}
where the vector fields $\xi_{1}^{a}$ and $\xi_{2}^{a}$ are the
generators of the gauge transformation $\Phi_{\lambda}$.


Comparing the representation (\ref{eq:Bruni-46-one}) and that in
terms of the generators ${}^{\cal X}\eta^{a}$ and 
${}^{\cal Y}\eta^{a}$ of the pull-back 
${\cal Y}^{*}_{\lambda}\circ\left({\cal X}_{\lambda}^{-1}\right)^{*}\;{}^{{\cal X}}\!Q$ 
($=\Phi_{\lambda}^{*}{}^{\cal X}\!Q$), we obtain explicit
correspondence between $\{\xi_{1}^{a},\xi_{2}^{a}\}$ and 
$\{{}^{\cal X}\eta^{a},{}^{\cal Y}\eta^{a}\}$ as follows: 
\begin{eqnarray}
  \xi_{1}^{a}
  =
  {}^{\cal Y}\eta^{a}
  -
  {}^{\cal X}\eta^{a},
  \quad
  \xi_{2}^{a}
  = 
  \left[
    {}^{\cal Y}\eta
    ,
    {}^{\cal X}\eta
  \right]^{a}.
  \label{eq:relation-between-xi-eta}
\end{eqnarray}
Further, because the gauge transformation $\Phi_{\lambda}$ is a
map within the background spacetime ${\cal M}_{0}$, the
generator should consist of vector fields on ${\cal M}_{0}$.


We can now derive the relation between the perturbations in the
two different gauges.
Up to second order, these relations are derived by substituting
(\ref{eq:Xgaueg-perturbation}) and
(\ref{eq:Xgaueg-perturbation}) into (\ref{eq:Bruni-46-one}):
\begin{eqnarray}
  \label{eq:Bruni-47-one}
  {}^{(1)}_{\;{\cal Y}}\!Q - {}^{(1)}_{\;{\cal X}}\!Q &=& 
  {\pounds}_{\xi_{1}}Q_{0}, \\
  \label{eq:Bruni-49-one}
  {}^{(2)}_{\;\cal Y}\!Q - {}^{(2)}_{\;\cal X}\!Q &=& 
  2 {\pounds}_{\xi_{(1)}} {}^{(1)}_{\;\cal X}\!Q 
  +\left\{{\pounds}_{\xi_{(2)}}+{\pounds}_{\xi_{(1)}}^{2}\right\} Q_{0}.
\end{eqnarray}


Here, we comment on the generic formula for the Taylor expansion
(\ref{eq:Bruni-46-one}).
In the case where we regard the pull-backs 
${\cal X}_{\lambda}^{*}$ and ${\cal Y}_{\lambda}^{*}$ of the
gauge choices are exponential maps, the product of two
exponential maps is also written by the exponential of the
infinite sum of the Lie derivative along infinitely many
generators through Baker-Campbell-Hausdorff 
formula\cite{Sopuerta-Bruni-Gualtieri-2004,N.Jacobson-1992}.
These infinitely many generators are constructed by the
commutators of ${}^{\cal X}\eta^{a}$ and ${}^{\cal Y}\eta^{a}$, which 
are regarded as higher-order derivatives of 
${}^{\cal X}\eta^{a}$ and ${}^{\cal Y}\eta^{a}$ in ${\cal N}$
and are regarded as the vector fields on ${\cal N}$.
The expression (\ref{eq:Bruni-46-one}) of the Taylor expansion
is just the expression up to $O(\lambda^{3})$ of the
Baker-Campbell-Hausdorff formula.
Of course, we may generalize the gauge choice 
${\cal X}_{\lambda}^{*}$ and ${\cal Y}_{\lambda}^{*}$ to more
general class of diffeomorphism than the exponential map.
Even in this case, the Taylor expansion (\ref{eq:Bruni-46-one})
is correct.
Since the gauge-transformation rules (\ref{eq:Bruni-47-one}) and
(\ref{eq:Bruni-49-one}) are direct consequences of the Taylor
expansion (\ref{eq:Bruni-46-one}), these gauge-transformation
rules are not changed even if we generalize the gauge choice 
${\cal X}_{\lambda}^{*}$ and ${\cal Y}_{\lambda}^{*}$, and we may
regard that two generator $\xi_{(1)}^{a}$ and $\xi_{(2)}^{a}$ in
Eqs.~(\ref{eq:Bruni-47-one}) are independent of each other.
Therefore, we may say that gauge-transformation rules
(\ref{eq:Bruni-47-one}) and (\ref{eq:Bruni-49-one}) are most
general gauge-transformation rules of the first and second
order, respectively.


\subsubsection{Gauge invariance}
\label{sec:Gauge-degree-of-freedom-in-perturbation-theory-gauge-inv}


We next introduce the concept of {\it gauge invariance}.
The gauge invariance considered in this paper is 
{\it order by order gauge invariance} proposed in
KN2009\cite{kouchan-second-cosmo-matter}.
We call the $k$th-order perturbation ${}^{(p)}_{{\cal X}}\!Q$ is
gauge invariant iff 
\begin{equation}
  {}^{(k)}_{\;\cal X}\!Q = {}^{(k)}_{\;\cal Y}\!Q
\end{equation}
for any gauge choice ${\cal X}_{\lambda}$ and
${\cal Y}_{\lambda}$. 
Through this concept of order by order gauge invariance, we can
decompose any perturbation of $Q$ into the gauge-invariant and
gauge-variant parts, as shown in KN2003\cite{kouchan-gauge-inv}.
In terms of these gauge-invariant variables, we can develop the
gauge-invariant perturbation theory.
However, this development is based on a non-trivial conjecture,
i.e., Conjecture \ref{conjecture:decomposition-conjecture} for
the linear order metric perturbation as explained below.


\subsection{Gauge-invariant variables}
\label{sec:gauge-invariant-variables}


Inspecting the gauge-transformation rules
(\ref{eq:Bruni-47-one}) and (\ref{eq:Bruni-49-one}), we define
gauge-invariant variables for metric perturbations and for
arbitrary matter fields.
First, we consider the metric perturbation and expand the metric 
$\bar{g}_{ab}$ on ${\cal M}$, which is pulled back to 
${\cal M}_{0}$ using a gauge choice  ${\cal X}_{\lambda}$ in the
form given in 
(\ref{eq:Bruni-35}),
\begin{eqnarray}
  {\cal X}^{*}_{\lambda}\bar{g}_{ab}
  &=&
  g_{ab} + \lambda {}_{{\cal X}}\!h_{ab} 
  + \frac{\lambda^{2}}{2} {}_{{\cal X}}\!l_{ab}
  + O^{3}(\lambda),
  \label{eq:metric-expansion}
\end{eqnarray}
where $g_{ab}$ is the metric on ${\cal M}_{0}$.
Of course, the expansion (\ref{eq:metric-expansion}) of the
metric depends entirely on the gauge choice 
${\cal X}_{\lambda}$.
Nevertheless, henceforth, we do not explicitly express the index
of the gauge choice ${\cal X}_{\lambda}$ if there is no
possibility of confusion.


Our starting point to construct gauge-invariant variables is the
following conjecture\cite{kouchan-gauge-inv} for the
linear-order metric perturbation $h_{ab}$ defined by
Eq.~(\ref{eq:metric-expansion}) : 
\begin{conjecture}
  \label{conjecture:decomposition-conjecture}
  If there is a tensor field $h_{ab}$ of the second rank, whose
  gauge transformation rule is
  \begin{eqnarray}
    {}_{{\cal Y}}\!h_{ab}
    -
    {}_{{\cal X}}\!h_{ab}
    =
    {\pounds}_{\xi_{(1)}}g_{ab},
    \label{eq:linear-metric-gauge-trans}
  \end{eqnarray}
  then there exist a tensor field ${\cal H}_{ab}$ and a vector
  field $X^{a}$ such that $h_{ab}$ is decomposed as
  \begin{eqnarray}
    h_{ab} =: {\cal H}_{ab} + {\pounds}_{X}g_{ab},
    \label{eq:linear-metric-decomp}
  \end{eqnarray}
  where ${\cal H}_{ab}$ and $X^{a}$ are transformed as
  \begin{equation}
    {}_{{\cal Y}}\!{\cal H}_{ab} - {}_{{\cal X}}\!{\cal H}_{ab} =  0, 
    \quad
    {}_{\quad{\cal Y}}\!X^{a} - {}_{{\cal X}}\!X^{a} = \xi^{a}_{(1)} 
    \label{eq:linear-metric-decomp-gauge-trans}
  \end{equation}
  under the gauge transformation (\ref{eq:Bruni-47-one}),
  respectively.
\end{conjecture}
In this conjecture, ${\cal H}_{ab}$ is gauge-invariant in the
sense as mentioned above, and we call ${\cal H}_{ab}$ as 
{\it gauge-invariant part} of the linear-order metric
perturbation $h_{ab}$.
On the other hand, the vector field $X^{a}$ in
Eq.~(\ref{eq:linear-metric-decomp}) is gauge dependent, and we
call $X^{a}$ as {\it gauge-variant part} of the metric
perturbation $h_{ab}$.


The main purpose of this paper is to prove Conjecture
\ref{conjecture:decomposition-conjecture} in some sense.
In the case of the cosmological perturbations on a homogeneous
and isotropic universe, we confirmed Conjecture
\ref{conjecture:decomposition-conjecture} is correct except
for some special modes of perturbations, and then we developed
the second-order cosmological perturbation theory in a
gauge-invariant manner\cite{kouchan-cosmo-second}.
On the other hand, in the case of the perturbation theory on a 
generic background spacetime, this conjecture was highly
non-trivial due to the non-trivial curvature of the background
spacetime.
We see this situation in detail in \S\ref{sec:K.Nakamura-2010-3}.
However, before going to the proof of Conjecture
\ref{conjecture:decomposition-conjecture}, we explain how the
higher-order gauge-invariant perturbation theory is developed
based on this conjecture, here.
Through this explanation, we emphasize the importance of
Conjecture \ref{conjecture:decomposition-conjecture}.


As shown in KN2003\cite{kouchan-gauge-inv}, the second-order
metric perturbations $l_{ab}$ are decomposed as
\begin{eqnarray}
  \label{eq:H-ab-in-gauge-X-def-second-1}
  l_{ab}
  =:
  {\cal L}_{ab} + 2 {\pounds}_{X} h_{ab}
  + \left(
      {\pounds}_{Y}
    - {\pounds}_{X}^{2} 
  \right)
  g_{ab},
\end{eqnarray}
where ${\cal L}_{ab}$ and $Y^{a}$ are the gauge-invariant and
gauge-variant parts of the second order metric perturbations,
i.e., 
\begin{eqnarray}
  \label{eq:second-metric-decomp-gauge-trans}
  {}_{{\cal Y}}\!{\cal L}_{ab} - {}_{{\cal X}}\!{\cal L}_{ab} = 0,
  \quad
  {}_{{\cal Y}}\!Y^{a} - {}_{{\cal X}}\!Y^{a}
  = \xi_{(2)}^{a} + [\xi_{(1)},X]^{a}.
\end{eqnarray}


Actually, using the gauge-variant part $X^{a}$ of the
linear-order metric perturbation $h_{ab}$, we consider the
tensor field $\hat{L}_{ab}$ defined by 
\begin{eqnarray}
  \label{eq:hatLab-def}
  \hat{L}_{ab} := l_{ab} - 2 {\pounds}_{X}h_{ab} + {\pounds}_{X}^{2}g_{ab}.
\end{eqnarray}
Through the gauge-transformation rules (\ref{eq:Bruni-49-one})
and (\ref{eq:linear-metric-decomp-gauge-trans}) for $l_{ab}$ and
$X^{a}$, respectively, the gauge-transformation rule for this
variable $\hat{L}_{ab}$ is given by 
\begin{eqnarray}
  {}_{{\cal Y}}\!\hat{L}_{ab} - {}_{{\cal X}}\!\hat{L}_{ab}
  = {\pounds}_{\sigma}g_{ab}, 
  \quad
  \sigma^{a} := \xi^{a}_{(2)} + \left[\xi_{(1)},X\right]^{a}.
\end{eqnarray}
This is identical to the gauge-transformation rule
(\ref{eq:linear-metric-gauge-trans}) in Conjecture
\ref{conjecture:decomposition-conjecture} and we may apply
Conjecture \ref{conjecture:decomposition-conjecture} to the 
variable $\hat{L}_{ab}$. 
Then, $\hat{L}_{ab}$ can be decomposed as
\begin{eqnarray}
  \hat{L}_{ab} = {\cal L}_{ab} + \pounds_{Y}g_{ab},
  \label{eq:second-hatLab-decomp}
\end{eqnarray}
where the gauge-transformation rules for ${\cal L}_{ab}$ and
$Y^{a}$ are given by
Eqs.~(\ref{eq:second-metric-decomp-gauge-trans}). 
Together with the definition (\ref{eq:hatLab-def}) of the
variable $\hat{L}_{ab}$, the decomposition
(\ref{eq:second-hatLab-decomp}) leads the decomposition
(\ref{eq:H-ab-in-gauge-X-def-second-1}) for the second-order
metric perturbation $l_{ab}$.


Furthermore, as shown in KN2003, using the first- and
second-order gauge-variant parts, $X^{a}$ and $Y^{a}$, of the
metric perturbations, the gauge-invariant variables for an
arbitrary tensor field $Q$ other than the metric are given by 
\begin{eqnarray}
  \label{eq:matter-gauge-inv-def-1.0}
  {}^{(1)}\!{\cal Q} &:=& {}^{(1)}\!Q - {\pounds}_{X}Q_{0}
  , \\ 
  \label{eq:matter-gauge-inv-def-2.0}
  {}^{(2)}\!{\cal Q} &:=& {}^{(2)}\!Q - 2 {\pounds}_{X} {}^{(1)}Q 
  - \left\{ {\pounds}_{Y} - {\pounds}_{X}^{2} \right\} Q_{0}
  .
\end{eqnarray}
It is straightforward to confirm that the variables
${}^{(p)}\!{\cal Q}$ defined by
(\ref{eq:matter-gauge-inv-def-1.0}) and
(\ref{eq:matter-gauge-inv-def-2.0}) are gauge invariant under
the gauge-transformation rules (\ref{eq:Bruni-47-one}) and
(\ref{eq:Bruni-49-one}), respectively.


Equations (\ref{eq:matter-gauge-inv-def-1.0}) and
(\ref{eq:matter-gauge-inv-def-2.0}) have an important 
implication.
To see this, we represent these equations as
\begin{eqnarray}
  \label{eq:matter-gauge-inv-decomp-1.0}
  {}^{(1)}\!Q &=& {}^{(1)}\!{\cal Q} + {\pounds}_{X}Q_{0}
  , \\ 
  \label{eq:matter-gauge-inv-decomp-2.0}
  {}^{(2)}\!Q  &=& {}^{(2)}\!{\cal Q} + 2 {\pounds}_{X} {}^{(1)}Q 
  + \left\{ {\pounds}_{Y} - {\pounds}_{X}^{2} \right\} Q_{0}
  .
\end{eqnarray}
These equations imply that any perturbation of first and second
order can always be decomposed into gauge-invariant and
gauge-variant parts as
Eqs.~(\ref{eq:matter-gauge-inv-decomp-1.0}) and
(\ref{eq:matter-gauge-inv-decomp-2.0}), respectively.


\subsection{Second-order gauge-invariant perturbation theory}
\label{sec:gauge-invariant-perturbation-theory}


When we consider the first- and the second-order perturbations
of the Einstein equation, we have to consider the perturbative
expansion of the Einstein tensor and the energy momentum
tensor. 
Now, we consider the perturbative expansion of the Einstein
tensor on ${\cal M}_{\lambda}$ as
\begin{equation}
  \bar{G}_{a}^{\;\;b}
  =
  G_{a}^{\;\;b}
  + \lambda {}^{(1)}\!G_{a}^{\;\;b} 
  + \frac{1}{2} \lambda^{2} {}^{(2)}\!G_{a}^{\;\;b} 
  + O(\lambda^{3}).
\end{equation}
As shown in KN2005\cite{kouchan-second}, the first- and the
second-order perturbation of the Einstein tensor are given by 
\begin{eqnarray}
  \label{eq:linear-Einstein}
  {}^{(1)}\!G_{a}^{\;\;b}
  &=&
  {}^{(1)}{\cal G}_{a}^{\;\;b}\left[{\cal H}\right]
  + {\pounds}_{X} G_{a}^{\;\;b}
  ,\\
  \label{eq:second-Einstein-2,0-0,2}
  {}^{(2)}\!G_{a}^{\;\;b}
  &=& 
  {}^{(1)}{\cal G}_{a}^{\;\;b}\left[{\cal L}\right]
  + {}^{(2)}{\cal G}_{a}^{\;\;b} \left[{\cal H}, {\cal H}\right]
  + 2 {\pounds}_{X} {}^{(1)}\!\bar{G}_{a}^{\;\;b}
  + \left\{ {\pounds}_{Y} - {\pounds}_{X}^{2} \right\} G_{a}^{\;\;b},
\end{eqnarray}
where
\begin{eqnarray}
  \label{eq:cal-G-def-linear}
  {}^{(1)}{\cal G}_{a}^{\;\;b}\left[A\right]
  &:=&
  {}^{(1)}\Sigma_{a}^{\;\;b}\left[A\right]
  - \frac{1}{2} \delta_{a}^{\;\;b} {}^{(1)}\Sigma_{c}^{\;\;c}\left[A\right]
  \label{eq:(1)Sigma-def-linear}
  , \\
  {}^{(1)}\Sigma_{a}^{\;\;b}\left[A\right]
  &:=&
  - 2 \nabla_{[a}^{}H_{d]}^{\;\;\;bd}\left[A\right]
  - A^{cb} R_{ac}
  , \\
  \label{eq:cal-G-def-second}
  {}^{(2)}{\cal G}_{a}^{\;\;b}\left[A, B\right]
  &:=&
  {}^{(2)}\Sigma_{a}^{\;\;b}\left[A, B\right]
  - \frac{1}{2} \delta_{a}^{\;\;b} {}^{(2)}\Sigma_{c}^{\;\;c}\left[A, B\right]
  , \\
  {}^{(2)}\Sigma_{a}^{\;\;b}\left[A, B\right]
  &:=& 
    2 R_{ad} B_{c}^{\;\;(b}A^{d)c}
  + 2 H_{[a}^{\;\;\;de}\left[A\right] H_{d]\;\;e}^{\;\;\;b}\left[B\right]
  + 2 H_{[a}^{\;\;\;de}\left[B\right] H_{d]\;\;e}^{\;\;\;b}\left[A\right]
  \nonumber\\
  &&
  + 2 A_{e}^{\;\;d} \nabla_{[a}H_{d]}^{\;\;\;be}\left[B\right]
  + 2 B_{e}^{\;\;d} \nabla_{[a}H_{d]}^{\;\;\;be}\left[A\right]
  \nonumber\\
  &&
  + 2 A_{c}^{\;\;b} \nabla_{[a}H_{d]}^{\;\;\;cd}\left[B\right]
  + 2 B_{c}^{\;\;b} \nabla_{[a}H_{d]}^{\;\;\;cd}\left[A\right]
  \label{eq:(2)Sigma-def-second}
  ,
\end{eqnarray}
and 
\begin{eqnarray}
  H_{ab}^{\;\;\;\;c}\left[A\right]
  &:=&
  \nabla_{(a}^{}A_{b)}^{\;\;\;c}
  - \frac{1}{2} \nabla^{c}_{}A_{ab}
  \label{eq:Habc-def-1}
  , \\
  H_{abc}\left[A\right] 
  &:=&
  g_{cd} H_{ab}^{\;\;\;\;d}\left[A\right]
  ,
  \quad
  H_{a}^{\;\;bc}\left[A\right] 
  := 
  g^{bd} H_{ad}^{\;\;\;\;c}\left[A\right]
  ,
  \quad
  \nonumber\\
  H_{a\;\;c}^{\;\;b}\left[A\right] 
  &:=& 
  g_{cd} H_{a}^{\;\;bd}\left[A\right].
  \label{eq:Habc-def-2}
\end{eqnarray}
We note that ${}^{(1)}{\cal G}_{a}^{\;\;b}\left[*\right]$ and
${}^{(2)}{\cal G}_{a}^{\;\;b}\left[*,*\right]$ in
Eqs.~(\ref{eq:linear-Einstein}) and
(\ref{eq:second-Einstein-2,0-0,2}) are the gauge-invariant parts
of the perturbative Einstein tensors, and
Eqs.~(\ref{eq:linear-Einstein}) and
(\ref{eq:second-Einstein-2,0-0,2}) have the same forms as 
Eqs.~(\ref{eq:matter-gauge-inv-decomp-1.0}) and
(\ref{eq:matter-gauge-inv-decomp-2.0}), respectively.


We also note that ${}^{(1)}{\cal G}_{a}^{\;\;b}\left[*\right]$
and ${}^{(2)}{\cal G}_{a}^{\;\;b}\left[*,*\right]$ defined by
Eqs.~(\ref{eq:cal-G-def-linear})--(\ref{eq:(2)Sigma-def-second})
satisfy the identities
\begin{eqnarray}
  \nabla_{a}
  {}^{(1)}{\cal G}_{b}^{\;\;a}\left[A\right]
  &=& 
  - H_{ca}^{\;\;\;\;a}\left[A\right] G_{b}^{\;\;c}
  + H_{ba}^{\;\;\;\;c}\left[A\right] G_{c}^{\;\;a}
  \label{eq:linear-order-divergence-of-calGab}
  , \\
  \nabla_{a}{}^{(2)}{\cal G}_{b}^{\;\;a}\left[A, B\right]
  &=& 
  - H_{ca}^{\;\;\;\;a}\left[A\right]
    {}^{(1)}\!{\cal G}_{b}^{\;\;c}\left[B\right]
  - H_{ca}^{\;\;\;\;a}\left[B\right]
    {}^{(1)}\!{\cal G}_{b}^{\;\;c}\left[A\right]
  \nonumber\\
  &&
  + H_{ba}^{\;\;\;\;e}\left[A\right]
    {}^{(1)}\!{\cal G}_{e}^{\;\;a}\left[B\right]
  + H_{ba}^{\;\;\;\;e}\left[B\right]
    {}^{(1)}\!{\cal G}_{e}^{\;\;a}\left[A\right]
  \nonumber\\
  &&
  - \left(
    H_{bad}\left[B\right] A^{dc} + H_{bad}\left[A\right] B^{dc}
  \right)
  G_{c}^{\;\;a}
  \nonumber\\
  &&
  + \left(
    H_{cad}\left[B\right] A^{ad} + H_{cad}\left[A\right] B^{ad}
  \right)
  G_{b}^{\;\;c},
  \label{eq:second-div-of-calGab-1,1}
\end{eqnarray}
for arbitrary tensor fields $A_{ab}$ and $B_{ab}$, respectively.
We can directly confirm these identities without specifying
arbitrary tensors $A_{ab}$ and $B_{ab}$ of the second rank,
respectively. 
These identities (\ref{eq:linear-order-divergence-of-calGab})
and (\ref{eq:second-div-of-calGab-1,1}) guarantee the first- and
second-order perturbations of the Bianchi identity
$\bar{\nabla}_{b}\bar{G}_{a}^{\;\;b}=0$.
This implies that our general framework of the second-order
gauge-invariant perturbation theory is self-consistent.


On the other hand, the energy momentum tensor on 
${\cal M}_{\lambda}$ is also expanded as 
\begin{eqnarray}
  \bar{T}_{a}^{\;\;b}
  =
  T_{a}^{\;\;b}
  + \lambda {}^{(1)}\!T_{a}^{\;\;b}
  + \frac{1}{2} \lambda^{2} {}^{(2)}\!T_{a}^{\;\;b}
  + O(\lambda^{3}).
\end{eqnarray}
According to Eqs.~(\ref{eq:matter-gauge-inv-decomp-1.0}) and
(\ref{eq:matter-gauge-inv-decomp-2.0}), we can also decompose
the first- and the second-order perturbations of the energy
momentum tensor ${}^{(1)}\!T_{a}^{\;\;b}$ and
${}^{(2)}\!T_{a}^{\;\;b}$ as 
\begin{eqnarray}
  {}^{(1)}\!T_{a}^{\;\;b}
  &=&
  {}^{(1)}\!{\cal T}_{a}^{\;\;b}
  + {\pounds}_{X}T_{a}^{\;\;b}
  , \\
  {}^{(2)}\!T_{a}^{\;\;b}
  &=&
  {}^{(2)}\!{\cal T}_{a}^{\;\;b}
  + 2 {\pounds}_{X}{}^{(1)}\!T_{a}^{\;\;b}
  + \left\{ {\pounds}_{Y} - {\pounds}_{X}^{2} \right\} T_{a}^{\;\;b}
  .
\end{eqnarray}
These decompositions are confirmed in the case of a perfect
fluid, an imperfect fluid, and a scalar field in
KN2009\cite{kouchan-second-cosmo-matter}.
Furthermore, in KN2009\cite{kouchan-second-cosmo-matter}, we
also showed that equations of motion for the matter field, which 
are derived from the divergence of the energy-momentum tensors,
are also decomposed into gauge-invariant and gauge-variant parts
as Eqs.~(\ref{eq:matter-gauge-inv-decomp-1.0}) and 
(\ref{eq:matter-gauge-inv-decomp-2.0}).
Therefore, we may say that the decomposition formulae
(\ref{eq:matter-gauge-inv-decomp-1.0}) and
(\ref{eq:matter-gauge-inv-decomp-2.0}) are universal.


Imposing order by order Einstein equations
\begin{eqnarray}
  G_{a}^{\;\;b} = 8\pi T_{a}^{\;\;b}, \quad
  {}^{(1)}\!G_{a}^{\;\;b} = 8\pi {}^{(1)}\!T_{a}^{\;\;b}, \quad
  {}^{(2)}\!G_{a}^{\;\;b} = 8\pi {}^{(2)}\!T_{a}^{\;\;b},
\end{eqnarray}
the first- and the second-order perturbation of the Einstein
equations are automatically given in gauge-invariant form as
\begin{eqnarray}
  {}^{(1)}\!{\cal G}_{a}^{\;\;b}\left[{\cal H}\right]
  =
  8\pi G {}^{(1)}{\cal T}_{a}^{\;\;b}
  , \quad
  {}^{(1)}\!{\cal G}_{a}^{\;\;b}\left[{\cal L}\right]
  + {}^{(2)}\!{\cal G}_{a}^{\;\;b}\left[{\cal H}, {\cal H}\right]
  =
  8\pi G \;\; {}^{(2)}{\cal T}_{a}^{\;\;b} 
  .
\end{eqnarray}
Furthermore, in KN2009\cite{kouchan-second-cosmo-matter}, we
also showed that the equations of motion for matter fields,
are automatically given in gauge-invariant form.   
Thus, we may say that any equation of order by order is
automatically gauge-invariant and we do not have to consider the
gauge degree of freedom at least in the level where we
concentrate only on the equations of the general relativistic
system.


We can also expect that the similar structure of equations of
the systems will be maintained in the any order perturbations
and our general framework be applicable to any order
general-relativistic perturbations.
Actually, decomposition formulae for the third-order
perturbations in two-parameter case which correspond to
Eqs.~(\ref{eq:matter-gauge-inv-decomp-1.0}) and
(\ref{eq:matter-gauge-inv-decomp-2.0}) are given in
KN2003\cite{kouchan-gauge-inv}.
Therefore, similar development is possible for the third-order
perturbations. 
Since we could not find any difficulties to extend higher-order
perturbations\cite{kouchan-gauge-inv} except for the necessity
of long cumbersome calculations, we can construct any order
perturbation theory in gauge-invariant manner, recursively.


We have to emphasize that the above general framework of the
higher-order gauge-invariant perturbation theory are independent
of the explicit form of the background metric $g_{ab}$, except
for Conjecture \ref{conjecture:decomposition-conjecture}, and
are valid not only in cosmological perturbation case but also
the other generic situations if Conjecture
\ref{conjecture:decomposition-conjecture} is true.
This implies that if we prove Conjecture
\ref{conjecture:decomposition-conjecture} for the generic
background spacetime, the above general framework is 
applicable to perturbation theories on any background 
spacetime.
This is the reason why we proposed Conjecture
\ref{conjecture:decomposition-conjecture} in
KN2009\cite{kouchan-second-cosmo-matter}.


Thus, Conjecture \ref{conjecture:decomposition-conjecture}
is the important premise of our general framework of
higher-order gauge-invariant perturbation theory. 
In the next section, we give a proof of Conjecture
\ref{conjecture:decomposition-conjecture} on the generic
background spacetime which admits ADM decomposition (see
Appendix \ref{sec:ADM-decomposition}).


\section{Decomposition of the linear-order metric perturbation}
\label{sec:K.Nakamura-2010-3}


Now, we give a proof of Conjecture
\ref{conjecture:decomposition-conjecture} on general background 
spacetimes which admit ADM decomposition (see Appendix
\ref{sec:ADM-decomposition}).
Therefore, the background spacetime ${\cal M}_{0}$ considered
here is $n+1$-dimensional spacetime which is described by the
direct product $\RF\times\Sigma$.
Here, $\RF$ is a time direction and $\Sigma$ is the spacelike
hypersurface with $\dim\Sigma = n$ embedded in ${\cal M}_{0}$. 
This means that ${\cal M}_{0}$ is foliated by the one-parameter
family of spacelike hypersurface $\Sigma(t)$, $t\in\RF$ is a time
function.
The metric on ${\cal M}_{0}$ is given as
Eq.~(\ref{eq:gab-decomp-dd-minus}), i.e.,  
\begin{eqnarray}
  \label{eq:gdb-decomp-dd-minus-main}
  g_{ab} &=& - \alpha^{2} (dt)_{a} (dt)_{b}
  + q_{ij}
  (dx^{i} + \beta^{i}dt)_{a}
  (dx^{j} + \beta^{j}dt)_{b},
\end{eqnarray}
where $\alpha$ is the lapse function, $\beta^{i}$ is the
shift vector, and $q_{ab}=q_{ij}(dx^{i})_{a}(dx^{i})_{b}$ is the
metric on $\Sigma(t)$.
The inverse of Eq.~(\ref{eq:gdb-decomp-dd-minus-main}) is given
by Eq.~(\ref{eq:gab-decomp-uu-minus}) in the Appendix
\ref{sec:ADM-decomposition}.


To consider the decomposition (\ref{eq:linear-metric-decomp}) of
$h_{ab}$, first, we consider the components of the metric
$h_{ab}$ as 
\begin{eqnarray}
  \label{eq:hab-ADM-decomp}
  h_{ab}
  =
  h_{tt} (dt)_{a}(dt)_{b}
  + 2 h_{ti} (dt)_{(a}(dx^{i})_{b)}
  + h_{ij} (dx^{i})_{a}(dx^{j})_{b}.
\end{eqnarray}
The components $h_{tt}$, $h_{ti}$, and $h_{ij}$ are regarded as
a scalar function, a vector field, and a tensor field on the 
spacelike hypersurface $\Sigma(t)$, respectively.
From the gauge-transformation rule
(\ref{eq:linear-metric-gauge-trans}), the components $\{h_{tt}$,
$h_{ti}$, $h_{ij}\}$ are transformed as
\begin{eqnarray}
  {}_{{\cal Y}}h_{tt}
  -
  {}_{{\cal X}}h_{tt}
  &=&
    2 \partial_{t}\xi_{t}
  - \frac{2}{\alpha}\left(
    \partial_{t}\alpha 
    + \beta^{i}D_{i}\alpha 
    - \beta^{j}\beta^{i}K_{ij}
  \right) \xi_{t}
  \nonumber\\
  && \quad
  - \frac{2}{\alpha} \left(
    \beta^{i}\beta^{k}\beta^{j} K_{kj}
    - \beta^{i} \partial_{t}\alpha
    + \alpha q^{ij} \partial_{t}\beta_{j}
  \right.
  \nonumber\\
  && \quad\quad\quad\quad
  \left.
    + \alpha^{2} D^{i}\alpha 
    - \alpha \beta^{k} D^{i} \beta_{k}
    - \beta^{i} \beta^{j} D_{j}\alpha 
  \right)\xi_{i}
  \label{eq:gauge-trans-of-htt-ADM-BG}
  , \\
  {}_{{\cal Y}}h_{ti}
  -
  {}_{{\cal X}}h_{ti}
  &=&
  \partial_{t}\xi_{i}
  + D_{i}\xi_{t}
  - \frac{2}{\alpha} \left(
    D_{i}\alpha 
    - \beta^{j}K_{ij}
  \right) \xi_{t}
  \nonumber\\
  && \quad
  - \frac{2}{\alpha} \left(
    - \alpha^{2} K^{j}_{\;\;i}
    + \beta^{j}\beta^{k} K_{ki}
    - \beta^{j} D_{i}\alpha 
    + \alpha D_{i}\beta^{j}
  \right) \xi_{j}
  \label{eq:gauge-trans-of-hti-ADM-BG}
  , \\
  {}_{{\cal Y}}h_{ij}
  -
  {}_{{\cal X}}h_{ij}
  &=&
  2 D_{(i}\xi_{j)}
  + \frac{2}{\alpha} K_{ij} \xi_{t}
  - \frac{2}{\alpha} \beta^{k} K_{ij} \xi_{k}
  \label{eq:gauge-trans-of-hij-ADM-BG}
  ,
\end{eqnarray}
where $K_{ij}$ is the extrinsic curvature of $\Sigma$ defined
Eq.~(\ref{extrinsic_definition-2}) in Appendix 
\ref{sec:ADM-decomposition} and $D_{i}$ is the covariant
derivative associate with the metric $q_{ij}$
($D_{i}q_{jk}=0$).


Apparently, the gauge-transformation rules
(\ref{eq:gauge-trans-of-htt-ADM-BG})--(\ref{eq:gauge-trans-of-hij-ADM-BG})
have the complicated form and it seems difficult to find the
decomposition as (\ref{eq:linear-metric-decomp}).
Therefore, we consider the proof of the decomposition
(\ref{eq:linear-metric-decomp}) from the simpler situations of
the background spacetime.
We consider the three cases.
In \S\ref{sec:trivial-simple-case}, the case where $\alpha=1$,
$\beta^{i}$, and $K_{ij}=0$ is considered.
In this simplest case, we may consider the non-trivial intrinsic 
curvature of $\Sigma$.
From the view point of a proof of Conjecture
\ref{conjecture:decomposition-conjecture}, this case is
trivial.
However, we find the outline of the proof of Conjecture
\ref{conjecture:decomposition-conjecture} through this case.
Second, in \S\ref{sec:synchorous-comoving-BG-case}, we
consider the case where $\alpha=1$, $\beta^{i}=0$, but
$K_{ij}\neq 0$.
This case includes not only many homogeneous background
spacetimes but also the Schwarzschild spacetime.
Furthermore, we note that the most non-trivial technical
part of the proof of Conjecture
\ref{conjecture:decomposition-conjecture} is given in this
case.
Finally, in \S\ref{sec:most-generic-case}, we consider the
most general case where $\alpha\neq 1$, $\beta^{1}\neq 0$, and
$K_{ij}\neq 0$ for completion.


\subsection{$\alpha=1$, $\beta_{i}=0$, and $K_{ij}=0$ case}
\label{sec:trivial-simple-case}

Here, we consider ${\cal M}_{0}$ satisfies the conditions
\begin{eqnarray}
  \alpha = 1, \quad \beta_{i}=0, \quad K_{ij} = 0.
  \label{eq:K.Nakamura-2010-2-simple-0-1}
\end{eqnarray}
In this case, the gauge-transformation rules
(\ref{eq:gauge-trans-of-htt-ADM-BG})--(\ref{eq:gauge-trans-of-hij-ADM-BG})
are given by 
\begin{eqnarray}
  {}_{{\cal Y}}h_{tt}
  -
  {}_{{\cal X}}h_{tt}
  &=&
  2 \partial_{t}\xi_{t}
  \label{eq:K.Nakamura-2010-2-simple-1-1}
  , \\
  {}_{{\cal Y}}h_{ti}
  -
  {}_{{\cal X}}h_{ti}
  &=&
  \partial_{t}\xi_{i}
  + D_{i}\xi_{t}
  \label{eq:K.Nakamura-2010-2-simple-1-2}
  , \\
  {}_{{\cal Y}}h_{ij}
  -
  {}_{{\cal X}}h_{ij}
  &=&
  2 D_{(i}\xi_{j)}
  \label{eq:K.Nakamura-2010-2-simple-1-3}
  .
\end{eqnarray}
To prove Conjecture \ref{conjecture:decomposition-conjecture}, 
we consider the decomposition of the symmetric tensor field on
$\Sigma$ reviewed in Appendix 
\ref{sec:S.Deser-1967-J.W.York.Jr-1973-J.W.York.Jr-1974}: 
\begin{eqnarray}
  \label{eq:K.Nakamura-2010-2-simple-1-4}
  h_{ti} &=& D_{i}h_{(VL)} + h_{(V)i}, \quad D^{i}h_{(V)i} = 0,
  \\
  \label{eq:K.Nakamura-2010-2-simple-1-5}
  h_{ij} &=& \frac{1}{n} q_{ij} h_{(L)} + h_{(T)ij}, \quad
  q^{ij} h_{(T)ij} = 0,
  \\
  \label{eq:K.Nakamura-2010-2-simple-1-6}
  h_{(T)ij} &=& (Lh_{(TV)})_{ij} + h_{(TT)ij},
  \quad
  D^{i}h_{(TT)ij} = 0,
  \\
  \label{eq:K.Nakamura-2010-2-simple-1-7}
  h_{(TV)i} &=& D_{i}h_{(TVL)} + h_{(TVV)i}, \quad
  D^{i}h_{(TVV)i} = 0,
\end{eqnarray}
where $(Lh_{(TV)})_{ij}$ is defined by [see
Eq.~(\ref{eq:J.W.York.Jr-1973-3-2}) in Appendix
\ref{sec:S.Deser-1967-J.W.York.Jr-1973-J.W.York.Jr-1974}] 
\begin{eqnarray}
  \label{eq:LhTVij-def}
  (Lh_{(TV)})_{ij} := D_{i}h_{(TV)j} + D_{j}h_{(TV)i} - \frac{2}{n}q_{ij}D^{l}h_{(TV)l}.
\end{eqnarray}
To derive gauge-transformation rules for
$\{h_{tt}$, $h_{(VL)}$, $h_{(V)i}$, $h_{(L)}$, $h_{(TV)i}$,
$h_{(TT)ij}\}$, or $\{h_{tt}$, $h_{(VL)}$, $h_{(V)i}$,
$h_{(L)}$, $h_{(TVL)}$, $h_{(TVV)i}$, $h_{(TT)ij}\}$, 
we decompose the generator $\xi_{i}$ as
\begin{eqnarray}
  \xi_{i} = D_{i}\xi_{(L)} + \xi_{(V)i}, \quad D^{i}\xi_{(V)i} = 0.
  \label{eq:K.Nakamura-2010-2-simple-1-9}
\end{eqnarray}


Through the decomposition
(\ref{eq:K.Nakamura-2010-2-simple-1-4}) and
(\ref{eq:K.Nakamura-2010-2-simple-1-9}), the
gauge-transformation rule
(\ref{eq:K.Nakamura-2010-2-simple-1-2}) is
given by 
\begin{eqnarray}
  {}_{{\cal Y}}h_{ti}
  -
  {}_{{\cal X}}h_{ti}
  &=&
  D_{i}\left(
    {}_{{\cal Y}}h_{(VL)} - {}_{{\cal X}}h_{(VL)}
  \right)
  +
  {}_{{\cal Y}}h_{(V)i}
  -
  {}_{{\cal X}}h_{(V)i}
  \nonumber\\
  &=&
  D_{i}\left(
    \partial_{t}\xi_{(L)}
    + \xi_{t}
  \right)
  + \partial_{t}\xi_{(V)i}
  ,
  \label{eq:K.Nakamura-2010-2-simple-1-10}
\end{eqnarray}
where we used $\partial_{t}D_{i}f=D_{i}\partial_{t}f$ for an
arbitrary scalar function $f$.
Taking the divergence of
Eq.~(\ref{eq:K.Nakamura-2010-2-simple-1-10}), we see that 
\begin{eqnarray}
  \Delta\left(
    {}_{{\cal Y}}h_{(VL)} - {}_{{\cal X}}h_{(VL)}
  \right)
  =
  \Delta\left(
    \partial_{t}\xi_{(L)}
    + \xi_{t}
  \right)
  + D^{i}\partial_{t}\xi_{(V)i}
  .
  \label{eq:K.Nakamura-2010-2-simple-1-12}
\end{eqnarray}
Since $K_{ij}=0$, $\alpha=1$, and $\beta_{i}=0$ in our case,
Eq.~(\ref{eq:derivative_extrinsic_minus}) yields 
\begin{eqnarray}
  \label{eq:derivative_extrinsic_minus-simple-1}
  K_{ij} = - \frac{1}{2} \partial_{t} q_{ij} = 0.
\end{eqnarray}
From Eqs.~(\ref{eq:n-dim-Christoffel}),
(\ref{eq:derivative_extrinsic_minus-simple-1}), and the property
$D^{i}\xi_{(V)i}=0$, we can easily verify that
\begin{eqnarray}
  D^{i}\partial_{t}\xi_{(V)i}
  =
  \partial_{t}D^{i}\xi_{(V)i}
  =
  0
  .
\end{eqnarray}
Then the gauge-transformation rule
(\ref{eq:K.Nakamura-2010-2-simple-1-12}) is given by 
\begin{eqnarray}
  \Delta\left(
    {}_{{\cal Y}}h_{(VL)} - {}_{{\cal X}}h_{(VL)}
  \right)
  =
  \Delta\left(
    \partial_{t}\xi_{(L)}
    + \xi_{t}
  \right)
  .
  \label{eq:K.Nakamura-2010-2-simple-1-12-2}
\end{eqnarray}
Here, we assume the existence of the Green function of the
Laplacian $\Delta:=D^{i}D_{i}$ and ignore the mode which belongs
the kernel of the Laplacian $\Delta$.
Then, we obtain
\begin{eqnarray}
  {}_{{\cal Y}}h_{(VL)} - {}_{{\cal X}}h_{(VL)}
  =
  \partial_{t}\xi_{(L)} + \xi_{t}
  .
  \label{eq:K.Nakamura-2010-2-simple-1-13}
\end{eqnarray}
Substituting Eq.~(\ref{eq:K.Nakamura-2010-2-simple-1-13}) into
Eq.~(\ref{eq:K.Nakamura-2010-2-simple-1-10}), we obtain 
\begin{eqnarray}
  {}_{{\cal Y}}h_{(V)i}
  -
  {}_{{\cal X}}h_{(V)i}
  =
  \partial_{t}\xi_{(V)i}
  .
  \label{eq:K.Nakamura-2010-2-simple-1-14}
\end{eqnarray}


Next, through Eq.~(\ref{eq:K.Nakamura-2010-2-simple-1-5}), the
trace part of Eq.~(\ref{eq:K.Nakamura-2010-2-simple-1-3}) is
given by 
\begin{eqnarray}
  q^{ij}{}_{{\cal Y}}h_{ij}
  -
  q^{ij}{}_{{\cal X}}h_{ij}
  =
  {}_{{\cal Y}}h_{(L)}
  -
  {}_{{\cal X}}h_{(L)}
  =
  2 D^{i}\xi_{i}
  = 
  2 \Delta\xi_{(L)}
  ,
  \label{eq:K.Nakamura-2010-2-simple-1-15}
\end{eqnarray}
where we used Eq.~(\ref{eq:K.Nakamura-2010-2-simple-1-9}).
Then, through the decomposition
(\ref{eq:K.Nakamura-2010-2-simple-1-6}), the gauge
transformation rule of the traceless part 
$h_{(T)ij}$ is given by 
\begin{eqnarray}
  {}_{{\cal Y}}h_{(T)ij}
  -
  {}_{{\cal X}}h_{(T)ij}
  &=&
  {}_{{\cal Y}}\!\left(Lh_{(TV)}\right)_{ij}
  -
  {}_{{\cal X}}\!\left(Lh_{(TV)}\right)_{ij}
  +
  {}_{{\cal Y}}h_{(TT)ij}
  -
  {}_{{\cal X}}h_{(TT)ij}
  \nonumber\\
  &=&
  \left(L\xi\right)_{ij}
  .
  \label{eq:K.Nakamura-2010-2-simple-1-16-2}
\end{eqnarray}
Taking the divergence of
Eq.~(\ref{eq:K.Nakamura-2010-2-simple-1-16-2}), we obtain 
\begin{eqnarray}
  {\cal D}^{jl}
  \left(
    {}_{{\cal Y}}h_{(TV)l} - {}_{{\cal X}}h_{(TV)l}
    - \xi_{l}
  \right)
  = 0,
  \label{eq:K.Nakamura-2010-2-simple-1-22}
\end{eqnarray}
where the derivative operator ${\cal D}^{ij}$ is defined by 
\begin{eqnarray}
  \label{eq:J.W.York.Jr-1973-1-2-kouchan-3-main}
  {\cal D}^{ij}
  :=
  q^{ij}\Delta + \left(1 - \frac{2}{n}\right) D^{i}D^{j} +
  R^{ij},
\end{eqnarray}
and its properties are discussed in Appendix
\ref{sec:S.Deser-1967-J.W.York.Jr-1973-J.W.York.Jr-1974}.
Here, we assume the existence of the Green function of the
derivative operator ${\cal D}^{ij}$ and ignore the modes which 
belong to the kernel of the derivative operator ${\cal D}^{ij}$.
Then, we obtain 
\begin{eqnarray}
  {}_{{\cal Y}}h_{(TV)l} - {}_{{\cal X}}h_{(TV)l}
  = \xi_{l}.
  \label{eq:K.Nakamura-2010-2-simple-1-23}
\end{eqnarray}
Substituting Eq.~(\ref{eq:K.Nakamura-2010-2-simple-1-23}) into
Eq.~(\ref{eq:K.Nakamura-2010-2-simple-1-16-2}), we obtain 
\begin{eqnarray}
  {}_{{\cal Y}}h_{(TT)ij} - {}_{{\cal X}}h_{(TT)ij} = 0.
  \label{eq:K.Nakamura-2010-2-simple-1-24}
\end{eqnarray}
Moreover, through the decomposition
(\ref{eq:K.Nakamura-2010-2-simple-1-7}), we obtain 
\begin{eqnarray}
  &&
  {}_{{\cal Y}}h_{(TVL)} - {}_{{\cal X}}h_{(TVL)}
  = \xi_{(L)}
  \label{eq:K.Nakamura-2010-2-simple-1-25}
  , \\
  &&
  {}_{{\cal Y}}h_{(TVV)i} - {}_{{\cal X}}h_{(TVV)i}
  = \xi_{(V)i}
  \label{eq:K.Nakamura-2010-2-simple-1-26}
  .
\end{eqnarray}
Here, we have also ignored the mode which belongs the kernel of
$\Delta$.


In summary, we have obtained the gauge-transformation rule of
variables $\{h_{tt}$, $h_{(VL)}$, $h_{(V)i}$, $h_{(L)}$,
$h_{(TV)i}$, $h_{(TT)ij}\}$, or $\{h_{tt}$, $h_{(VL)}$,
$h_{(V)i}$, $h_{(L)}$, $h_{(TVL)}$, $h_{(TVV)i}$, $h_{(TT)ij}\}$
as
\begin{eqnarray}
  &&
  {}_{{\cal Y}}h_{tt}
  -
  {}_{{\cal X}}h_{tt}
  =
  2 \partial_{t}\xi_{t}
  \label{eq:K.Nakamura-2010-2-simple-1-27}
  , \\
  &&
  {}_{{\cal Y}}h_{(VL)} - {}_{{\cal X}}h_{(VL)}
  =
  \partial_{t}\xi_{(L)} + \xi_{t}
  \label{eq:K.Nakamura-2010-2-simple-1-28}
  , \\
  &&
  {}_{{\cal Y}}h_{(V)i}
  -
  {}_{{\cal X}}h_{(V)i}
  =
  \partial_{t}\xi_{(V)i}
  \label{eq:K.Nakamura-2010-2-simple-1-29}
  , \\
  &&
  {}_{{\cal Y}}h_{(L)}
  -
  {}_{{\cal X}}h_{(L)}
  =
  2 D^{i}\xi_{i}
  =
  2 \Delta\xi_{(L)}
  \label{eq:K.Nakamura-2010-2-simple-1-29-2}
  , \\
  &&
  {}_{{\cal Y}}h_{(TVL)} - {}_{{\cal X}}h_{(TVL)}
  = \xi_{(L)}
  \label{eq:K.Nakamura-2010-2-simple-1-30}
  , \\
  &&
  {}_{{\cal Y}}h_{(TVV)i} - {}_{{\cal X}}h_{(TVV)i}
  = \xi_{(V)i}
  \label{eq:K.Nakamura-2010-2-simple-1-31}
  , \\
  &&
  {}_{{\cal Y}}h_{(TT)ij} - {}_{{\cal X}}h_{(TT)ij} = 0.
  \label{eq:K.Nakamura-2010-2-simple-1-32}
\end{eqnarray}


Now, we construct gauge-invariant variables.
First, the gauge-transformation rule
(\ref{eq:K.Nakamura-2010-2-simple-1-32}) shows $h_{(TT)ij}$ is
gauge invariant by itself:
\begin{eqnarray}
  \chi_{ij} := h_{(TT)ij}.
  \label{eq:K.Nakamura-2010-2-simple-1-33}
\end{eqnarray}
From Eq.~(\ref{eq:K.Nakamura-2010-2-simple-1-6}), this variable
$\chi_{ij}$ satisfy the transverse-traceless condition 
$D^{i}\chi_{ij} = 0 = q^{ij}\chi_{ij}$.


Second, from the gauge-transformation rules
(\ref{eq:K.Nakamura-2010-2-simple-1-29}) and
(\ref{eq:K.Nakamura-2010-2-simple-1-31}), we can define the
gauge-invariant variable for the vector mode as
\begin{eqnarray}
  \nu_{i}
  :=
  h_{(V)i} - \partial_{t}h_{(TVV)i}.
  \label{eq:K.Nakamura-2010-2-simple-1-35}
\end{eqnarray}
Actually, gauge-transformation rule for the variable $\nu_{i}$
is given by 
\begin{eqnarray}
  {}_{{\cal Y}}\!\nu_{i}
  -
  {}_{{\cal X}}\!\nu_{i}
  &=&
  \left(
    {}_{{\cal Y}}h_{(V)i}
    - \partial_{t}{}_{{\cal Y}}h_{(TVV)i}
  \right)
  -
  \left(
    {}_{{\cal X}}h_{(V)i}
    - \partial_{t}{}_{{\cal X}}h_{(TVV)i}
  \right)
  \nonumber\\
  &=&
  \partial_{t}\xi_{(V)i}
  - \partial_{t}\xi_{(V)i}
  = 0.
  \label{eq:K.Nakamura-2010-2-simple-1-34}
\end{eqnarray}
From the definition, the gauge-invariant variable $\nu_{i}$
satisfy the transverse condition $D^{i}\nu_{i}=0$ as a result.


Next, we consider scalar modes.
First, from the gauge-transformation rules
(\ref{eq:K.Nakamura-2010-2-simple-1-29-2}) and
(\ref{eq:K.Nakamura-2010-2-simple-1-30}), we see that the
variable $\Psi$ defined by 
\begin{eqnarray}
  - 2 n \Psi := h_{(L)} - 2 \Delta h_{(TVL)}
  \label{eq:K.Nakamura-2010-2-simple-1-37}
\end{eqnarray}
is gauge invariant.
Actually, the gauge-transformation rule for $\Psi$
is given by 
\begin{eqnarray}
  {}_{{\cal Y}}\Psi - {}_{{\cal X}}\Psi
  &=&
  - \frac{1}{2n}
  \left(
    {}_{{\cal Y}}h_{(L)} - 2 \Delta{}_{{\cal Y}}h_{(TVL)}
  \right)
  + \frac{1}{2n}
  \left(
    {}_{{\cal X}}h_{(L)} - 2 \Delta{}_{{\cal X}}h_{(TVL)}
  \right)
  =
  0
  .
  \label{eq:K.Nakamura-2010-2-simple-1-36}
\end{eqnarray}
To define another gauge-invariant variables for scalar modes, we 
consider the gauge-transformation rule
(\ref{eq:K.Nakamura-2010-2-simple-1-28}) and
(\ref{eq:K.Nakamura-2010-2-simple-1-30}).
We find that the variable $\hat{X}_{t}$ defined by
\begin{eqnarray}
  \hat{X}_{t} := h_{(VL)} - \partial_{t}h_{(TVL)} 
  \label{eq:K.Nakamura-2010-2-simple-1-38}
\end{eqnarray}
is transformed as 
\begin{eqnarray}
  {}_{{\cal Y}}\hat{X}_{t}
  -
  {}_{{\cal X}}\hat{X}_{t}
  &=&
  \xi_{t}
  .
  \label{eq:K.Nakamura-2010-2-simple-1-39}
\end{eqnarray}
Using this variable $\hat{X}_{t}$, we define the gauge-invariant
combination as 
\begin{eqnarray}
  - 2 \Phi := h_{tt} - 2 \partial_{t}\hat{X}_{t}.
  \label{eq:K.Nakamura-2010-2-simple-1-40}
\end{eqnarray}
Actually, we can easily check the variable $\Phi$ is gauge
invariant under the gauge-transformation rules
(\ref{eq:K.Nakamura-2010-2-simple-1-27}) and
(\ref{eq:K.Nakamura-2010-2-simple-1-39}).


In terms of the gauge-invariant variables defined by
Eqs.~(\ref{eq:K.Nakamura-2010-2-simple-1-33}), 
(\ref{eq:K.Nakamura-2010-2-simple-1-35}),
(\ref{eq:K.Nakamura-2010-2-simple-1-37}), and
(\ref{eq:K.Nakamura-2010-2-simple-1-40}), and the variable
$\hat{X}_{t}$ defined by Eq.~(\ref{eq:K.Nakamura-2010-2-simple-1-38}),
the original set $\{h_{tt}$, $h_{ti}$, $h_{ij}\}$ of the
components of the linear metric perturbation is given by 
\begin{eqnarray}
  h_{tt} &=& - 2 \Phi + 2 \partial_{t}\hat{X}_{t}
  \label{eq:K.Nakamura-2010-2-simple-1-41}
  , \\
  h_{ti}
  &=&
  \nu_{i}
  + \partial_{t}h_{(TV)i} + D_{i}\hat{X}_{t},
  \label{eq:K.Nakamura-2010-2-simple-1-42}
  , \\
  h_{ij}
  &=&
  - 2 q_{ij} \Psi
  + \chi_{ij}
  + D_{i}h_{(TV)j}
  + D_{j}h_{(TV)i}
  .
  \label{eq:K.Nakamura-2010-2-simple-1-43}
\end{eqnarray}
On the other hand, the components of
Eq.~(\ref{eq:linear-metric-decomp}) with 
\begin{eqnarray}
  X_{a} =: X_{t}(dt)_{a} + X_{i}(dx^{i})_{a}
\end{eqnarray}
are given by
\begin{eqnarray}
  h_{tt}
  &=&
  {\cal H}_{tt}
  + 2 \partial_{t}X_{t}
  \label{eq:K.Nakamura-2010-2-simple-1-51}
  , \\
  h_{ti}
  &=&
  {\cal H}_{ti}
  + \partial_{t}X_{i}
  + D_{i}X_{t}
  \label{eq:K.Nakamura-2010-2-simple-1-52}
  , \\
  h_{ij}
  &=&
  {\cal H}_{ij}
  + D_{i}X_{j}
  + D_{j}X_{i}
  .
  \label{eq:K.Nakamura-2010-2-simple-1-52-2}
\end{eqnarray}
Since the variable ${\cal H}_{tt}$, ${\cal H}_{ti}$, and 
${\cal H}_{ij}$ are gauge-invariant and the gauge-transformation
rules for the variable $X_{t}$ and $X_{i}$ are given by 
\begin{eqnarray}
  \label{eq:gauge-trans-Xt-Xi-simple-1}
  {}_{{\cal Y}}X_{t} - {}_{{\cal X}}X_{t} = \xi_{t}, 
  \quad
  {}_{{\cal Y}}X_{i} - {}_{{\cal X}}X_{i} = \xi_{i}, 
\end{eqnarray}
respectively, we may naturally identify the variables as
follows:
\begin{eqnarray}
  &&
  {\cal H}_{tt} = - 2 \Phi,
  \quad
  {\cal H}_{ti} = \nu_{i},
  \quad
  {\cal H}_{ij} = - 2 q_{ij} \Psi + \chi_{ij},
  \label{eq:gauge-inv-identification-simple-1}
  ,\\
  &&
  X_{t} = \hat{X}_{t},
  \quad
  X_{i} = h_{(TV)i}
  .
  \label{eq:gauge-var-identification-simple-1}
\end{eqnarray}
The gauge-transformation rules
(\ref{eq:K.Nakamura-2010-2-simple-1-23}) [or equivalently
Eqs.~(\ref{eq:K.Nakamura-2010-2-simple-1-30}) and
(\ref{eq:K.Nakamura-2010-2-simple-1-31})] and
(\ref{eq:K.Nakamura-2010-2-simple-1-39}) support these
identifications.
Thus,
Eqs.~(\ref{eq:K.Nakamura-2010-2-simple-1-41})--(\ref{eq:K.Nakamura-2010-2-simple-1-43})
show that the linear-order metric perturbation $h_{ab}$ is
decomposed into gauge-invariant and gauge-variant parts as
Eq.~(\ref{eq:linear-metric-decomp}) in the case for the
background spacetime with $\alpha=1$, $\beta^{i}=0$, and
$K_{ij}=0$.


\subsection{$\alpha=1$, $\beta^{i}=0$, but $K_{ij}\neq 0$ case}
\label{sec:synchorous-comoving-BG-case}


Here, we also consider the case where the background metric is
described by Eq.~(\ref{eq:gdb-decomp-dd-minus-main}) with
$\alpha = 1$, $\beta^{i} = 0$, but the background spacetime has
the non-trivial extrinsic curvature $K_{ij}\neq 0$.
In this case, the extrinsic curvature $K_{ij}$ is proportional to
the time derivative of the metric $q_{ij}$ on $\Sigma$
\begin{eqnarray}
  \label{eq:K.Nakamura-2010-2-simple-4-15}
  K_{ij} = - \frac{1}{2} \partial_{t}q_{ij}
\end{eqnarray}
from Eq.~(\ref{eq:derivative_extrinsic_minus}) in Appendix
\ref{sec:ADM-decomposition} and gauge-transformation rules 
(\ref{eq:gauge-trans-of-htt-ADM-BG})--(\ref{eq:gauge-trans-of-hij-ADM-BG})
for the components $\{h_{tt}$, $h_{ti}$, $h_{ij}\}$ are given by 
\begin{eqnarray}
  \label{eq:gauge-trans-of-htt-ADM-BG-case2}
  {}_{{\cal Y}}h_{tt}
  -
  {}_{{\cal X}}h_{tt}
  &=&
  2 \partial_{t}\xi_{t}
  , \\
  \label{eq:gauge-trans-of-hti-ADM-BG-case2}
  {}_{{\cal Y}}h_{ti}
  -
  {}_{{\cal X}}h_{ti}
  &=&
  \partial_{t}\xi_{i}
  + D_{i}\xi_{t}
  + 2 K^{j}_{\;\;i} \xi_{j}
  , \\
  \label{eq:gauge-trans-of-hij-ADM-BG-case2}
  {}_{{\cal Y}}h_{ij}
  -
  {}_{{\cal X}}h_{ij}
  &=&
  2 D_{(i}\xi_{j)}
  + 2 K_{ij} \xi_{t}
  .
\end{eqnarray}


Inspecting gauge-transformation rules 
(\ref{eq:gauge-trans-of-hti-ADM-BG-case2})--(\ref{eq:gauge-trans-of-hij-ADM-BG-case2}), 
we first define a new symmetric tensor field $\hat{H}_{ab}$
whose components are defined by 
\begin{eqnarray}
  \hat{H}_{tt} := h_{tt}, \quad
  \hat{H}_{ti} := h_{ti}, \quad
  \hat{H}_{ij} := h_{ij} - 2 K_{ij}\hat{X}_{t}.
  \label{eq:hatH-def-case2}
\end{eqnarray}
Here, we assume the existence of the variable $\hat{X}_{t}$
whose gauge-transformation rule is given by 
\begin{eqnarray}
  {}_{{\cal Y}}\hat{X}_{t}
  -
  {}_{{\cal X}}\hat{X}_{t}
  =
  \xi_{t}
  .
  \label{eq:K.Nakamura-2010-2-simple-4-11}
\end{eqnarray}
The existence of the variable $\hat{X}_{t}$ is confirmed later
soon.
Similar technique is given by T.~S.~Pereira et
al.\cite{Pereira:2007yy} in the perturbations on
Bianchi type I cosmology.
Now, the components $\hat{H}_{ti}$ and $\hat{H}_{ij}$ are
regarded as a vector and a symmetric tensor on $\Sigma(t)$,
respectively.
Then, we may apply the decomposition in Appendix
\ref{sec:S.Deser-1967-J.W.York.Jr-1973-J.W.York.Jr-1974} to
$\hat{H}_{ti}$ and $\hat{H}_{ij}$: 
\begin{eqnarray}
  \label{eq:K.Nakamura-2010-2-simple-4-7}
  \hat{H}_{ti} &=& D_{i}h_{(VL)} + h_{(V)i}, \quad D^{i}h_{(V)i} = 0,
  \\
  \label{eq:K.Nakamura-2010-2-simple-4-8}
  \hat{H}_{ij} &=& \frac{1}{n} q_{ij} h_{(L)} + h_{(T)ij}, \quad
  q^{ij} h_{(T)ij} = 0,
  \\
  \label{eq:K.Nakamura-2010-2-simple-4-9}
  h_{(T)ij} &=& \left(Lh_{(TV)}\right)_{ij} + h_{(TT)ij},
  \quad
  D^{i}h_{(TT)ij} = 0,
  \\
  \label{eq:K.Nakamura-2010-2-simple-4-10}
  h_{(TV)i} &=& D_{i}h_{(TVL)} + h_{(TVV)i}, \quad
  D^{i}h_{(TVV)i} = 0.
\end{eqnarray}


Through Eqs.~(\ref{eq:K.Nakamura-2010-2-simple-4-7}) and
(\ref{eq:hatH-def-case2}), the gauge-transformation rule
(\ref{eq:gauge-trans-of-hti-ADM-BG-case2}) is given by 
\begin{eqnarray}
  D_{i}\left(
    {}_{{\cal Y}}h_{(VL)} - {}_{{\cal X}}h_{(VL)}
  \right)
  +
  {}_{{\cal Y}}h_{(V)i}
  -
  {}_{{\cal X}}h_{(V)i}
  =
  \partial_{t}\xi_{i}
  + D_{i}\xi_{t}
  + 2 K^{j}_{\;\;i} \xi_{j}
  .
  \label{eq:K.Nakamura-2010-2-simple-4-12}
\end{eqnarray}
Taking the divergence of this gauge-transformation rule
(\ref{eq:K.Nakamura-2010-2-simple-4-12}), we obtain
\begin{eqnarray}
  \Delta \left(
    {}_{{\cal Y}}h_{(VL)} - {}_{{\cal X}}h_{(VL)}
  \right)
  =
  D^{i}\partial_{t}\xi_{i}
  + \Delta \xi_{t}
  + 2 D^{i}\left(K^{j}_{\;\;i} \xi_{j}\right)
  ,
  \label{eq:K.Nakamura-2010-2-simple-4-14}
\end{eqnarray}
where we used the divergenceless property of the variable
$h_{(V)i}$.


Now, we evaluate the term $D^{i}\partial_{t}\xi_{i}$ in
Eq.~(\ref{eq:K.Nakamura-2010-2-simple-4-14}).
First, we note that the time-derivative of the
intrinsic metric $\partial_{t}q_{ij}$ is given by
Eq.~(\ref{eq:K.Nakamura-2010-2-simple-4-15}), i.e.,  
\begin{eqnarray}
  \label{eq:K.Nakamura-2010-2-simple-4-16}
  \partial_{t}q_{ij} = - 2 K_{ij}.
\end{eqnarray}
Keeping the relation (\ref{eq:K.Nakamura-2010-2-simple-4-16}) in
our mind, we can evaluate $D_{i}\partial_{t}\xi_{j}$ as
\begin{eqnarray}
  D_{i}\partial_{t}\xi_{j}
  &=&
  \partial_{t}D_{i}\xi_{j}
  - \xi_{k} \left(
      D_{i}K_{j}^{\;\;k}
    + D_{j}K_{i}^{\;\;k}
    - D^{k}K_{ij}
  \right)
  .
  \label{eq:K.Nakamura-2010-2-simple-4-18}
\end{eqnarray}
From Eq.~(\ref{eq:K.Nakamura-2010-2-simple-4-18}), we further
evaluate the term $D^{i}\partial_{t}\xi_{i}$ as
\begin{eqnarray}
  D^{i}\partial_{t}\xi_{i}
  &=&
    \partial_{t}\left(D^{i}\xi_{i}\right)
  + \xi_{k} D^{k}K
  - 2 D_{i}\left(K^{ij} \xi_{j}\right)
  .
  \label{eq:K.Nakamura-2010-2-simple-4-20}
\end{eqnarray}
Through Eq.~(\ref{eq:K.Nakamura-2010-2-simple-4-20}), 
the gauge-transformation rule
(\ref{eq:K.Nakamura-2010-2-simple-4-14}) is given by 
\begin{eqnarray}
  \Delta \left(
    {}_{{\cal Y}}h_{(VL)} - {}_{{\cal X}}h_{(VL)}
  \right)
  =
  \partial_{t}\left(D^{i}\xi_{i}\right)
  + \Delta \xi_{t}
  + \xi_{k} D^{k}K
  .
  \label{eq:K.Nakamura-2010-2-simple-4-21}
\end{eqnarray}
Here again, it is convenient to introduce the decomposition of
$\xi_{i}$ as Eq.~(\ref{eq:K.Nakamura-2010-2-simple-1-9}), i.e.,
\begin{eqnarray}
  \xi_{i} = D_{i}\xi_{(L)} + \xi_{(V)i}, \quad D^{i}\xi_{(V)i} = 0.
  \label{eq:K.Nakamura-2010-2-simple-4-22}
\end{eqnarray}
Through Eq.~(\ref{eq:K.Nakamura-2010-2-simple-4-20}), the
gauge-transformation rule
(\ref{eq:K.Nakamura-2010-2-simple-4-21}) is given by  
\begin{eqnarray}
  \Delta \left(
    {}_{{\cal Y}}h_{(VL)} - {}_{{\cal X}}h_{(VL)}
    - \xi_{t}
  \right)
  &=&
  \Delta\partial_{t}\xi_{(L)}
  + 2 D_{i}\left( K^{ij} D_{j}\xi_{(L)} \right)
  + D^{k}K \xi_{(V)k}
  .
  \label{eq:K.Nakamura-2010-2-simple-4-24}
\end{eqnarray}
Since we ignore the modes which belong to the kernel of the
operator $\Delta$, we obtain the gauge-transformation rule for
the variable $h_{(VL)}$ as 
\begin{eqnarray}
  {}_{{\cal Y}}h_{(VL)} - {}_{{\cal X}}h_{(VL)}
  &=&
    \xi_{t}
  + \partial_{t}\xi_{(L)}
  +
  \Delta^{-1}
  \left[
    2 D_{i}\left(
        K^{ij} D_{j}\xi_{(L)}
    \right)
    + D^{k}K \xi_{(V)k}
  \right]
  .
  \label{eq:K.Nakamura-2010-2-simple-4-25}
\end{eqnarray}
Substituting Eq.~(\ref{eq:K.Nakamura-2010-2-simple-4-25}) into
Eq.~(\ref{eq:K.Nakamura-2010-2-simple-4-12}), we obtain the
gauge-transformation rule for the variable $h_{(V)}i$ as follows:  
\begin{eqnarray}
  {}_{{\cal Y}}h_{(V)i}
  -
  {}_{{\cal X}}h_{(V)i}
  &=&
    \partial_{t}\xi_{(V)i}
  + 2 K^{j}_{\;\;i} D_{j}\xi_{(L)}
  + 2 K^{j}_{\;\;i} \xi_{(V)j}
  \nonumber\\
  &&
  - D_{i}\Delta^{-1}
  \left[
    2 D_{i}\left( K^{ij} D_{j}\xi_{(L)} \right)
    + D^{k}K \xi_{(V)k}
  \right]
  .
  \label{eq:K.Nakamura-2010-2-simple-4-27}
\end{eqnarray}
The divergenceless property of
Eq.~(\ref{eq:K.Nakamura-2010-2-simple-4-27}) can be easily
checked through Eq.~(\ref{eq:K.Nakamura-2010-2-simple-4-20}).


From Eqs.~(\ref{eq:gauge-trans-of-hij-ADM-BG-case2})
and (\ref{eq:K.Nakamura-2010-2-simple-4-11}), the gauge
transformation rule for $\hat{H}_{ij}$ is given as
\begin{eqnarray}
  {}_{{\cal Y}}\!\hat{H}_{ij} 
  -
  {}_{{\cal X}}\!\hat{H}_{ij} 
  &=&
  \left(
    {}_{{\cal Y}}\!h_{ij} - 2 K_{ij}{}_{{\cal Y}}\!\hat{X}_{t}
  \right)
  -
  \left(
    {}_{{\cal X}}\!h_{ij} - 2 K_{ij}{}_{{\cal X}}\!\hat{X}_{t}
  \right)
  =
  2 D_{(i}\xi_{j)}
  .
\end{eqnarray}
In terms of the decomposition
(\ref{eq:K.Nakamura-2010-2-simple-4-8})--(\ref{eq:K.Nakamura-2010-2-simple-4-10}),
the gauge-transformation rules for the variables $h_{(L)}$ and
$h_{(T)ij}$ is derived from 
\begin{eqnarray}
  \frac{1}{n} q_{ij} \left(
    {}_{{\cal Y}}h_{(L)}
    -
    {}_{{\cal X}}h_{(L)}
  \right)
  + 
  \left(
    {}_{{\cal Y}}h_{(T)ij}
    -
    {}_{{\cal X}}h_{(T)ij}
  \right)
  &=&
  2 D_{(i}\xi_{j)}
  .
  \label{eq:K.Nakamura-2010-2-simple-4-31}
\end{eqnarray}
The trace part of the gauge transformation rule
(\ref{eq:K.Nakamura-2010-2-simple-4-31}) is given by 
\begin{eqnarray}
  {}_{{\cal Y}}h_{(L)}
  -
  {}_{{\cal X}}h_{(L)}
  &=&
  2 D^{i}\xi_{i}
  ,
  \label{eq:K.Nakamura-2010-2-simple-4-32}
\end{eqnarray}
and the traceless part of the gauge transformation rule
(\ref{eq:K.Nakamura-2010-2-simple-4-31}) is given by 
\begin{eqnarray}
  {}_{{\cal Y}}h_{(T)ij}
  -
  {}_{{\cal X}}h_{(T)ij}
  &=&
  \left(L\xi\right)_{ij}
  .
  \label{eq:K.Nakamura-2010-2-simple-4-32-2}
\end{eqnarray}
Applying Eq.~(\ref{eq:K.Nakamura-2010-2-simple-4-9}),
the gauge-transformation rules
(\ref{eq:K.Nakamura-2010-2-simple-4-32-2}) are given by 
\begin{eqnarray}
  \left(
    L\left(
      {}_{{\cal Y}}h_{(TV)} - {}_{{\cal X}}h_{(TV)}
    \right)
  \right)_{ij}
  + \left(
    {}_{{\cal Y}}h_{(TT)ij} - {}_{{\cal X}}h_{(TT)ij}
  \right)
  =
  \left(L\xi\right)_{ij}
  .
  \label{eq:K.Nakamura-2010-2-simple-4-33}
\end{eqnarray}
Taking the divergence of
Eq.~(\ref{eq:K.Nakamura-2010-2-simple-4-33}), we obtain 
\begin{eqnarray}
  {\cal D}^{jl} \left[
    {}_{{\cal Y}}h_{(TV)l} - {}_{{\cal X}}h_{(TV)l}
    - \xi_{l}
  \right]
  =
  0
  .
  \label{eq:K.Nakamura-2010-2-simple-4-36}
\end{eqnarray}
Since we ignore the modes which belong to the kernel of the
elliptic derivative operator ${\cal D}^{jl}$ in this paper, we
obtain 
\begin{eqnarray}
  {}_{{\cal Y}}h_{(TV)l} - {}_{{\cal X}}h_{(TV)l} = \xi_{l}.
  \label{eq:K.Nakamura-2010-2-simple-4-37}
\end{eqnarray}
Applying the decomposition formulae 
(\ref{eq:K.Nakamura-2010-2-simple-4-10}) and
(\ref{eq:K.Nakamura-2010-2-simple-4-22}), the
gauge-transformation rules for the variable $h_{(TVL)}$ and 
$h_{(TVV)}$ are given by 
\begin{eqnarray}
  \label{eq:K.Nakamura-2010-2-simple-4-38}
  {}_{{\cal Y}}h_{(TVL)} - {}_{{\cal X}}h_{(TVL)} = \xi_{(L)l}, \\
  \label{eq:K.Nakamura-2010-2-simple-4-39}
  {}_{{\cal Y}}h_{(TVV)l} - {}_{{\cal X}}h_{(TVV)l} = \xi_{(V)l}
\end{eqnarray}
since we ignore the modes which belong to the kernel of $\Delta$
in this paper. 
Further, substituting
Eq.~(\ref{eq:K.Nakamura-2010-2-simple-4-37}) into
Eq.~(\ref{eq:K.Nakamura-2010-2-simple-4-33}), we obtain the
gauge-transformation rule for $h_{(TT)ij}$ as follows
\begin{eqnarray}
  {}_{{\cal Y}}h_{(TT)ij} - {}_{{\cal X}}h_{(TT)ij}
  =
  0
  .
  \label{eq:K.Nakamura-2010-2-simple-4-38-2}
\end{eqnarray}


Thus, the gauge transformation rule for the variables $h_{tt}$,
$h_{(VL)}$, $h_{(V)i}$, $h_{(L)}$, $h_{(TV)i}$, ($h_{(TVL)}$ and
$h_{(TVV)i}$), and $h_{(TT)ij}$ are summarized as : 
\begin{eqnarray}
  {}_{{\cal Y}}h_{tt}
  -
  {}_{{\cal X}}h_{tt}
  &=&
  2 \partial_{t}\xi_{t}
  \label{eq:K.Nakamura-2010-2-simple-4-39-2}
  , \\
  {}_{{\cal Y}}h_{(VL)} - {}_{{\cal X}}h_{(VL)}
  &=&
    \partial_{t}\xi_{(L)}
  + \xi_{t}
  +
  \Delta^{-1}
  \left[
    2 D_{i}\left( K^{ij} D_{j}\xi_{(L)} \right)
    + D^{k}K \xi_{(V)k}
  \right]
  \label{eq:K.Nakamura-2010-2-simple-4-40}
  , \\
  {}_{{\cal Y}}h_{(V)i}
  -
  {}_{{\cal X}}h_{(V)i}
  &=&
    \partial_{t}\xi_{(V)i}
  + 2 K^{j}_{\;\;i} D_{j}\xi_{(L)}
  + 2 K^{j}_{\;\;i} \xi_{(V)j}
  \nonumber\\
  && \quad
  - D_{i}\Delta^{-1}
  \left[
    2 D_{i} \left( K^{ij} D_{j}\xi_{(L)} \right)
    + D^{k}K \xi_{(V)k}
  \right]
  \label{eq:K.Nakamura-2010-2-simple-4-41}
  , \\
  {}_{{\cal Y}}h_{(L)}
  -
  {}_{{\cal X}}h_{(L)}
  &=&
  2 D^{i}\xi_{i}
  \label{eq:K.Nakamura-2010-2-simple-4-42}
  , \\
  {}_{{\cal Y}}h_{(TV)l}
  -
  {}_{{\cal X}}h_{(TV)l}
  &=&
  \xi_{l}
  \label{eq:K.Nakamura-2010-2-simple-4-43}
  , \\
  {}_{{\cal Y}}h_{(TT)ij} - {}_{{\cal X}}h_{(TT)ij}
  &=&
  0
  .
  \label{eq:K.Nakamura-2010-2-simple-4-44}
\end{eqnarray}
The equation (\ref{eq:K.Nakamura-2010-2-simple-4-43}) 
is equivalent to the gauge-transformation rules
\begin{eqnarray}
  \label{eq:K.Nakamura-2010-2-simple-4-45}
  {}_{{\cal Y}}h_{(TVL)} - {}_{{\cal X}}h_{(TVL)}
  &=&
  \xi_{(L)}
  , \\
  \label{eq:K.Nakamura-2010-2-simple-4-46}
  {}_{{\cal Y}}h_{(TVV)l} - {}_{{\cal X}}h_{(TVV)l}
  &=&
  \xi_{(V)l}
  .
\end{eqnarray}


Now, we construct gauge-invariant variables.
First, Eq.~(\ref{eq:K.Nakamura-2010-2-simple-4-44}) shows that
the variable $h_{(TT)ij}$ is itself gauge invariant. 
Therefore, we define the transverse-traceless gauge-invariant
tensor as
\begin{eqnarray}
  \label{eq:K.Nakamura-2010-2-simple-4-47}
  \chi_{ij} := h_{(TT)ij}.
\end{eqnarray}
Second, through Eqs.~(\ref{eq:K.Nakamura-2010-2-simple-4-41}),
(\ref{eq:K.Nakamura-2010-2-simple-4-45}), and 
(\ref{eq:K.Nakamura-2010-2-simple-4-46}), we can define the
gauge-invariant variable for vector mode as 
\begin{eqnarray}
  \nu_{i}
  &=&
  h_{(V)i}
  - \partial_{t}h_{(TVV)i}
  - 2 K^{j}_{\;\;i} \left( D_{j}h_{(TVL)} + h_{(TVV)j} \right)
  \nonumber\\
  && \quad
  + D_{i}\Delta^{-1}
  \left[
    2 D_{i}\left( K^{ij} D_{j}h_{(TVL)} \right)
    + D^{k}K h_{(TVV)k}
  \right]
  \label{eq:K.Nakamura-2010-2-simple-4-48}
  .
\end{eqnarray}
Actually, it is straightforward to confirm that the variable $\nu_{i}$
defined by Eq.~(\ref{eq:K.Nakamura-2010-2-simple-4-48}) is gauge
invariant. Further, we can also confirm the divergenceless
property of the variable $\nu_{i}$, i.e., $D^{i}\nu_{i} = 0$
through the definition (\ref{eq:K.Nakamura-2010-2-simple-4-48}) 
and the formula (\ref{eq:K.Nakamura-2010-2-simple-4-20}).


Next, we consider the gauge invariant variables for scalar
modes.
Before doing this, we first construct the variable $\hat{X}_{t}$
in Eq.~(\ref{eq:hatH-def-case2}).
Inspecting gauge-transformation rules
(\ref{eq:K.Nakamura-2010-2-simple-4-40}), 
(\ref{eq:K.Nakamura-2010-2-simple-4-45}), and
(\ref{eq:K.Nakamura-2010-2-simple-4-46}), we consider the
combination 
\begin{eqnarray}
  \hat{X}_{t}
  &:=&
  h_{(VL)}
  - \partial_{t}h_{(TVL)}
  -
  \Delta^{-1}
  \left[
      2 D_{i}\left( K^{ij}D_{j}h_{(TVL)} \right)
    +   D^{k}K h_{(TVV)k}
  \right]
  .
  \label{eq:K.Nakamura-2010-2-simple-4-53}
\end{eqnarray}
The variable $\hat{X}_{t}$ defined by
Eq.~(\ref{eq:K.Nakamura-2010-2-simple-4-53}) does satisfy the
gauge-transformation rule (\ref{eq:K.Nakamura-2010-2-simple-4-11}).
We also define the variable $\hat{X}_{i}$ by 
\begin{eqnarray}
  \label{eq:K.Nakamura-2010-2-simple-4-56}
  \hat{X}_{i} := h_{(TV)i} = D_{i}h_{(TVL)} + h_{(TVV)i}
  .
\end{eqnarray}
From this gauge-transformation rule
(\ref{eq:K.Nakamura-2010-2-simple-4-43}), or equivalently
Eqs.~(\ref{eq:K.Nakamura-2010-2-simple-4-45}) and
(\ref{eq:K.Nakamura-2010-2-simple-4-46}), the 
gauge-transformation rule for the variable $\hat{X}_{i}$ defined
by Eq.~(\ref{eq:K.Nakamura-2010-2-simple-4-56}) is given by 
\begin{eqnarray}
  \label{eq:K.Nakamura-2010-2-simple-4-57}
  {}_{{\cal Y}}\hat{X}_{i}
  -
  {}_{{\cal X}}\hat{X}_{i}
  =
  \xi_{i}
  .
\end{eqnarray}


Now, we define the gauge invariant variables for the scalar
mode.
First, inspecting gauge-transformation rules
(\ref{eq:K.Nakamura-2010-2-simple-4-11}) and 
(\ref{eq:K.Nakamura-2010-2-simple-4-39-2}), we define the
variable $\Psi$ by 
\begin{eqnarray}
  \label{eq:K.Nakamura-2010-2-simple-4-58}
  - 2 \Phi &:=& h_{tt} - 2 \partial_{t}\hat{X}_{t}.
\end{eqnarray}
Actually, we can easily confirm that the variable $\Phi$ is
gauge invariant.
Second, inspecting gauge-transformation rules
(\ref{eq:K.Nakamura-2010-2-simple-4-42}) and
(\ref{eq:K.Nakamura-2010-2-simple-4-57}), we define the
gauge-invariant variable $\Psi$ by 
\begin{eqnarray}
  \label{eq:K.Nakamura-2010-2-simple-4-60}
  - 2 n \Psi := h_{(L)} - 2 D^{i}\hat{X}_{i}.
\end{eqnarray}


In terms of the gauge-invariant variables $\chi_{ij}$,
$\nu_{i}$, $\Phi$, and $\Psi$, which are defined by
Eq.~(\ref{eq:K.Nakamura-2010-2-simple-4-47}),
(\ref{eq:K.Nakamura-2010-2-simple-4-48}),
(\ref{eq:K.Nakamura-2010-2-simple-4-58}), and
(\ref{eq:K.Nakamura-2010-2-simple-4-60}), respectively, 
and the gauge-variant variables $\hat{X}_{t}$ and $\hat{X}_{i}$
which are defined by
Eqs.~(\ref{eq:K.Nakamura-2010-2-simple-4-53}) and
(\ref{eq:K.Nakamura-2010-2-simple-4-56}), respectively, the
original set $\{h_{tt}$, $h_{ti}$, $h_{ij}\}$ of the components
of the linear metric perturbation is given by 
\begin{eqnarray}
  \label{eq:K.Nakamura-2010-2-simple-4-71}
  h_{tt} &=& - 2 \Phi + 2 \partial_{t}\hat{X}_{t}, \\
  h_{ti}
  &=&
    \nu_{i}
  + D_{i}\hat{X}_{t}
  + \partial_{t}\hat{X}_{i}
  + 2 K^{j}_{\;\;i} \hat{X}_{j}
  \label{eq:K.Nakamura-2010-2-simple-4-72}
  , \\
  h_{ij}
  &=&
  - 2 \Psi q_{ij} 
  + \chi_{ij}
  + D_{i}\hat{X}_{j} + D_{j}\hat{X}_{i}
  + 2 K_{ij}\hat{X}_{t}
  \label{eq:K.Nakamura-2010-2-simple-4-73}
  .
\end{eqnarray}
On the other hand, we consider the decomposition formula
(\ref{eq:linear-metric-decomp}).
The components of the expression
(\ref{eq:linear-metric-decomp}) in the situation of this
subsection with
\begin{eqnarray}
  X_{a} =: X_{t}(dt)_{a} + X_{i}(dx^{i})_{a}
\end{eqnarray}
are given by 
\begin{eqnarray}
  h_{tt}
  &=&
  {\cal H}_{tt}
  +
  2 \partial_{t}X_{t}
  \label{eq:K.Nakamura-2010-2-simple-4-78}
  , \\
  h_{ti}
  &=&
  {\cal H}_{ti}
  + \partial_{t}X_{i}
  + D_{i}X_{t}
  + 2 K^{j}_{\;\;i} X_{j}
  \label{eq:K.Nakamura-2010-2-simple-4-79}
  , \\
  h_{ij}
  &=&
  {\cal H}_{ij}
  + D_{i}X_{j}
  + D_{j}X_{i}
  + 2 K_{ij} X_{t}
  \label{eq:K.Nakamura-2010-2-simple-4-80}
  .
\end{eqnarray}
Comparing
Eqs.~(\ref{eq:K.Nakamura-2010-2-simple-4-71})--(\ref{eq:K.Nakamura-2010-2-simple-4-73})
and
Eqs.~(\ref{eq:K.Nakamura-2010-2-simple-4-78})--(\ref{eq:K.Nakamura-2010-2-simple-4-80}),
we easily see a natural choice of the components of the
gauge-invariant part ${\cal H}_{ab}$ and the components of the 
gauge-variant parts $X_{a}$ are given by 
\begin{eqnarray}
  \label{eq:calHab-component-identification-case2}
  {\cal H}_{tt} = - 2 \Phi, \quad
  {\cal H}_{ti} = \nu_{i}, \quad
  {\cal H}_{ij} = - 2 \Psi q_{ij} + \chi_{ij},
\end{eqnarray}
and 
\begin{eqnarray}
  \label{eq:Xt-Xi-component-identification-case2}
  X_{t} = \hat{X}_{t}, \quad X_{i} = \hat{X}_{i}.
\end{eqnarray}
Of course, we may add the Killing vectors associated with the
metric $q_{ab}$ to the definition of $X_{a}$.
The gauge-transformation rules
(\ref{eq:K.Nakamura-2010-2-simple-4-11}) and
(\ref{eq:K.Nakamura-2010-2-simple-4-57}) support these
identifications. 
Thus,
Eqs.~(\ref{eq:calHab-component-identification-case2}) and
(\ref{eq:Xt-Xi-component-identification-case2}) show that the
linear-order metric perturbation $h_{ab}$ is also decomposed
into gauge-invariant and gauge-variant parts as
Eq.~(\ref{eq:linear-metric-decomp}) even in the case for the 
background spacetime with $\alpha=1$, $\beta^{i}=0$, but
$K_{ij}\neq 0$. 
These results are already reported in the previous
letter\cite{kouchan-decomp-letter-version} by the present
author.


\subsection{The case for arbitrary $\alpha$, $\beta_{i}$, and $K_{ij}$}
\label{sec:most-generic-case}


Now, we consider the most generic case of the metric
(\ref{eq:gdb-decomp-dd-minus-main}) where $\alpha\neq 1$,
$\beta^{i}\neq 0$, and $K_{ij}\neq 0$.
Considering the components of the metric perturbation $h_{ab}$
as Eq.~(\ref{eq:hab-ADM-decomp}), the gauge-transformation rules
for these components are given by
Eqs.~(\ref{eq:gauge-trans-of-htt-ADM-BG})--(\ref{eq:gauge-trans-of-hij-ADM-BG}). 
To do this, we first assume that the existence of the variables
$\hat{X}_{t}$ and $\hat{X}_{i}$ whose gauge-transformation rules
are given by   
\begin{eqnarray}
  &&
  {}_{{\cal Y}}\hat{X}_{t}
  -
  {}_{{\cal X}}\hat{X}_{t}
  =
  \xi_{t}
  \label{eq:K.Nakamura-2010-note-B-14}
  , \\
  &&
  {}_{{\cal Y}}\hat{X}_{i}
  -
  {}_{{\cal X}}\hat{X}_{i}
  =
  \xi_{i}
  \label{eq:K.Nakamura-2010-note-B-15}
  .
\end{eqnarray}
This assumption is confirmed through the construction of the
gauge-invariant variables for the linear-order metric
perturbation below.
Inspecting gauge-transformation rules
(\ref{eq:gauge-trans-of-htt-ADM-BG})--(\ref{eq:gauge-trans-of-hij-ADM-BG}),
we define the symmetric tensor field $\hat{H}_{ab}$ whose
components are given by 
\begin{eqnarray}
  \hat{H}_{tt}
  &:=&
  h_{tt} 
  + \frac{2}{\alpha}\left(
    \partial_{t}\alpha 
    + \beta^{i}D_{i}\alpha 
    - \beta^{j}\beta^{i}K_{ij}
  \right) \hat{X}_{t}
  \nonumber\\
  && \quad
  + \frac{2}{\alpha} \left(
    \beta^{i}\beta^{k}\beta^{j} K_{kj}
    - \beta^{i} \partial_{t}\alpha
    + \alpha q^{ij} \partial_{t}\beta_{j}
  \right.
  \nonumber\\
  && \quad\quad\quad\quad
  \left.
    + \alpha^{2} D^{i}\alpha 
    - \alpha \beta^{k} D^{i} \beta_{k}
    - \beta^{i} \beta^{j} D_{j}\alpha 
  \right)\hat{X}_{i}
  \label{eq:hatHtt-def-generic}
  , \\
  \hat{H}_{ti}
  &:=&
  h_{ti}
  + \frac{2}{\alpha} \left(
    D_{i}\alpha 
    - \beta^{j}K_{ij}
  \right) \hat{X}_{t}
  \nonumber\\
  && \quad
  + \frac{2}{\alpha} \left(
    - \alpha^{2} K^{j}_{\;\;i}
    + \beta^{j}\beta^{k} K_{ki}
    - \beta^{j} D_{i}\alpha 
    + \alpha D_{i}\beta^{j}
  \right) \hat{X}_{j}
  \label{eq:hatHti-def-generic}
  , \\
  \hat{H}_{ij}
  &:=&
  h_{ij}
  - \frac{2}{\alpha} K_{ij} \hat{X}_{t}
  + \frac{2}{\alpha} \beta^{k} K_{ij} \hat{X}_{k}
  \label{eq:hatHij-def-generic}
  .
\end{eqnarray}
The gauge transformation rules
(\ref{eq:gauge-trans-of-htt-ADM-BG})--(\ref{eq:gauge-trans-of-hij-ADM-BG})
and our assumptions (\ref{eq:K.Nakamura-2010-note-B-14}) and
(\ref{eq:K.Nakamura-2010-note-B-15}) give the
gauge-transformation rules of the components of $\hat{H}_{ab}$
as follows:
\begin{eqnarray}
  {}_{{\cal Y}}\!\hat{H}_{tt}
  -
  {}_{{\cal X}}\!\hat{H}_{tt}
  &=&
  2 \partial_{t}\xi_{t}
  ,
  \label{eq:gauge-trans-hatHtt}
  \\
  {}_{{\cal Y}}\!\hat{H}_{ti}
  -
  {}_{{\cal X}}\!\hat{H}_{ti}
  &=&
  \partial_{t}\xi_{i}
  + D_{i}\xi_{t}
  ,
  \label{eq:gauge-trans-hatHti}
  \\
  {}_{{\cal Y}}\!\hat{H}_{ij}
  -
  {}_{{\cal X}}\!\hat{H}_{ij}
  &=&
  2 D_{(i}\xi_{j)}
  .
  \label{eq:gauge-trans-hatHij}
\end{eqnarray}


Since the components $\hat{H}_{it}$ and $\hat{H}_{ij}$ are
regarded as a vector and a symmetric tensor on $\Sigma(t)$,
respectively, we may apply the decomposition reviewed in
Appendix \ref{sec:S.Deser-1967-J.W.York.Jr-1973-J.W.York.Jr-1974} to
$\hat{H}_{ti}$ and $\hat{H}_{ij}$:
\begin{eqnarray}
  \label{eq:K.Nakamura-2010-2-generic-4-7}
  \hat{H}_{ti} &=& D_{i}h_{(VL)} + h_{(V)i}, \quad D^{i}h_{(V)i} = 0,
  \\
  \label{eq:K.Nakamura-2010-2-generic-4-8}
  \hat{H}_{ij} &=& \frac{1}{n} q_{ij} h_{(L)} + h_{(T)ij}, \quad
  q^{ij} h_{(T)ij} = 0,
  \\
  \label{eq:K.Nakamura-2010-2-generic-4-9}
  h_{(T)ij} &=& \left(Lh_{(TV)}\right)_{ij} + h_{(TT)ij},
  \quad
  D^{i}h_{(TT)ij} = 0,
  \\
  \label{eq:K.Nakamura-2010-2-generic-4-10}
  h_{(TV)i} &=& D_{i}h_{(TVL)} + h_{(TVV)i}, \quad
  D^{i}h_{(TVV)i} = 0.
\end{eqnarray}
The gauge-transformation rules (\ref{eq:gauge-trans-hatHti}) and
(\ref{eq:gauge-trans-hatHij}) gives the gauge-transformation
rules for the variables $h_{(VL)}$, $h_{(V)i}$, $h_{(L)}$,
$h_{(T)ij}$, $h_{(TV)i}$ (or equivalently $h_{(TVL)}$ and
$h_{(TVV)i}$), and $h_{(TT)ij}$ as in the previous subsection.


First, we consider the gauge-transformation rule
(\ref{eq:gauge-trans-hatHti}) in terms of the decomposition
(\ref{eq:K.Nakamura-2010-2-generic-4-7}):
\begin{eqnarray}
  \label{eq:gauge-trans-hatHti-decomp}
  {}_{{\cal Y}}\!\hat{H}_{ti}
  -
  {}_{{\cal X}}\!\hat{H}_{ti}
  &=&
  D_{i}\left(
    {}_{{\cal Y}}\!h_{(VL)}
    -
    {}_{{\cal X}}\!h_{(VL)}
  \right)
  + \left(
    {}_{{\cal Y}}\!h_{(V)i}
    -
    {}_{{\cal Y}}\!h_{(V)i}
  \right)
  =
  \partial_{t}\xi_{i}
  + D_{i}\xi_{t}
  .
\end{eqnarray}
Taking the divergence of this gauge-transformation rule and
through the property $D^{i}h_{(V)i}=0$, we obtain 
\begin{eqnarray}
  \Delta\left(
    {}_{{\cal Y}}\!h_{(VL)}
    -
    {}_{{\cal X}}\!h_{(VL)}
  \right)
  =
  D^{i}\partial_{t}\xi_{i}
  +
  \Delta\xi_{t}
  \label{eq:gauge-trans-div-hatHti-decomp}
  .
\end{eqnarray}


To evaluate the first term $D^{i}\partial_{t}\xi_{i}$ in
Eq.~(\ref{eq:gauge-trans-div-hatHti-decomp}),
we note that $\partial_{t}q_{ij}$ is given by 
\begin{eqnarray}
  \partial_{t}q_{ij}
  =
  - 2 \alpha K_{ij} + 2 D_{(i}\beta_{j)}
  \label{eq:K.Nakamura-2010-note-B-21}
\end{eqnarray}
from Eq.~(\ref{eq:derivative_extrinsic_minus}) in Appendix
\ref{sec:ADM-decomposition}. 
Keep this equation in our mind, we consider the derivative
$D_{j}\partial_{t}\xi_{i}$ as in the previous subsection.
This is given by 
\begin{eqnarray}
  D_{j}\partial_{t}\xi_{i}
  &=&
  \partial_{t}D_{j}\xi_{i}
  - D_{i}\left(
    \alpha K_{kj} - D_{(k}\beta_{j)}
  \right) \xi^{k}
  \nonumber\\
  && \quad
  - D_{j}\left(
    \alpha K_{ki} - D_{(k}\beta_{i)}
  \right) \xi^{k}
  + D_{k}\left(
    \alpha K_{ij} - D_{(i}\beta_{j)}
  \right) \xi^{k}
  \label{eq:K.Nakamura-2010-note-B-23}
  .
\end{eqnarray}
From Eq.~(\ref{eq:K.Nakamura-2010-note-B-23}), we can easily
derive $D^{i}\partial_{t}\xi_{i}$ as
\begin{eqnarray}
  D^{i}\partial_{t}\xi_{i}
  &=&
      \partial_{t}D^{i}\xi_{i}
  - 2 \left(
    \alpha K^{ij} - D^{(i}\beta^{j)}
  \right) D_{j}\xi_{i}
  \nonumber\\
  && \quad
  - 2 D_{i}\left(
    \alpha K^{li} - D^{(l}\beta^{i)}
  \right) \xi_{l}
  + D^{l}\left(
    \alpha K - D^{i}\beta_{i}
  \right) \xi_{l}
  .
  \label{eq:K.Nakamura-2010-note-B-25}
\end{eqnarray}
Further, we consider the decomposition of the component
$\xi_{i}$ as Eq.~(\ref{eq:K.Nakamura-2010-2-simple-4-22}), i.e.,  
\begin{eqnarray}
  \xi_{i} = D_{i}\xi_{(L)} + \xi_{(V)i}, \quad D^{i}\xi_{(V)i} = 0.
  \label{eq:K.Nakamura-2010-note-B-25-2}
\end{eqnarray}


Through the decomposition (\ref{eq:K.Nakamura-2010-note-B-25-2})
and the formula (\ref{eq:K.Nakamura-2010-note-B-25}), the
gauge-transformation rule
(\ref{eq:gauge-trans-div-hatHti-decomp}) is given by
\begin{eqnarray}
  \Delta\left(
    {}_{{\cal Y}}\!h_{(VL)}
    -
    {}_{{\cal X}}\!h_{(VL)}
    - \xi_{t}
  \right)
  &=&
  \Delta\partial_{t}\xi_{(L)}
  - 2 \left(
    \alpha K^{ij} - D^{(i}\beta^{j)}
  \right) D_{j}\xi_{(V)i}
  \nonumber\\
  && \quad
  - 2 D_{i}\left(
    \alpha K^{li} - D^{(l}\beta^{i)}
  \right) \xi_{(V)l}
  \nonumber\\
  && \quad
  + D^{l}\left(
    \alpha K - D^{i}\beta_{i}
  \right) \xi_{(V)l}
  ,
  \label{eq:K.Nakamura-2010-note-B-26}
\end{eqnarray}
where we have used Eq.~(\ref{eq:K.Nakamura-2010-note-B-25})
twice.
Ignoring the mode which belongs to the kernel of the derivative
operator $\Delta$, we obtain 
\begin{eqnarray}
  {}_{{\cal Y}}h_{(VL)}
  - {}_{{\cal X}}h_{(VL)}
  &=&
  \xi_{t}
  + \partial_{t}\xi_{(L)}
  \nonumber\\
  && 
  +
  \Delta^{-1}
  \left[
    - 2 \left(
      \alpha K^{ij} - D^{(i}\beta^{j)}
    \right) D_{j}\xi_{(V)i}
  \right.
  \nonumber\\
  && \quad\quad\quad
  \left.
    + \left\{
      D^{l}\left(
        \alpha K - D^{i}\beta_{i}
      \right)
    \right.
  \right.
  \nonumber\\
  && \quad\quad\quad\quad\quad
  \left.
    \left.
      - 2 D_{i}\left(
        \alpha K^{li} - D^{(l}\beta^{i)}
      \right)
    \right\} \xi_{(V)l}
  \right]
  \label{eq:K.Nakamura-2010-note-B-28}
  .
\end{eqnarray}
Substituting Eq.~(\ref{eq:K.Nakamura-2010-note-B-28}) into
Eq.~(\ref{eq:gauge-trans-hatHti-decomp}) we obtain
\begin{eqnarray}
  {}_{{\cal Y}}\!h_{(V)i}
  -
  {}_{{\cal Y}}\!h_{(V)i}
  &=&
  \partial_{t}\xi_{(V)i}
  - D_{i}\Delta^{-1}
  \left[
    - 2 \left(
      \alpha K^{ij} - D^{(i}\beta^{j)}
    \right) D_{j}\xi_{(V)i}
  \right.
  \nonumber\\
  && \quad\quad\quad\quad\quad\quad\quad\quad
  \left.
    + \left\{
      D^{l}\left(
        \alpha K - D^{i}\beta_{i}
      \right)
    \right.
  \right.
  \nonumber\\
  && \quad\quad\quad\quad\quad\quad\quad\quad\quad
  \left.
    \left.
      - 2 D_{i}\left(
        \alpha K^{li} - D^{(l}\beta^{i)}
      \right)
    \right\} \xi_{(V)l}
  \right]
  \label{eq:K.Nakamura-2010-note-B-30}
  .
\end{eqnarray}


The gauge-transformation rules for $h_{L}$ and $h_{(T)ij}$ are
given from Eq.~(\ref{eq:gauge-trans-hatHij}).
Since we consider the decomposition
(\ref{eq:K.Nakamura-2010-2-generic-4-8}), the
gauge-transformation rule (\ref{eq:gauge-trans-hatHij}) is given
by  
\begin{eqnarray}
  {}_{{\cal Y}}\!\hat{H}_{ij}
  -
  {}_{{\cal X}}\!\hat{H}_{ij}
  &=&
  \frac{1}{n} q_{ij} \left(
    {}_{{\cal Y}}\!h_{(L)}
    -
    {}_{{\cal X}}\!h_{(L)}
  \right)
  + \left(
    {}_{{\cal Y}}\!h_{(T)ij}
    -
    {}_{{\cal X}}\!h_{(T)ij}
  \right)
  =
  2 D_{(i}\xi_{j)}
  \label{eq:K.Nakamura-2010-note-B-32}
  .
\end{eqnarray}
Taking the trace of Eq.~(\ref{eq:K.Nakamura-2010-note-B-32}), we
obtain 
\begin{eqnarray}
  {}_{{\cal Y}}h_{(L)} - {}_{{\cal X}}h_{(L)}
  &=&
  2 D^{i}\xi_{i}
  .
  \label{eq:K.Nakamura-2010-note-B-33}
\end{eqnarray}
The traceless part of Eq.~(\ref{eq:K.Nakamura-2010-note-B-32})
is given by 
\begin{eqnarray}
  {}_{{\cal Y}}h_{(T)ij} - {}_{{\cal X}}h_{(T)ij}
  &=&
  \left(L\xi\right)_{ij}
  .
  \label{eq:K.Nakamura-2010-note-B-34}
\end{eqnarray}
Note that the variable $h_{(T)ij}$ is also decomposed as
Eq.~(\ref{eq:K.Nakamura-2010-2-generic-4-9}) and the 
gauge-transformation rules for the variable $h_{(T)_{ij}}$ is
given by
\begin{eqnarray}
  {}_{{\cal Y}}h_{(T)ij}
  -
  {}_{{\cal X}}h_{(T)ij}
  &=&
  \left(L\left({}_{{\cal Y}}\!h_{(TV)}-{}_{{\cal X}}\!h_{(TV)}\right)\right)_{ij}
  + {}_{{\cal Y}}h_{(TT)ij} - {}_{{\cal X}}h_{(TT)ij}
  \nonumber\\
  &=&
  \left(L\xi\right)_{ij}
  .
  \label{eq:K.Nakamura-2010-note-B-35}
\end{eqnarray}
Taking the divergence of
Eq.~(\ref{eq:K.Nakamura-2010-note-B-35}), we obtain 
\begin{eqnarray}
  {\cal D}^{jl}
  \left(
    {}_{{\cal Y}}h_{(TV)l} - {}_{{\cal X}}h_{(TV)l} - \xi_{l}
  \right)
  =
  0
  .
  \label{eq:K.Nakamura-2010-note-B-37}
\end{eqnarray}
Since we ignore the modes which belong to the kernel of 
${\cal D}^{jl}$, we obtain  
\begin{eqnarray}
  \label{eq:K.Nakamura-2010-note-B-38}
  {}_{{\cal Y}}h_{(TV)l} - {}_{{\cal X}}h_{(TV)l} = \xi_{l}.
\end{eqnarray}
Through the decomposition formula
(\ref{eq:K.Nakamura-2010-2-generic-4-10}) and 
(\ref{eq:K.Nakamura-2010-note-B-25-2}), we easily derive 
\begin{eqnarray}
  \label{eq:K.Nakamura-2010-note-B-41}
  {}_{{\cal Y}}h_{(TVL)} - {}_{{\cal X}}h_{(TVL)} = \xi_{(L)}, \\
  \label{eq:K.Nakamura-2010-note-B-42}
  {}_{{\cal Y}}h_{(TVV)i} - {}_{{\cal X}}h_{(TVV)i} = \xi_{(V)i},
\end{eqnarray}
where we ignore the mode which belong to the kernel of
$\Delta$. 
Substituting Eq.~(\ref{eq:K.Nakamura-2010-note-B-38}) into
(\ref{eq:K.Nakamura-2010-note-B-35}), we obtain 
\begin{eqnarray}
  \label{eq:K.Nakamura-2010-note-B-40}
  {}_{{\cal Y}}h_{(TT)ij} - {}_{{\cal X}}h_{(TT)ij} = 0.
\end{eqnarray}


In summary, we have obtained the gauge-transformation rules for
the variables $\hat{H}_{tt}$, $h_{(VL)}$, $h_{(V)i}$, $h_{(L)}$,
$h_{(T)ij}$, $h_{(TV)i}$, $h_{(TVL)}$, $h_{(TVV)i}$, and
$h_{(TT)ij}$ as follows:
\begin{eqnarray}
  &&
  {}_{{\cal Y}}\!\hat{H}_{tt}
  -
  {}_{{\cal X}}\!\hat{H}_{tt}
  =
  2 \partial_{t}\xi_{t}
  \label{eq:K.Nakamura-2010-note-B-43}
  , \\
  &&
  {}_{{\cal Y}}h_{(VL)}
  - 
  {}_{{\cal X}}h_{(VL)}
  =
  \xi_{t}
  + \partial_{t}\xi_{(L)}
  \nonumber\\
  && \quad\quad\quad\quad\quad\quad\quad\quad\quad
  +
  \Delta^{-1}
  \left[
    - 2 \left(
      \alpha K^{ij} - D^{(i}\beta^{j)}
    \right) D_{j}\xi_{(V)i}
  \right.
  \nonumber\\
  && \quad\quad\quad\quad\quad\quad\quad\quad\quad\quad\quad\quad\quad
  \left.
    + \left\{
      D^{l}\left(
        \alpha K - D^{i}\beta_{i}
      \right)
    \right.
  \right.
  \nonumber\\
  && \quad\quad\quad\quad\quad\quad\quad\quad\quad\quad\quad\quad\quad\quad\quad
  \left.
    \left.
      - 2 D_{i}\left(
        \alpha K^{li} - D^{(l}\beta^{i)}
      \right)
    \right\} \xi_{(V)l}
  \right]
  \label{eq:K.Nakamura-2010-note-B-44}
  , \\
  &&
  {}_{{\cal Y}}\!h_{(V)i}
  -
  {}_{{\cal X}}\!h_{(V)i}
  =
  \partial_{t}\xi_{(V)i}
  - D_{i}\Delta^{-1}
  \left[
    - 2 \left(
      \alpha K^{kj} - D^{(k}\beta^{j)}
    \right) D_{j}\xi_{(V)k}
  \right.
  \nonumber\\
  && \quad\quad\quad\quad\quad\quad\quad\quad\quad\quad\quad\quad\quad\quad\quad\quad
  \left.
    + \left\{
      D^{l}\left(
        \alpha K - D^{k}\beta_{k}
      \right)
    \right.
  \right.
  \nonumber\\
  && \quad\quad\quad\quad\quad\quad\quad\quad\quad\quad\quad\quad\quad\quad\quad\quad\quad
  \left.
    \left.
      - 2 D_{k}\left(
        \alpha K^{lk} - D^{(l}\beta^{k)}
      \right)
    \right\} \xi_{(V)l}
  \right]
  \label{eq:K.Nakamura-2010-note-B-45}
  , \\
  &&
  {}_{{\cal Y}}h_{(L)} - {}_{{\cal X}}h_{(L)}
  =
  2 D^{i}\xi_{i}
  \label{eq:K.Nakamura-2010-note-B-46}
  , \\
  &&
  {}_{{\cal Y}}h_{(T)ij} - {}_{{\cal X}}h_{(T)ij}
  =
  \left(L\xi\right)_{ij}
  \label{eq:K.Nakamura-2010-note-B-47}
  , \\
  &&
  {}_{{\cal Y}}h_{(TV)l} - {}_{{\cal X}}h_{(TV)l}
  = \xi_{l}
  \label{eq:K.Nakamura-2010-note-B-48}
  ,\\
  \label{eq:K.Nakamura-2010-note-B-49}
  &&
  {}_{{\cal Y}}h_{(TVL)} - {}_{{\cal X}}h_{(TVL)}
  = \xi_{(L)}
  , \\
  \label{eq:K.Nakamura-2010-note-B-50}
  &&
  {}_{{\cal Y}}h_{(TVV)i} - {}_{{\cal X}}h_{(TVV)i}
  = \xi_{(V)i}
  , \\
  \label{eq:K.Nakamura-2010-note-B-50-1}
  &&
  {}_{{\cal Y}}h_{(TT)ij} - {}_{{\cal X}}h_{(TT)ij} = 0.
\end{eqnarray}


Here, we note that the gauge transformation rule
(\ref{eq:K.Nakamura-2010-note-B-47}) coincides with the gauge
transformation rule (\ref{eq:K.Nakamura-2010-note-B-15}) for the
variable $\hat{X}_{i}$.
Then, we may identify the variable $\hat{X}_{i}$ with
$h_{(TV)i}$:
\begin{eqnarray}
  \label{eq:K.Nakamura-2010-note-B-51}
  \hat{X}_{i} := h_{(TV)i}.
\end{eqnarray}
Thus, we have confirmed the existence of the variable
$\hat{X}_{i}$. 
Next, we show the existence of the variable $\hat{X}_{t}$ whose
gauge-transformation rule is given by
Eq.~(\ref{eq:K.Nakamura-2010-note-B-14}). 
To do this, we consider the gauge transformation rules
(\ref{eq:K.Nakamura-2010-note-B-44}),
(\ref{eq:K.Nakamura-2010-note-B-49}), and
(\ref{eq:K.Nakamura-2010-note-B-50}).
Inspecting these gauge transformation rules, we find the
definition of $\hat{X}_{t}$ as
\begin{eqnarray}
  \hat{X}_{t} 
  &:=&
  h_{(VL)}
  - \partial_{t}h_{(TVL)}
  \nonumber\\
  && 
  -
  \Delta^{-1}
  \left[
    - 2 \left(
      \alpha K^{ij} - D^{(i}\beta^{j)}
    \right) D_{j}h_{(TVV)i}
  \right.
  \nonumber\\
  && \quad\quad\quad
  \left.
    + \left\{
      D^{l}\left(
        \alpha K - D^{i}\beta_{i}
      \right)
      - 2 D_{i}\left(
        \alpha K^{li} - D^{(l}\beta^{i)}
      \right)
    \right\} h_{(TVV)l}
  \right]
  .
  \label{eq:K.Nakamura-2010-note-B-52}
\end{eqnarray}
Actually, the gauge transformation rule for $\hat{X}_{t}$
defined by Eq.~(\ref{eq:K.Nakamura-2010-note-B-52}) is given by 
Eq.~(\ref{eq:K.Nakamura-2010-note-B-14}). 
This is desired property for the variable $\hat{X}_{t}$.
Thus, we have confirm the existence of the variables
$\hat{X}_{t}$ and $\hat{X}_{i}$ which was assumed in the
definitions
(\ref{eq:hatHtt-def-generic})--(\ref{eq:hatHij-def-generic}) of 
the components of the tensor field $\hat{H}_{ab}$.


Now, we construct gauge invariant variables for the linear-order
metric perturbation. 
First, the gauge transformation rule
(\ref{eq:K.Nakamura-2010-note-B-50-1}) shows that $h_{(TT)ij}$
is gauge invariant by itself and we define the gauge-invariant 
transverse-traceless tensor by
\begin{eqnarray}
  \label{eq:K.Nakamura-2010-note-B-55}
  \chi_{ij} := h_{(TT)ij}.
\end{eqnarray}


Inspecting the gauge-transformation rules
(\ref{eq:K.Nakamura-2010-note-B-45}) and
(\ref{eq:K.Nakamura-2010-note-B-50}), we define the vector
mode $\nu_{i}$ by 
\begin{eqnarray}
  \nu_{i}
  &:=&
  h_{(V)i}
  - \partial_{t}h_{(TVV)i}
  \nonumber\\
  &&
  + D_{i}\Delta^{-1}
  \left[
    - 2 \left(
      \alpha K^{kj} - D^{(k}\beta^{j)}
    \right) D_{j}h_{(TVV)k}
  \right.
  \nonumber\\
  && \quad\quad\quad\quad\quad
  \left.
    + \left\{
      D^{l}\left(
        \alpha K - D^{k}\beta_{k}
      \right)
    \right.
  \right.
  \nonumber\\
  && \quad\quad\quad\quad\quad\quad\quad
  \left.
    \left.
      - 2 D_{k}\left(
        \alpha K^{lk} - D^{(l}\beta^{k)}
      \right)
    \right\} h_{(TVV)l}
  \right]
  .
  \label{eq:K.Nakamura-2010-note-B-56}
\end{eqnarray}
Actually, we can easily confirm that the variable $\nu_{i}$ is
gauge invariant, i.e., ${}_{{\cal Y}}\nu_{i}-{}_{{\cal X}}\nu_{i}=0$.
Through the divergenceless property $D^{i}h_{(V)i}=0$ for the
variable $h_{(V)i}$, we easily derive
\begin{eqnarray}
  D^{i}\nu_{i}
  &=&
  - D^{i}\partial_{t}h_{(TVV)i}
  + \left(
    - 2 \alpha K^{kj} + 2 D^{(k}\beta^{j)}
  \right) D_{j}h_{(TVV)k}
  \nonumber\\
  &&
  + \left\{
    - 2 D_{k}\left(
      \alpha K^{lk} - D^{(l}\beta^{k)}
    \right)
    + D^{l}\left(
      \alpha K - D^{k}\beta_{k}
    \right)
  \right\} h_{(TVV)l}
  .
  \label{eq:K.Nakamura-2010-note-B-59}
\end{eqnarray}
Further, through the formula
(\ref{eq:K.Nakamura-2010-note-B-25}) for the variable
$h_{(TVV)i}$ and the divergenceless property of the variable
$h_{(TVV)i}$, we easily see the divergenceless property 
$D^{i}\nu_{i} = 0$.


Next, we consider the scalar modes. 
First, inspecting gauge-transformation rules
(\ref{eq:K.Nakamura-2010-note-B-14}) and
(\ref{eq:gauge-trans-hatHtt}), we define the scalar variable
$\Phi$ by
\begin{eqnarray}
  - 2 \Phi
  :=
  \hat{H}_{tt}
  - 2 \partial_{t}\hat{X}_{t}
  .
  \label{eq:K.Nakamura-2010-note-B-62}
\end{eqnarray}
Actually, we can easily confirm that this variable $\Phi$ is
gauge invariant.
Inspecting the gauge-transformation rules
(\ref{eq:K.Nakamura-2010-note-B-15}) and 
(\ref{eq:K.Nakamura-2010-note-B-33}), we define another
gauge-invariant variable $\Psi$ by   
\begin{eqnarray}
  - 2 n \Psi 
  :=
  h_{(L)} - 2 D^{i}\hat{X}_{i}
  .
  \label{eq:K.Nakamura-2010-note-B-64}
\end{eqnarray}
We can easily confirm the gauge invariance of the variable
$\Psi$ through the gauge-transformation rules
(\ref{eq:K.Nakamura-2010-note-B-15}) and
(\ref{eq:K.Nakamura-2010-note-B-33}).


In summary, we have defined gauge invariant variables as
follows: 
\begin{eqnarray}
  - 2 \Phi
  &:=&
  \hat{H}_{tt}
  - 2 \partial_{t}\hat{X}_{t}
  \label{eq:K.Nakamura-2010-note-B-66}
  , \\
  - 2 n \Psi 
  &:=&
  h_{(L)} - 2 D^{i}\hat{X}_{i}
  \label{eq:K.Nakamura-2010-note-B-67}
  , \\
  \nu_{i}
  &:=&
  h_{(V)i}
  - \partial_{t}h_{(TVV)i}
  \nonumber\\
  &&
  + D_{i}\Delta^{-1}
  \left[
    - 2 \left(
      \alpha K^{kj} - D^{(k}\beta^{j)}
    \right) D_{j}h_{(TVV)k}
  \right.
  \nonumber\\
  && \quad\quad\quad\quad\quad
  \left.
    + \left\{
      D^{l}\left(
        \alpha K - D^{k}\beta_{k}
      \right)
    \right.
  \right.
  \nonumber\\
  && \quad\quad\quad\quad\quad\quad\quad
  \left.
    \left.
      - 2 D_{k}\left(
        \alpha K^{lk} - D^{(l}\beta^{k)}
      \right)
    \right\} h_{(TVV)l}
  \right]
  \label{eq:K.Nakamura-2010-note-B-68}
  , \\
  \chi_{ij} &:=& h_{(TT)ij}.
  \label{eq:K.Nakamura-2010-note-B-69}
\end{eqnarray}
In terms of these gauge-invariant variables and the variables
$\hat{X}_{t}$ and $\hat{X}_{i}$, which are defined by
Eqs.~(\ref{eq:K.Nakamura-2010-note-B-52}) and
(\ref{eq:K.Nakamura-2010-note-B-51}), respectively, the original
components $\{h_{tt}$, $h_{ti}$, $h_{ij}\}$ of the metric
perturbation $h_{ab}$ is given by 
\begin{eqnarray}
  h_{tt}
  &=&
  - 2 \Phi
  + 2 \partial_{t}\hat{X}_{t}
  - \frac{2}{\alpha}\left(
    \partial_{t}\alpha 
    + \beta^{i}D_{i}\alpha 
    - \beta^{j}\beta^{i}K_{ij}
  \right) \hat{X}_{t}
  \nonumber\\
  && \quad
  - \frac{2}{\alpha} \left(
    \beta^{i}\beta^{k}\beta^{j} K_{kj}
    - \beta^{i} \partial_{t}\alpha
    + \alpha q^{ij} \partial_{t}\beta_{j}
  \right.
  \nonumber\\
  && \quad\quad\quad\quad
  \left.
    + \alpha^{2} D^{i}\alpha 
    - \alpha \beta^{k} D^{i} \beta_{k}
    - \beta^{i} \beta^{j} D_{j}\alpha 
  \right)\hat{X}_{i}
  \label{eq:K.Nakamura-2010-note-B-73}
  , \\
  h_{ti}
  &=&
  \nu_{i}
  + D_{i}\hat{X}_{t} 
  + \partial_{t}\hat{X}_{i}
  - \frac{2}{\alpha} \left(
    D_{i}\alpha 
    - \beta^{j}K_{ij}
  \right) \hat{X}_{t}
  \nonumber\\
  && 
  - \frac{2}{\alpha} \left(
    - \alpha^{2} K^{j}_{\;\;i}
    + \beta^{j}\beta^{k} K_{ki}
    - \beta^{j} D_{i}\alpha 
    + \alpha D_{i}\beta^{j}
  \right) \hat{X}_{j}
  \label{eq:K.Nakamura-2010-note-B-74}
  , \\
  h_{ij}
  &=&
  - 2 \Psi q_{ij}
  + \chi_{ij}
  + D_{i}\hat{X}_{j}
  + D_{j}\hat{X}_{i}
  + \frac{2}{\alpha} K_{ij} \hat{X}_{t}
  - \frac{2}{\alpha} \beta^{k} K_{ij} \hat{X}_{k}
  \label{eq:K.Nakamura-2010-note-B-75}
  .
\end{eqnarray}
On the other hand, the component representations of the
decomposition formula (\ref{eq:linear-metric-decomp}) with 
\begin{eqnarray}
  X_{a} =: X_{t}(dt)_{a} + X_{i}(dx^{i})_{a}
\end{eqnarray}
are given by 
\begin{eqnarray}
  h_{tt}
  &=&
  {\cal H}_{tt}
  + 2 \partial_{t}X_{t}
  - \frac{2}{\alpha}\left(
    \partial_{t}\alpha 
    + \beta^{i}D_{i}\alpha 
    - \beta^{k}\beta^{i}K_{ij}
  \right) X_{t}
  \nonumber\\
  &&
  - \frac{2}{\alpha} \left(
    \beta^{i}\beta^{k}\beta^{j} K_{kj}
    - \beta^{i} \partial_{t}\alpha
    + \alpha q^{ij} \partial_{t}\beta_{j}
  \right.
  \nonumber\\
  && \quad\quad\quad
  \left.
    + \alpha^{2} D^{i}\alpha 
    - \alpha \beta^{k} D^{i} \beta_{k}
    - \beta^{i} \beta^{j} D_{j}\alpha 
  \right) X_{i}
  \label{eq:K.Nakamura-2010-note-B-80}
  , \\
  h_{ti}
  &=&
  {\cal H}_{ti}
  + \partial_{t}X_{i}
  + D_{i}X_{t}
  - \frac{2}{\alpha} \left(
    D_{i}\alpha 
    - \beta^{j}K_{ij}
  \right) X_{t}
  \nonumber\\
  &&
  - \frac{2}{\alpha} \left(
    - \alpha^{2} K^{j}_{\;\;i}
    + \beta^{j}\beta^{k} K_{ki}
    - \beta^{j} D_{i}\alpha 
    + \alpha D_{i}\beta^{j}
  \right) X_{j}
  \label{eq:K.Nakamura-2010-note-B-81}
  , \\
  h_{ij}
  &=&
  {\cal H}_{ij}
  + D_{i}X_{j}
  + D_{j}X_{i}
  + \frac{2}{\alpha}K_{ij} X_{t}
  - \frac{2}{\alpha} \beta^{k} K_{ij} X_{k}
  \label{eq:K.Nakamura-2010-note-B-82}
  .
\end{eqnarray}
Comparing
Eqs.~(\ref{eq:K.Nakamura-2010-note-B-80})--(\ref{eq:K.Nakamura-2010-note-B-82})
with
Eqs.~(\ref{eq:K.Nakamura-2010-note-B-73})--(\ref{eq:K.Nakamura-2010-note-B-75}),
we may identify the components of the gauge-invariant variables
${\cal H}_{ab}$ so that 
\begin{eqnarray}
  {\cal H}_{tt} := - 2 \Phi
  , \quad
  {\cal H}_{ti} := \nu_{i}
  , \quad
  {\cal H}_{ij} := - 2 \Psi q_{ij} + \chi_{ij}
\end{eqnarray}
and the components of the gauge-variant variables $X_{a}$ so
that
\begin{eqnarray}
  X_{t} := \hat{X}_{t}, \quad
  X_{i} := \hat{X}_{i}.
\end{eqnarray}
Thus, the decomposition formula
(\ref{eq:linear-metric-decomp}) is correct for the linear-order
perturbation on a generic background spacetime, if we assume the
existence of two Green function of the derivative operators
$\Delta:=D^{i}D_{i}$ and ${\cal D}^{ij}$ which is defined by
Eq.~(\ref{eq:J.W.York.Jr-1973-1-2-kouchan-3-main}).
In other words, in the above proof, we ignore the modes which
belong to the kernel of these derivative operators $\Delta$ and
${\cal D}^{jl}$.
To take these modes into account, the different treatments are
necessary.


\section{Comparison with the FRW background case}
\label{sec:K.Nakamura-2010-4}


In this section, we consider the comparison with the case where
the background spacetime ${\cal M}_{0}$ is a homogeneous and
isotropic universe which is discussed in
KN2007\cite{kouchan-cosmo-second}. 
This case corresponds to the case $\alpha=1$, $\beta^{i}=0$ and
$K_{ij}=-Hq_{ij}$, where $H=\partial_{t}a/a$ and $a$ is the scale
factor of the universe.


In the paper KN2007\cite{kouchan-cosmo-second}, we consider the
decomposition of the components $h_{ti}$ and $h_{ij}$ of the
metric perturbation $h_{ab}$ as 
\begin{eqnarray}
  \label{eq:KN2007-Prog-4.2}
  h_{ti}
  &=&
  \tilde{D}_{i}\tilde{h}_{(VL)} + \tilde{h}_{(V)i}
  , \quad
  \tilde{D}^{i}\tilde{h}_{(V)i} = 0
  , \\
  \label{eq:KN2007-Prog-4.3}
  h_{ij}
  &=&
  a^{2} \tilde{h}_{(L)} \gamma_{ij} + a^{2} \tilde{h}_{(T)ij}
  , \quad
  \gamma^{ij}\tilde{h}_{(T)ij}=0, \\
  \label{eq:KN2007-Prog-4.4}
  \tilde{h}_{(T)ij}
  &=&
  \left(
    \tilde{D}_{i}\tilde{D}_{j} - \frac{1}{n} \gamma_{ij} \tilde{\Delta}
  \right) \tilde{h}_{(TL)} 
  + 2 \tilde{D}_{(i}\tilde{h}_{(TV)j)}
  + \tilde{h}_{(TT)ij}
  , \\
  \label{eq:KN2007-Prog-4.5}
  \tilde{D}^{i}\tilde{h}_{(TV)i} &=& 0
  , \quad
  \tilde{D}^{i}\tilde{h}_{(TT)ij} = 0,
\end{eqnarray}
where $q_{ij}=a^{2}\gamma_{ij}$, $\gamma_{ij}$ is the metric
on a maximally symmetric space, $\tilde{D}_{i}$ is the covariant 
derivative associated with the metric $\gamma_{ij}$, and
$\tilde{\Delta}:=\tilde{D}^{i}\tilde{D}_{i}$. 
This decomposition is slightly different from the decomposition
(\ref{eq:K.Nakamura-2010-2-simple-4-7})--(\ref{eq:K.Nakamura-2010-2-simple-4-10})
with the definition (\ref{eq:hatH-def-case2}) of the variable
$\hat{H}_{ab}$.
Furthermore, as noted in KN2007\cite{kouchan-cosmo-second},
there should exist Green functions of the derivative operators
$\tilde{\Delta}$, $\tilde{\Delta}+2K$, and $\tilde{\Delta} + 3K$
to guarantee the one to one correspondence of the set $\{h_{tt}$,
$h_{ti}$, $h_{ij}\}$ and $\{\{h_{tt}$, $h_{(VL)}$, $h_{(L)}$,
$h_{(TL)}\}$, $\{h_{(V)i}$, $h_{(TV)i}\}$, $h_{(TT)ij}\}$, where
$K$ is the curvature constant on the maximally symmetric space.
The special modes which belong to the kernel of the derivative
operators $\tilde{\Delta}$, $\tilde{\Delta}+(n-1)K$, and
$\tilde{\Delta} + nK$ were not included in the consideration of
the paper KN2007\cite{kouchan-cosmo-second}.
On the other hand, in this paper, we ignore the modes which
belong to the kernel of the derivative operator $\Delta$ and
${\cal D}^{ij}$.
In this section, we briefly discuss these correspondence.


First, we note that the decomposition (\ref{eq:KN2007-Prog-4.2})
of the component $h_{ti}$ of the metric perturbation $h_{ab}$ is
equivalent to (\ref{eq:K.Nakamura-2010-2-simple-4-7}).
Although the tiny difference between
Eq.~(\ref{eq:KN2007-Prog-4.2}) and
(\ref{eq:K.Nakamura-2010-2-simple-4-7}) is in the definition of
the covariant derivatives $D_{i}$ (associated with the metric
$q_{ij}$) and $\tilde{D}_{i}$ (associated with the metric
$\gamma_{ij}:=(1/a^{2})q_{ij}$), we may say that
$\tilde{h}_{(VL)}$ and $\tilde{h}_{(V)i}$ in
Eq.~(\ref{eq:KN2007-Prog-4.2}) are identical with $h_{(VL)}$ and
$h_{(V)i}$ in Eq.~(\ref{eq:K.Nakamura-2010-2-simple-4-7}),
respectively.


We also note that the trace parts of these two decompositions
are almost equivalent.
Actually, since the extrinsic curvature $K_{ij}$ on the
background $\Sigma$ is proportional to the intrinsic metric
$q_{ij}$ in this case, we easily see that 
\begin{eqnarray}
  q^{ij}h_{ij}
  &=&
  q^{ij}\hat{H}_{ij} + 2 q^{ij} K_{ij} \hat{X}_{t}
  \nonumber\\
  &=&
  h_{(L)} - 2 n H \left(
    h_{(VL)}
    - \partial_{t}h_{(TVL)}
    + 2 H h_{(TVL)}
  \right)
  ,
\end{eqnarray}
where we used Eq.~(\ref{eq:K.Nakamura-2010-2-simple-4-53}). 
On the other hand, the trance part of $h_{ij}$ given by
Eq.~(\ref{eq:KN2007-Prog-4.3}) is $n\tilde{h}_{(L)}$.
Thus, the variable $\tilde{h}_{(L)}$ in
Eq.~(\ref{eq:KN2007-Prog-4.3}) corresponds to the variables in
this paper as
\begin{eqnarray}
  \tilde{h}_{(L)}
  =
  \frac{1}{n} h_{(L)}
  - 2 H \left(
    h_{(VL)}
    - \partial_{t}h_{(TVL)}
    + 2 H h_{(TVL)}
  \right)
  .
\end{eqnarray}


Since the extrinsic curvature $K_{ij}$ in
Eq.~(\ref{eq:hatH-def-case2}) is proportional to the intrinsic
metric $q_{ij}$ in this case, the main difference between
decompositions
(\ref{eq:K.Nakamura-2010-2-simple-4-8})--(\ref{eq:K.Nakamura-2010-2-simple-4-10})
and (\ref{eq:KN2007-Prog-4.3})--(\ref{eq:KN2007-Prog-4.5}) are
in the traceless part.
The traceless part $\tilde{h}_{(T)ij}$ in
Eq.~(\ref{eq:KN2007-Prog-4.4}) is also given by 
\begin{eqnarray}
  \tilde{h}_{(T)ij}
  &=&
  \left(
    \tilde{D}_{i}\tilde{D}_{j} - \frac{1}{n} \gamma_{ij} \tilde{\Delta}
  \right) h_{(TL)} 
  + 2 \tilde{D}_{(i}\tilde{h}_{(TV)j)}
  + \tilde{h}_{(TT)ij}
  \nonumber\\
  &=&
  a^{2} D_{i}\left(
    \frac{1}{2} D_{j}\tilde{h}_{(TL)} + \tilde{h}_{(TV)j}
  \right)
  + a^{2} D_{j}\left(
    \frac{1}{2} D_{i}\tilde{h}_{(TL)} + \tilde{h}_{(TV)i}
  \right)
  \nonumber\\
  &&
  - \frac{2}{n} q_{ij} a^{2} D^{k}\left(
    \frac{1}{2} D_{k}\tilde{h}_{(TL)} + \tilde{h}_{(TV)k}
  \right)
  + a^{2} \tilde{h}_{(TT)ij}
  ,
\end{eqnarray}
where we used 
\begin{eqnarray}
  0 = \tilde{D}^{k}\tilde{h}_{(TV)k} = a^{2} D^{k}\tilde{h}_{(TV)k}.
\end{eqnarray}
Comparing (\ref{eq:K.Nakamura-2010-2-simple-4-9}), we obtain the
correspondence of the variables 
\begin{eqnarray}
  h_{(TV)i}
  &=&
  \frac{1}{2} a^{2} D_{j}\tilde{h}_{(TL)} + a^{2} \tilde{h}_{(TV)j}
  , \quad
  h_{(TT)ij}
  =
  a^{2} \tilde{h}_{(TT)ij}
  , \\
  h_{(TVL)}
  &=&
  \frac{1}{2} a^{2} \tilde{h}_{(TL)}
  , \quad
  h_{(TVV)i}
  =
  a^{2} \tilde{h}_{(TV)i}
  .
\end{eqnarray}
Therefore, in the case of the homogeneous isotropic universe,
the decomposition
(\ref{eq:K.Nakamura-2010-2-simple-4-7})--(\ref{eq:K.Nakamura-2010-2-simple-4-10})
is equivalent to the decomposition
(\ref{eq:KN2007-Prog-4.2})--(\ref{eq:KN2007-Prog-4.5}).


However, in the case of the generic background spacetime, the
decomposition
(\ref{eq:KN2007-Prog-4.2})--(\ref{eq:KN2007-Prog-4.5}) is
ill-defined.
Actually, if we regard that the decomposition
(\ref{eq:KN2007-Prog-4.2})--(\ref{eq:KN2007-Prog-4.5}) is that
for the generic background spacetime, we cannot separate
$\tilde{h}_{(TL)}$ and $\tilde{h}_{(TV)j}$ due to the
non-trivial curvature terms of the background ${\cal M}_{0}$ as
pointed out by Deser\cite{S.Deser-1967}.
These curvature terms come from the commutation relation between
the covariant derivative $D_{i}$ and the derivative operator 
${\cal D}^{ij}$.
This is why we apply the decomposition
(\ref{eq:K.Nakamura-2010-2-simple-4-7})--(\ref{eq:K.Nakamura-2010-2-simple-4-10})
instead of
(\ref{eq:KN2007-Prog-4.2})--(\ref{eq:KN2007-Prog-4.5}).


Finally, we consider the correspondence of the special modes
which we ignore in this paper and
KN2007\cite{kouchan-cosmo-second}. 
Trivially, the above operator 
$\tilde{\Delta}:=\tilde{D}^{i}\tilde{D}_{i}$ corresponds to the
Laplacian $\Delta$ in this paper. 
The above derivative operator $\tilde{\Delta} + (n-1)K$
corresponds to the derivative operator ${\cal D}^{ij}$.
In the case of the maximally symmetric $n$-space, the
Riemann curvature and Ricci curvature are given by 
\begin{eqnarray}
  \label{eq:curvature-tensors-in-maximally-symmetric-n-space}
  {}^{(n)}R_{ijkl} = 2 K q_{k[i}q_{j]l}
  = 2 K q_{k[i}q_{j]l},
  \quad
  R_{ik} = q^{jl}{}^{(n)}R_{ijkl}
  = (n-1) K q_{ik}.
\end{eqnarray}
In this case, the derivative operator ${\cal D}^{ij}$ defined by
Eq.~(\ref{eq:J.W.York.Jr-1973-1-2-kouchan-3-main}) is given by 
\begin{eqnarray}
  {\cal D}^{ij}
  =
  q^{ij}\left(
    \Delta + (n - 1) K
  \right) + \left(1 - \frac{2}{n}\right) D^{i}D^{j}.
\end{eqnarray}
When the operator ${\cal D}^{ij}$ acts on an arbitrary
transverse vector field $v_{i}$ ($D^{i}v_{i}=0$), we easily see
that 
\begin{eqnarray}
  {\cal D}^{ij}v_{j} = \left( \Delta + (n - 1) K \right) v^{i}.
\end{eqnarray}
Finally, we point out that the above derivative operator
$\tilde{\Delta}+nK$ appears in the case where the derivative
operator ${\cal D}_{j}^{\;\;l}$ acts on the gradient $D_{l}f$ of
an arbitrary scalar function $f$. 
Actually, we easily see that 
\begin{eqnarray}
  {\cal D}^{jl} D_{l}f
  &=&
  2 \frac{n-1}{n} \left[
      D^{j}\Delta
    + \frac{n}{n-1} R^{jl} D_{l}
  \right] f
  \label{eq:K.Nakamura-2010-note-B-92}
  .
\end{eqnarray}
In the case of maximally symmetric $n$-space, curvature tensors
are given by
Eqs.~(\ref{eq:curvature-tensors-in-maximally-symmetric-n-space})
and the derivative operator ${\cal D}^{jl} D_{l}$ is
given by 
\begin{eqnarray}
  {\cal D}^{jl} D_{l}f
  &=&
  2 \frac{n-1}{n} D^{j} \left(
    \Delta + n K 
  \right) f
  \label{eq:K.Nakamura-2010-note-B-95}
  .
\end{eqnarray}
When we solve the equation 
\begin{eqnarray}
  {\cal D}^{jl} D_{l}f = g^{j}, 
\end{eqnarray}
we have to use the Green function $\Delta$ and
$\Delta + n K$.
These are the reason for the fact that the Green functions
$\Delta^{-1}$, $\left(\Delta+(n-1)K\right)^{-1}$, and
$\left(\Delta+nK\right)^{-1}$ were necessary to guarantee the 
one-to-one correspondence between the components $\{h_{ti}$,
$h_{ij}\}$ and $\{\tilde{h}_{(VL)}$, $\tilde{h}_{(V)i}$,
$\tilde{h}_{(L)}$, $\tilde{h}_{(TL)}$, $\tilde{h}_{(TV)i}$,
$\tilde{h}_{(TT)ij}\}$ in
Eqs.~(\ref{eq:KN2007-Prog-4.2})--(\ref{eq:KN2007-Prog-4.5}).
In other words, we may say that the special modes belong to the
kernel of the derivative operators $\Delta$ and ${\cal D}^{ij}$
which are ignored in this paper are equivalent to the special
modes which belong to the kernel of the derivative operators
$\Delta$, $\Delta+(n-1)K$, and $\Delta+nK$ which are ignored in
the paper KN2007\cite{kouchan-cosmo-second}.


\section{Summary and discussions}
\label{sec:summary}


In summary, after reviewing the general framework of the
higher-order gauge-invariant perturbation theory in general
relativity, we prove Conjecture
\ref{conjecture:decomposition-conjecture} for generic background
spacetime which admits ADM decomposition.
In this proof, we assumed the existence of Green functions of
the elliptic derivative operators $\Delta$ and ${\cal D}^{ij}$.
Roughly speaking, Conjecture
\ref{conjecture:decomposition-conjecture} states that we know
the procedure to decompose the linear-order metric perturbation
$h_{ab}$ into its gauge-invariant part ${\cal H}_{ab}$ and
gauge-variant part $X_{a}$.
In the cosmological perturbation case, this conjecture is
confirmed and the second-order cosmological perturbation theory
was developed in our series of
papers\cite{kouchan-cosmo-second,kouchan-second-cosmo-matter,kouchan-second-cosmo-consistency}. 
However, as reviewed in
\S\ref{sec:General-framework-of-the-gauge-invariant-perturbation-theory},
Conjecture \ref{conjecture:decomposition-conjecture} was the only
non-trivial part when we consider the general framework of
gauge-invariant perturbation theory on generic background
spacetimes.
Although there may exist many approaches to prove Conjecture
\ref{conjecture:decomposition-conjecture}, in this paper, we
just proposed a proof for generic background spacetimes.


As noted above, in our proof, we assume the existence of the
Green functions for the elliptic derivative operators $\Delta$
and ${\cal D}^{ij}$.
This assumption implies that we have ignored the modes which
belong to the kernel of these derivative operators.
Within the arguments in this paper, there is no information for
the treatment of these mode.
To discuss these modes, different treatments of perturbations
are necessary.
We call this problem as {\it zero-mode problem}.
The situation is similar to the cosmological perturbation case
as noted in \S\ref{sec:K.Nakamura-2010-4} and zero-mode problem
exists even in the cosmological perturbation case.
In the cosmological perturbation case, zero-mode means the modes 
which belong to the kernel of the derivative operator $\Delta$,
$\Delta + (n-1)K$, and $\Delta + n K$, where $K$ is the
curvature constant of the maximally symmetric space in cosmology
and $n$ is the dimension of this maximally symmetric space.


This zero-mode problem in cosmological perturbations also
corresponds to the $l=0$ and $l=1$ mode problem in perturbation
theory on spherically symmetric background spacetimes.
In the perturbation theory on spherically symmetric background
spacetimes, we consider the similar decomposition to
Eqs.~(\ref{eq:KN2007-Prog-4.3})--(\ref{eq:KN2007-Prog-4.5}) and
the indices $i,j,...$ in these equations correspond to the
indices of the components of a tensor field on $S^{2}$.
Since $S^{2}$ is a 2-dimensional maximally symmetric space with
the positive curvature, we may regard $n=2$ and $K=1$.
Then, the above three derivative operators are given by
$\Delta$, $\Delta + 1$, and $\Delta + 2$.
Since the eigenvalue of the Laplacian $\Delta$ on $S^{2}$ is
given by $\Delta = -l(l+1)$, we may say that the modes with
$l=0$ and $l=1$ belong to the kernel of the derivative operator
$\Delta$, $\Delta + (n-1) K$, and $\Delta + n K$.
Therefore, we may say that the problem concerning about the
modes with $l=0$ and $l=1$ in the perturbations on spherically
symmetric background spacetime is the same problem as the
zero-mode problem mentioned above.


Thus, the arguments in this paper shows that zero-mode problem
generally appears in many perturbation theories in general
relativity we have seen that the appearance of this zero-mode
problem from general point of view.
To resolve this zero-mode problem, carefully discussions on
domains of functions for perturbations will be necessary. 
We leave this zero-mode problem as a future work.


Although we should take care of the zero-mode problem, we have
almost completed the general framework of the higher-order
gauge-invariant perturbation theory in general relativity.
The proof of Conjecture
\ref{conjecture:decomposition-conjecture} shown in this paper
gives rise to the possibility of the application of our general
framework for the higher-order gauge-invariant perturbation
theory not only to cosmological
perturbations\cite{kouchan-cosmo-second,kouchan-second-cosmo-matter,kouchan-second-cosmo-consistency}
but also to perturbations of black hole spacetimes or
perturbations of general relativistic stars.
Therefore, we may say that the wide applications of our
gauge-invariant perturbation theory are opened due to the
discussions in this paper.
We also leave these development of gauge-invariant perturbation
theories for these background spacetimes as future works.


\section*{Acknowledgments}


The author deeply acknowledged to Professor Robert Manuel Wald
for valuable discussions when the author visited to Chicago
University in 2004.
This work is motivated by the discussions at that time.
The author also thanks Professor Masa-Katsu Fujimoto in National
Astronomical Observatory of Japan for his various support.


\appendix
\section{ADM decomposition}
\label{sec:ADM-decomposition}


Here, we briefly review the ADM
decomposition\cite{R.Arnowitt-S.Deser-C.W.Misner-1962,Wald-book}.


We consider the $n+1$-dimensional spacetime $({\cal M},g_{ab})$.
The topology of ${\cal M}$ is given by 
${\cal M}=\RF^{1}\times\Sigma$, where $\Sigma$ is the
$n$-dimensional manifold. 
This means that the entire ${\cal M}$ is foliated by the
one-parameter family of the manifolds $\Sigma_{t}$ where $t$ is
the parameter along $\RF^{1}$ in $\RF^{1}\times\Sigma$.
Here, we note that it is not necessary to impose that the entire
spacetime ${\cal M}$ is decomposed into $\RF^{1}\times\Sigma$ in
the global sense.
However, in this section, we impose that there exists a
one-parameter family of $n$-dimensional submanifolds
$\Sigma_{t}$ in a $n+1$-dimensional manifold ${\cal M}$, for
simplicity.


In general relativity, $t$ is regarded as the time function on
${\cal M}$ and the decomposition of ${\cal M}$ into
$\RF^{1}\times\Sigma$ is regarded as the $n+1$-decomposition of
the spacetime ${\cal M}$ into the space $\Sigma$ and time
$\RF^{1}$. 
From the view point of this decomposition, we can describe the
${\cal M}$ by the time-evolution of $\Sigma$, i.e., the geometry 
of ${\cal M}$ is described in terms of the geometry of $\Sigma$
which is embedded in ${\cal M}$.
Any geometrical quantities on ${\cal M}$ is given in terms of
the geometry of $\Sigma$ and its ``time-evolution''.


First, we consider the $n+1$-decomposition of the metric
$g_{ab}$ on ${\cal M}$.
Let $t^{a}$ be a vector field on ${\cal M}$ satisfying
\begin{equation}
  t^{a}\nabla_{a}t=1,
  \label{eq:ta-def}
\end{equation}
i.e.,
\begin{equation}
  t^{a} = \left(\frac{\partial}{\partial t}\right)^{a}.
\end{equation}
Let us call the direction along which the function $t$
increases as the future direction and $t^{a}$ defined by
(\ref{eq:ta-def}) is called future-directed, the direction along
which the function $t$ decrease as the past direction.


On the other hand, let us denote the unit normal to $\Sigma_{t}$
by $n^{a}$, which is hypersurface orthogonal.
The normalization condition for $n^{a}$ is given by 
\begin{equation}
  n^{a}n_{a} = -1.
  \label{eq:norm-of-na}
\end{equation}
The metric $g_{ab}$ on ${\cal M}$ induces a metric $q_{ab}$ on
$\Sigma_{t}$. 
On each $\Sigma_{t}$, $q_{ab}$ is given by 
\begin{equation}
  q_{ab} := g_{ab} + n_{a} n_{b}.
  \label{eq:q_ab_def}
\end{equation}
The overall signature of $n^{a}$ is chosen so that $n^{a}$ is
future-directed.
Then there is a positive function $\alpha$ so that 
\begin{equation}
  \label{eq:lapse-function-def}
  n^{a}\nabla_{a}t =: \frac{1}{\alpha}.
\end{equation}
This positive function $\alpha$ is called the lapse function
with respect to $t^{a}$. 
Further, we impose the $n_{a}$ is the hypersurface orthogonal to
the hypersurfaces $\Sigma_{t}$, which implies that
$n_{a}\propto\nabla_{a}t$. 
Due to the normalization condition (\ref{eq:norm-of-na}), we
easily see that 
\begin{equation}
  \label{eq:normal-form-and-lapse-function-minus}
  n_{a} = - \alpha \nabla_{a}t = - \alpha \left(dt\right)_{a}.
\end{equation}
From Eq.~(\ref{eq:normal-form-and-lapse-function-minus}), we
decompose the vector field $t^{a}$ into its normal and
tangential parts to $\Sigma_{t}$
\begin{eqnarray}
  \alpha = - t^{a} n_{a}, \quad \beta_{a} := q_{ab} t^{b},
\end{eqnarray}
where $\beta_{a}$ is called the shift vector with respect to
$t^{a}$. 
Equivalently, the vector field $t^{a}$ is decomposed as
\begin{equation}
  t^{a} = t^{b}\delta_{b}^{a} 
  = \beta^{a} + \alpha n^{a}.
  \label{eq:t-is-decomposedby-beta-and-n}
\end{equation}
From Eq.~(\ref{eq:ta-def}) and the definition
(\ref{eq:lapse-function-def}) of the lapse function $\alpha$, we
easily see that
\begin{equation}
  1 = t^{a}\nabla_{a}t = \alpha n^{a}\nabla_{a}t + \beta^{a}\nabla_{a}t
  = 1 + \beta^{a}\nabla_{a}t,
\end{equation}
which yield
\begin{equation}
  \beta^{a}\nabla_{a}t = 0,
  \label{eq:no-time-component-in-beta}
\end{equation}
i.e., $\beta^{a}$ has no component along $(\partial/\partial t)^{a}$.
Further, due to the normalization (\ref{eq:norm-of-na}) of the
vector $n_{a}$, the lapse function $\alpha$ is also given by
\begin{equation}
  \alpha^{2} = - \frac{1}{g^{ab}(\nabla_{a}t)(\nabla_{b}t)}.
\end{equation}
The decomposition (\ref{eq:t-is-decomposedby-beta-and-n}) of the
vector field $t^{a}=\left(\partial/\partial t\right)^{a}$ is also 
yields
\begin{equation}
  n^{a} = \frac{1}{\alpha} \left[
    t^{a} - \beta^{a}
  \right]
  .
  \label{eq:nua-decomp-3}
\end{equation}


Introducing the spatial coordinate so that $(t, x^{i})$ is the
spacetime coordinate, i.e., the coordinate basis of the tangent
space on the spacetime is the set 
\begin{equation}
  \left\{ (dt)_{a}, (dx^{i})_{a}\right\}, \quad
  \left\{ \left(\frac{\partial}{\partial t}\right)^{a}, 
    \left(\frac{\partial}{\partial x^{i}}\right)^{a} \right\},
    \label{eq:coordinate-basis-2}
\end{equation}
$\beta^{a}$ is given by
\begin{equation}
  \beta^{a} = \beta^{i} \left(\frac{\partial}{\partial x^{i}}\right)^{a},
\end{equation}
because of Eq.~(\ref{eq:no-time-component-in-beta}).
Then, (\ref{eq:nua-decomp-3}) is given in terms of the coordinate
system
\begin{equation}
  n^{a} = \frac{1}{\alpha} \left[
    \left(\frac{\partial}{\partial t}\right)^{a} 
    - \beta^{i}\left(\frac{\partial}{\partial x^{i}}\right)^{a} 
  \right]
  .
  \label{eq:nua-decomp-component}
\end{equation}
From 
\begin{equation}
  \nabla_{a}t
  = \delta_{a}^{\;\;b} \nabla_{b}t
  = \left(q_{a}^{\;\;c} - n_{a}n^{c}\right) \nabla_{c}t
  q_{a}^{\;\;b}\nabla_{b} t - n_{a} \frac{1}{\alpha}.
\end{equation}
and (\ref{eq:normal-form-and-lapse-function-minus}), we can also
see that
\begin{equation}
  q_{a}^{\;\;b}\nabla_{b}t = 0.
\end{equation}
This implies 
\begin{equation}
  q^{ab}\nabla_{b}t = q^{ba}\nabla_{b}t = 0,
\end{equation}
which yields $q^{ab}$ as no component along $(\partial/\partial t)^{a}$.
This means that the induced inverse metric $q^{ab}$ has the
following component representation
\begin{equation}
  q^{ab} = q^{ij} \left(\frac{\partial}{\partial x^{i}}\right)^{a} 
  \left(\frac{\partial}{\partial x^{j}}\right)^{b}
  .
  \label{eq:component-representation-of-quab-minus}
\end{equation}
Together with the component representation
(\ref{eq:nua-decomp-3}) and
(\ref{eq:component-representation-of-quab-minus}), we obtain the
spacetime inverse metric $g^{ab}$ in terms of the coordinate 
system $(t,x^{i}$: 
\begin{eqnarray}
  g^{ab} &=& - n^{a} n^{b} + q^{ab}, \\
  &=& 
  \frac{\epsilon}{\alpha^{2}} 
  \left\{
    \left(\frac{\partial}{\partial t}\right)^{a}
    - \beta^{i}
    \left(\frac{\partial}{\partial x^{i}}\right)^{a}
  \right\}
  \left\{
    \left(\frac{\partial}{\partial t}\right)^{b}
    - \beta^{j}
    \left(\frac{\partial}{\partial x^{j}}\right)^{b}
  \right\}
  \nonumber\\
  && \quad
  + q^{ij}
  \left(\frac{\partial}{\partial x^{i}}\right)^{a}
  \left(\frac{\partial}{\partial x^{j}}\right)^{b}.
  \label{eq:gab-decomp-uu-minus}
\end{eqnarray}
Through the relation 
\begin{equation}
  g_{ab}g^{bc} = \delta_{a}^{\;\;c},
\end{equation}
the straight forward calculation leads the coordinate
representation of the metric $g_{ab}$, which is given by
\begin{eqnarray}
  \label{eq:gab-decomp-dd-minus}
  g_{ab} &=& - \alpha^{2} (dt)_{a} (dt)_{b}
  + q_{ij}
  (dx^{i} + \beta^{i}dt)_{a}
  (dx^{j} + \beta^{j}dt)_{b},
\end{eqnarray}
where $(dt,dx^{i})$ is the coordinate basis on ${\cal M}$ (more 
precisely, on an open set ${\cal U}\subset{\cal M}$), and
$q_{ij}$ is the inverse matrix of $q^{ij}$ in
Eq.~(\ref{eq:gab-decomp-uu-minus}), i.e.,
\begin{equation}
  q_{ij}q^{jk} =: q_{i}^{\;\;k} = \delta_{i}^{\;\;k}
\end{equation}
and $\delta_{i}^{\;\;k}$ is the $n$-dimensional Kronecker's delta.
In terms of the coordinate basis (\ref{eq:gab-decomp-dd-minus}), 
the unit normal vector $n^{a}$ is given by
\begin{eqnarray}
  n_{a} &=& - \alpha (dt)_{a}, \\
  n^{a} &=& g^{ab} n_{b} = \frac{1}{\alpha}
  \left(
    \frac{\partial}{\partial t}
    - \beta^{j} \frac{\partial}{\partial x^{j}}
  \right)^{a}.
\end{eqnarray}


Here, we summarize the components of the spacetime metric
$g_{ab}$ and inverse metric $g^{ab}$ on ${\cal M}$ as follows:
\begin{eqnarray}
  &&
  g_{tt} = - \alpha^{2} + q_{ij} \beta^{i} \beta^{j}, \quad
  g_{ti} = g_{i t} = q_{ij} \beta^{j} = \beta_{i}, \quad 
  g_{ij} = q_{ij},
  \\
  &&
  g^{tt} = - \frac{1}{\alpha^{2}}, \quad
  g^{ti} = g^{i t} = \frac{1}{\alpha^{2}} \beta^{i}, \quad
  g^{ij} = q^{ij} - \frac{1}{\alpha^{2}}\beta^{i}\beta^{j}. 
\end{eqnarray}


Next, we consider the connection between the covariant
derivative $\nabla_{a}$ on $({\cal M},g_{ab})$ and the covariant
derivative $D_{a}$ on $(\Sigma,q_{ab})$.
This correspondence is given from the Christoffel symbol
$\Gamma^{i}_{jk}$ associated with the metric $g_{ab}$ in the 
coordinate system (\ref{eq:gab-decomp-dd-minus}):
\begin{eqnarray}
  \Gamma^{t}_{tt} &=& \frac{1}{\alpha}\partial_{t}\alpha 
  + \frac{1}{\alpha}\beta^{i}D_{i}\alpha 
  + \frac{1}{2\alpha^{2}}\beta^{k}\beta^{i}\left\{
    \partial_{t} q_{ij} - 2 D_{(i}\beta_{j)}
  \right\}
  ,
  \label{eq:3+1-general-christoffel-minus-1}
  \\
  \Gamma^{t}_{i t} &=& \frac{1}{\alpha}D_{i}\alpha 
  + \frac{1}{2\alpha^{2}} \beta^{j}
  \left\{
    \partial_{t} q_{ij} - 2 D_{(i}\beta_{j)}
  \right\}
  ,
  \label{eq:3+1-general-christoffel-minus-2}
  \\
  \Gamma^{t}_{ij} &=& \frac{1}{2\alpha^{2}}
  \left\{
    \partial_{t} q_{ij} - 2 D_{(i}\beta_{j)}
  \right\}
  \label{eq:3+1-general-christoffel-minus-3}
  , \\ 
  \Gamma^{i}_{tt} &=& 
  - \frac{1}{2\alpha^{2}} \beta^{i}\beta^{k}\beta^{j}
  \left\{
    \partial_{t} q_{kj} - 2 D_{(k}\beta_{j)}
  \right\}
  - \frac{1}{\alpha} \beta^{i} \partial_{t}\alpha
  + q^{ij} \partial_{t}\beta_{j}
  \nonumber\\
  && \quad
  + \alpha D^{i}\alpha 
  - \beta^{k} D^{i} \beta_{k}
  - \frac{1}{\alpha} \beta^{i} \beta^{j} D_{j}\alpha, 
  \label{eq:3+1-general-christoffel-minus-4}
  \\ 
  \Gamma^{i}_{j t} &=& \frac{1}{2}
  q^{ik}
  \left\{
    \partial_{t} q_{kj} - 2 D_{(k}\beta_{j)}
  \right\}
  - \frac{1}{2\alpha^{2}} \beta^{i}\beta^{k} 
  \left\{
    \partial_{t} q_{kj} - 2 D_{(k}\beta_{j)}
  \right\}
  \nonumber\\
  && \quad
  - \frac{1}{\alpha} \beta^{i} D_{j}\alpha 
  + D_{j}\beta^{i},
  \label{eq:3+1-general-christoffel-minus-5}
  \\
  \Gamma^{i}_{jk} &=& 
  - \frac{1}{2\alpha^{2}} \beta^{i}
  \left\{
    \partial_{t} q_{kj} - 2 D_{(k}\beta_{j)}
  \right\}
  + {}^{(n)}\!\Gamma^{i}_{jk}, 
  \label{eq:3+1-general-christoffel-minus-6}
\end{eqnarray}
where ${}^{(n)}\!\Gamma^{i}_{jk}$ is the Christoffel
symbol associated with the metric $q_{ij}$:
\begin{eqnarray}
  \label{eq:n-dim-Christoffel}
  {}^{(n)}\!\Gamma^{i}_{jk}
  =
  \frac{1}{2} q^{il}\left(
    \partial_{j}q_{lk} + \partial_{k}q_{lj} - \partial_{l}q_{jk}
  \right)
  .
\end{eqnarray}
It is also convenient to introduce the extrinsic curvature
\begin{equation}
  \label{extrinsic_definition-2}
  K_{ab} = - q_{a}^{\;\;c} q_{b}^{\;\;d} \nabla_{c}n_{d}.
\end{equation}
In the coordinate system, on which the metric is given
by (\ref{eq:gab-decomp-dd-minus}), this extrinsic curvature is
given by 
\begin{equation}
  \label{eq:derivative_extrinsic_minus}
  K_{ij} = - \frac{1}{2\alpha}
  \left[\frac{\partial}{\partial t} q_{ij} -
  D_{i}\beta_{j} - D_{j}\beta_{i}\right].
\end{equation}
Through this component of the extrinsic curvature $K_{ij}$, the
above components
(\ref{eq:3+1-general-christoffel-minus-1})--(\ref{eq:3+1-general-christoffel-minus-6})
of the Christoffel symbols $\Gamma^{a}_{bc}$ are given by 
\begin{eqnarray}
  \Gamma^{t}_{tt} &=& \frac{1}{\alpha}\partial_{t}\alpha 
  + \frac{1}{\alpha}\beta^{i}D_{i}\alpha 
  - \frac{1}{\alpha}\beta^{k}\beta^{i}K_{ij}
  ,
  \label{eq:3+1-general-christoffel-minus-1-2}
  \\
  \Gamma^{t}_{i t} &=& \frac{1}{\alpha}D_{i}\alpha 
  - \frac{1}{\alpha} \beta^{j}K_{ij}
  ,
  \label{eq:3+1-general-christoffel-minus-2-2}
  \\
  \Gamma^{t}_{ij} &=& - \frac{1}{\alpha}K_{ij}
  , 
  \label{eq:3+1-general-christoffel-minus-3-2}
  \\ 
  \Gamma^{i}_{tt} &=& 
  \frac{1}{\alpha} \beta^{i}\beta^{k}\beta^{j} K_{kj}
  - \frac{1}{\alpha} \beta^{i} \partial_{t}\alpha
  + q^{ij} \partial_{t}\beta_{j}
  \nonumber\\
  && \quad
  + \alpha D^{i}\alpha 
  - \beta^{k} D^{i} \beta_{k}
  - \frac{1}{\alpha} \beta^{i} \beta^{j} D_{j}\alpha
  , 
  \label{eq:3+1-general-christoffel-minus-4-2}
  \\ 
  \Gamma^{i}_{j t} &=& - \alpha K^{i}_{\;\;j}
  + \frac{1}{\alpha} \beta^{i}\beta^{k} K_{kj}
  - \frac{1}{\alpha} \beta^{i} D_{j}\alpha 
  + D_{j}\beta^{i}
  ,
  \label{eq:3+1-general-christoffel-minus-5-2}
  \\
  \Gamma^{i}_{jk} &=& 
  \frac{1}{\alpha} \beta^{i} K_{kj}
  + {}^{(n)}\!\Gamma^{i}_{jk}
  \label{eq:3+1-general-christoffel-minus-6-2}
  .
\end{eqnarray}


\section{Covariant orthogonal decomposition of symmetric tensors}
\label{sec:S.Deser-1967-J.W.York.Jr-1973-J.W.York.Jr-1974}


Since the each order metric perturbation is regarded as a
symmetric tensor on the background spacetime 
$({\cal M}_{0},g_{ab})$ through an appropriate gauge choice, the
covariant decomposition of symmetric tensors is useful and
actually used in the main text. 
Here, we review of the covariant decomposition of symmetric
tensors of the second rank on an curved Riemannian manifold
based on the work by York\cite{J.W.York-1973-1974}.


In the generic curved Riemannian space $(\Sigma,q_{ab})$
($\dim\Sigma=n$), one can decompose an arbitrary vector or
one-form into its transverse and longitudinal parts as
\begin{eqnarray}
  \label{eq:J.W.York.Jr-1973-1-2-kouchan-1}
  A_{a} = A_{i}(dx^{i})_{a} = \left(D_{i}A_{(L)} + A_{(V)i}\right)(dx^{i})_{a},
  \quad
  D^{i}A_{(V)i} = 0,
\end{eqnarray}
where $D_{i}$ is the covariant derivative associated with the
metric $q_{ab}=q_{ij}(dx^{i})_{a}(dx^{j})_{b}$. 
$A_{(L)}$ is called the longitudinal part or the scalar part and
$A_{(V)i}$ is called the transverse part or vector part of the
vector field $A_{a}$ on $(\Sigma,q_{ab})$, respectively.


Moreover, this decomposition is not only covariant with respect
to arbitrary coordinate transformations, it is also orthogonal
in the natural global scalar product.
To clarify this orthogonality,
York\cite{J.W.York-1973-1974} introduced the inner
product for the vector fields on $\Sigma$.
This is, for any two vectors $V^{a}$ and $W^{a}$, we have
\begin{eqnarray}
  \label{eq:J.W.York.Jr-1973-1-2}
  \int_{\Sigma} \epsilon_{q} V^{a}W^{b}q_{ab},
\end{eqnarray}
where $\epsilon_{q}$ denotes the volume element which makes the
integral invariant and the integration extends over the entire
manifold $(\Sigma,q_{ab})$.
In terms of this inner product, the orthogonality of the vector
fields $V^{a}=D^{a}V_{(L)}:=q^{ab}D_{b}V_{(L)}$ and
$W^{a}=q^{ab}V_{(V)b}$ with $D^{a}V_{(V)a}=0$ is given by 
\begin{eqnarray}
  \int_{\Sigma} \epsilon_{q} D_{a}V_{(L)} V_{(V)b} q^{ab}
  = 
  \int_{\partial\Sigma} s_{a} V_{(L)} V_{(V)b} q^{ab}
  -
  \int_{\Sigma} \epsilon_{q} V_{(L)} D_{a}V_{(V)b} q^{ab}
  ,
  \label{eq:J.W.York.Jr-1973-1-2-kouchan-2}
\end{eqnarray}
where $s_{a}$ is the volume element of the $(n-1)$-dimensional
boundary $\partial\Sigma$ of $\Sigma$.
Since the second term of
Eq.~(\ref{eq:J.W.York.Jr-1973-1-2-kouchan-2}) vanishes due to
the condition $D^{a}V_{(V)a}=0$, the inner product $(V,W)$
vanishes if $V_{(L)}$ and $V_{(V)b}$ satisfy some appropriate
boundary conditions at the boundary $\partial\Sigma$ of $\Sigma$
so that the first term of
Eq.~(\ref{eq:J.W.York.Jr-1973-1-2-kouchan-1}) vanishes.
In this sense, the scalar part (the first term in
Eq.~(\ref{eq:J.W.York.Jr-1973-1-2-kouchan-1})) and the vector
part (the second term in
Eq.~(\ref{eq:J.W.York.Jr-1973-1-2-kouchan-1})) orthogonal to
each other.
Geometrically, the decomposition of $1$-forms, and more
generally $p$-forms, leads via de Rham's theorem to a
characterization of topological invariants of $\Sigma$ (i.e.,
Betti Numbers)\cite{G.DeRham-1960}.


In this appendix, it is assumed that the $n$-dimensional space 
$\Sigma$ is {\it closed} (compact manifolds without boundary)
following York's discussions. 
The choice of closed spaces is made for mathematical
convenience but the decomposition discussed here is also valid
for any other $n$-dimensional spaces $\Sigma$ with the boundary
$\partial\Sigma$ with some appropriate boundary conditions at
$\partial\Sigma$.
Through this assumption, in this appendix, we consider the
TT-decomposition (transverse traceless decomposition) of a
symmetric tensor $\psi^{ab}$ on $\Sigma$, which is defined by 
\begin{eqnarray}
  \label{eq:J.W.York.Jr-1973-2-2}
  \psi^{ab} = \psi_{TT}^{ab} + \psi_{L}^{ab} + \psi_{Tr}^{ab},
\end{eqnarray}
where the longitudinal part is 
\begin{eqnarray}
  \label{eq:J.W.York.Jr-1973-3-2}
  \psi_{L}^{ab} := D^{a}W^{b} + D^{b}W^{a} -
  \frac{2}{n} q^{ab} D_{c}W^{c}
  =: (LW)^{ab}
\end{eqnarray}
and the trace part is
\begin{eqnarray}
  \label{eq:J.W.York.Jr-1973-4-2}
  \psi_{Tr}^{ab} := \frac{1}{n} \psi q^{ab},
  \quad
  \psi := q_{cd} \psi^{cd}.
\end{eqnarray}


Let us suppose that both an arbitrary symmetric tensor field
$\psi^{ab}$ and the metric $q_{ab}$ are $C^{\infty}$ tensor
fields on $\Sigma$.
First, we define $\psi_{TT}^{ab}$ in accordance with
Eq.~(\ref{eq:J.W.York.Jr-1973-2-2}) by
\begin{eqnarray}
  \label{eq:J.W.York.Jr-1973-7-2}
  \psi_{TT}^{ab} := \psi^{ab} - \frac{1}{n} \psi g^{ab} - (LW)^{ab}.
\end{eqnarray}
We note that the tensor $\psi_{TT}^{ab}$ is traceless, i.e., 
\begin{eqnarray}
  \label{eq:J.W.York.Jr-1973-8-2}
  q_{ab}\psi_{TT}^{ab} = 0
\end{eqnarray}
by its construction (\ref{eq:J.W.York.Jr-1973-7-2}).
Further, we require the transversality on the tensor field
$\psi_{TT}^{ab}$, i.e., 
\begin{eqnarray}
  \label{eq:J.W.York.Jr-1973-9-2}
  D_{b}\psi_{TT}^{ab} = 0.
\end{eqnarray}
Equation (\ref{eq:J.W.York.Jr-1973-9-2}) leads to a covariant
equation of the vector field $W^{a}$ in
Eq.~(\ref{eq:J.W.York.Jr-1973-7-2}) as 
\begin{eqnarray}
  \label{eq:J.W.York.Jr-1973-10-2}
  D_{a}(LW)^{ab}
  =
  D_{a}\left(\psi^{ab} - \frac{1}{n} \psi q^{ab}\right).
\end{eqnarray}
The explicit expression of (\ref{eq:J.W.York.Jr-1973-10-2}) is
given by 
\begin{eqnarray}
  \label{eq:J.W.York.Jr-1973-10-2-explicit}
  {\cal D}^{bc} W_{c}
  =
  D_{a}\left(\psi^{ab} - \frac{1}{n} \psi q^{ab}\right),
\end{eqnarray}
where the derivative operator ${\cal D}^{bc}$ is defined by 
\begin{eqnarray}
  \label{eq:J.W.York.Jr-1973-1-2-kouchan-3}
  {\cal D}^{bc}
  :=
  q^{bc}\Delta + \left(1 - \frac{2}{n}\right) D^{b}D^{c} +
  R^{bc}
  , \quad
  \Delta := D^{a}D_{a}
  .
\end{eqnarray}


The basic properties of
Eq.~(\ref{eq:J.W.York.Jr-1973-10-2-explicit}) are also discussed
by York\cite{J.W.York-1973-1974}. 
The operator ${\cal D}^{ab}$ defined by
Eq.~(\ref{eq:J.W.York.Jr-1973-1-2-kouchan-3}) is linear and
second order by its definition.
As discussed by York, this operator is strongly elliptic,
negative-definite, self-adjoint, and its ``harmonic'' functions
are always orthogonal to the source (right-hand side) in
Eq.~(\ref{eq:J.W.York.Jr-1973-10-2-explicit}).
Here, ``harmonic'' functions of ${\cal D}^{ab}$ means functions
which belong the kernel of the operator ${\cal D}^{ab}$.
Moreover, he showed that
Eq.~(\ref{eq:J.W.York.Jr-1973-10-2-explicit}) will always
possess solutions $W^{a}$ which is unique up to conformal Killing
vectors.
Due to these situation, in this paper, we assume that the Green
function $({\cal D}^{-1})_{ab}$ defined by 
\begin{eqnarray}
  \label{eq:J.W.York.Jr-1973-kouchan-12}
  ({\cal D}^{-1})_{ab}{\cal D}^{bc}
  = {\cal D}^{bc}({\cal D}^{-1})_{ab}
  = \delta_{a}^{\;\;c}
\end{eqnarray}
exists through appropriate boundary conditions at the boundary
$\partial\Sigma$ of $\Sigma$.  
Although York's discussions are for the case of the {\it closed}
space $\Sigma$, we review his discussions here.
In this review, we explicitly write the boundary terms which are
neglected by the closed boundary condition to keep the
extendibility to non-closed $\Sigma$ case of discussions in our
mind.


The ellipticity of an operator depends only upon its 
{\it principal part}, i.e., the highest derivatives acting on
the unknown quantities which it contains.
To see the ellipticity of an operator, we consider the
replacement of the each derivative operator $D_{a}$ occurring in 
its principal part by an arbitrary vector $V_{a}$.
Through this replacement, the principal part of the operator
defines a linear transformation ${\bf \sigma}_{v}$.
The operator is said to be elliptic if ${\bf \sigma}_{v}$ is an
isomorphism\cite{M.Berger-D.Ebin-1969}.
In the present case, 
\begin{eqnarray}
  \label{eq:J.W.York.Jr-1974-2.4}
  \left[{\bf \sigma}_{v}({\cal D})\right]^{ab}
  = V^{b}V^{a} + q^{ab}V_{c}V^{c}.
\end{eqnarray}
Here, ${\bf \sigma}_{v}$ operates on vector $X_{a}$ and defines
a vector-space isomorphism when the determinant of 
${\bf \sigma}_{v}$ is non-vanishing for all non-vanishing
$V^{a}$.
The fact that $\det{\bf \sigma}_{v}\neq 0$ here is verified, for
example, by choosing  
$V^{a}=\left(\partial/\partial x^{\mu}\right)^{a}$ in a local 
Cartesian frame $\{x^{\mu}\}$.
The operator is said to be {\it strongly elliptic} if all the
eigenvalues of ${\bf \sigma}_{v}$ are nonvanishing and have the 
same sign.
This is easily checked and ${\cal D}^{ab}$ is strongly elliptic.


To show that ${\cal D}^{ab}$ is negative definite, we consider
the inner product (\ref{eq:J.W.York.Jr-1973-1-2}) of the vector
field ${\cal D}W^{a}:={\cal D}^{ab}W_{b}$ and $W^{a}$:
\begin{eqnarray}
  \int_{\Sigma} \epsilon_{q} q_{ab} W^{a}{\cal D}^{bc}W_{c}
  &=&
  \int_{\Sigma} \epsilon_{q} q_{ab} W^{a} \left(
    D_{c}(LW)^{bc}
  \right)
  \nonumber\\
  &=&
  \int_{\Sigma} \epsilon_{q}
  \left\{
    D_{c}\left(
      W_{b} (LW)^{bc}
    \right)
    - \frac{1}{2} (LW)_{bc} (LW)^{bc}
  \right\}
  \nonumber\\
  &=&
  \int_{\partial\Sigma} s_{c} W_{b} (LW)^{bc}
  - \frac{1}{2}
  \int_{\Sigma} \epsilon_{q}
   (LW)_{bc} (LW)^{bc}
  \label{eq:J.W.York.Jr-1973-13-2}
  ,
\end{eqnarray}
where we use the fact that the tensor $(LW)^{bc}$ is symmetric
and traceless.
Eq.~(\ref{eq:J.W.York.Jr-1973-13-2}) shows that the operator
${\cal D}^{ab}$ has the negative eigenvalues in the case where
the first term (boundary term) in
Eq.~(\ref{eq:J.W.York.Jr-1973-13-2}) is neglected, unless
$(LW)^{bc}=0$. 
The self-adjointness of the operator ${\cal D}^{ab}$ is follows
from a similar argument in which one integrates by parts twice:
\begin{eqnarray}
  \int_{\Sigma} \epsilon_{q} q_{ab} V^{a}({\cal D}W)^{b}
  &=&
  \int_{\Sigma} \epsilon_{q} q_{ab} V^{a}\left(
    D_{c}(LW)^{bc}
  \right)
  \nonumber\\
  &=&
  \int_{\Sigma} \epsilon_{q} \left[
    D_{c}\left( q_{ab} V^{a} (LW)^{bc} \right)
    - \nabla_{c}V_{b}(LW)^{bc}
  \right]
  \nonumber\\
  &=&
  \int_{\Sigma} \epsilon_{q} \left[
    D_{c}\left( q_{ab} V^{a} (LW)^{bc} \right)
    - \frac{1}{2}\left(LV\right)^{bc}\left(LW\right)_{bc}
  \right]
  \nonumber\\
  &=&
  \int_{\Sigma} \epsilon_{q} \left[
    D_{c}\left( q_{ab} V^{a} (LW)^{bc} \right)
    - \left(LV\right)^{bc}D_{c}W_{b}
  \right]
  \nonumber\\
  &=&
  \int_{\Sigma} \epsilon_{q} \left[
    D_{c}\left( q_{ab} V^{a} (LW)^{bc} \right)
    - D_{c}\left(\left(LV\right)^{bc}W_{b}\right)
    + W_{b} \nabla_{c}\left(LV\right)^{bc}
  \right]
  \nonumber\\
  &=&
  \int_{\Sigma} \epsilon_{q} \left[
    D_{c}\left( q_{ab} V^{a} (LW)^{bc} \right)
    - D_{c}\left(\left(LV\right)^{bc}W_{b}\right)
    + W_{b} {\cal D}^{bc}V_{c}
  \right]
  \nonumber\\
  &=&
  \int_{\partial\Sigma} s_{c} \left[
    V_{b} (LW)^{bc} - \left(LV\right)^{bc}W_{b}
  \right]
  +
  \int_{\Sigma} \epsilon_{q} W_{b} {\cal D}^{bc}V_{c}
  \label{eq:J.W.York.Jr-1973-15-2}
\end{eqnarray}
for any vectors $V$ and $W$, where we use the fact that the
tensor $(LW)^{ab}$ and $(LV)^{ab}$ are symmetric and traceless. 
Eq.~(\ref{eq:J.W.York.Jr-1973-15-2}) shows that the operator
${\cal D}^{ab}$ is self-adjoint if the first term (boundary
term) in Eq.~(\ref{eq:J.W.York.Jr-1973-15-2}) is neglected.


When we can neglect the boundary terms in
Eq.~(\ref{eq:J.W.York.Jr-1973-13-2}), the right-hand side of
(\ref{eq:J.W.York.Jr-1973-13-2}) can vanish only if $(LW)^{ab}=0$.
This means either $W^{a}=0$ or $W^{a}$ is a conformal Killing
vector (or Killing vector) of the metric $q_{ab}$.
The condition for a conformal Killing vector is, of course, not
satisfied for an arbitrary metric but this is given by
\begin{eqnarray}
  \label{eq:J.W.York.Jr-1973-16-2}
  {\pounds}_{W}q_{ab} = \lambda q_{ab}
\end{eqnarray}
for some scalar function $\lambda$, where ${\pounds}_{W}$
denotes the Lie derivative along $W$.
Taking the trace of both sides, we find
\begin{eqnarray}
  \label{eq:J.W.York.Jr-1973-18-2}
  \lambda = \frac{2}{n} \nabla_{c}W^{c}.
\end{eqnarray}
Therefore, $W^{a}$ is a conformal Killing vector if and only if
\begin{eqnarray}
  \label{eq:J.W.York.Jr-1973-19-2}
  \nabla^{a}W^{b} + \nabla^{b}W^{a}
  - \frac{2}{3} q^{ab} \nabla_{c}W^{c}
  \equiv (LW)^{ab} = 0.
\end{eqnarray}
It follows that the only nontrivial solutions of 
${\cal D}^{ab}W_{b}=0$ are conformal Killing vectors if they
exist.
Hence the nontrivial ``harmonic'' functions of ${\cal D}^{ab}$
are conformal Killing vectors.
We shall now show that even if these ``harmonic'' solutions
exist, they are always orthogonal to the right-hand side of
(\ref{eq:J.W.York.Jr-1973-10-2}) and, hence, can cause no
difficulties in solving equation (\ref{eq:J.W.York.Jr-1973-10-2})
by an eigen function expansion.


Denote the conformal Killing vectors by $W^{a}=C^{a}$, where by
definition $(LC)^{ab}=0$.
Form the scalar product of the right-hand side of
(\ref{eq:J.W.York.Jr-1973-10-2}) with $C$ and integrate by parts
to find
\begin{eqnarray}
  &&
  \int_{\Sigma} \epsilon_{q}
  q_{ac}
  D_{b}\left(\psi^{ab}-\frac{1}{n}q^{ab}\psi\right) C^{c}
  \nonumber\\
  &=&
  \int_{\Sigma} \epsilon_{q}
  \left[
    D_{b}\left(
      q_{ac} \left(\psi^{ab}-\frac{1}{n}q^{ab}\psi\right) C^{c}
    \right)
    -
    \left(\psi^{ab}-\frac{1}{n}q^{ab}\psi\right) D_{b}C_{a}
  \right]
  \nonumber\\
  &=&
  \int_{\partial\Sigma} s_{b} \left(\psi^{ab}-\frac{1}{n}q^{ab}\psi\right) C_{a}
  - \frac{1}{2} \int_{\Sigma} \epsilon_{q}
  \left(\psi^{ab}-\frac{1}{n}q^{ab}\psi\right) (LC)_{ab}
  =
  0
  ,
  \label{eq:J.W.York.Jr-1973-20-2}
\end{eqnarray}
where we use the fact that $\psi^{ab}-\frac{1}{n}q^{ab}\psi$ is
symmetric and traceless and we also neglect the boundary term.
Hence the source in
Eq.~(\ref{eq:J.W.York.Jr-1973-10-2-explicit}) is in the domain
of $({\cal D}^{-1})^{ab}$ and $({\cal D}^{-1})^{ab}$ gives the
solution to Eq.~(\ref{eq:J.W.York.Jr-1973-10-2-explicit}) even
in the presence of conformal symmetries.


These results also show that the solution to
Eq.~(\ref{eq:J.W.York.Jr-1973-10-2-explicit}) must be unique up
to conformal Killing vector fields.
Since only $(LW)^{ab}$ enters in the definition
(\ref{eq:J.W.York.Jr-1973-7-2}) of $\psi_{TT}^{ab}$, conformal
Killing vectors cannot affect $\psi_{TT}^{ab}$.


The orthogonality of $\psi_{TT}^{ab}$, $(LW)^{ab}$, and
$\frac{1}{n}\psi q^{ab}$ is easily demonstrated.
We see readily that $\frac{1}{n}\psi q^{ab}$ is pointwise
orthogonal to $(LW)^{ab}$ and to $\psi_{TT}^{ab}$, as
$(LW)^{ab}$ and $\psi^{ab}_{TT}$ are both trace-free.
To show that $\psi_{TT}^{ab}$ and $(LV)^{ab}$ are orthogonal for
any vector $V$ and any TT tensor, we have only to show that 
\begin{eqnarray}
  \int_{\Sigma} \epsilon_{q} q_{ac}q_{bd} (LW)^{ab}\psi_{TT}^{cd}
  &=&
  \int_{\Sigma} \epsilon_{q} 2 D_{a}W_{b}\psi_{TT}^{ab}
  \nonumber\\
  &=&
  \int_{\Sigma} \epsilon_{q} \left(
    D_{a}\left(2 W_{b}\psi_{TT}^{ab}\right)
    - 2 W_{b} D_{a}\psi_{TT}^{ab}
  \right)
  \nonumber\\
  &=&
  \int_{\partial\Sigma} s_{a}\left(2 W_{b}\psi_{TT}^{ab}\right)
  - \int_{\Sigma} \epsilon_{q} \left(
    2 W_{b} D_{a}\psi_{TT}^{ab}
  \right)
  =
  0
  ,
  \label{eq:J.W.York.Jr-1973-20.5}
\end{eqnarray}
where we use the fact that the tensor $\psi_{TT}^{ab}$ is
symmetric, traceless, and transverse
(\ref{eq:J.W.York.Jr-1973-9-2}).
We also neglect the boundary term in
Eq.~(\ref{eq:J.W.York.Jr-1973-20.5}).
Thus, we conclude that the decomposition defined by
(\ref{eq:J.W.York.Jr-1973-7-2}) exists, is unique, and is
orthogonal.


One can further decompose the vector $W^{a}$ uniquely into 
its transverse and longitudinal parts with respect to the metric
$q_{ab}$. 
This splitting is orthogonal, as in
Eq.~(\ref{eq:J.W.York.Jr-1973-1-2-kouchan-1}).


Since the above discussions are for closed spaces $\Sigma$,
careful discussions on the boundary terms which are neglected in 
the closed $\Sigma$ is necessary if we extend the above
arguments to non-closed $\Sigma$ case.
However, we do not go into these detailed issues.
Instead, in the main text, we assume that the existence of the
Green function of the derivative operator ${\cal D}^{ab}$ and
use the transverse-traceless decomposition for an arbitrary
symmetric tensor on $\Sigma$ discussed here.



\end{document}